\documentclass[onecolumn]{aastex63}

\usepackage{rotating}
\usepackage[utf8]{inputenc}
\usepackage[a4paper, total={6in, 8in}]{geometry}
\usepackage{natbib}
\usepackage{longtable}
\usepackage[page]{appendix}
\newcommand{\ksec}[1]{\, \\ \textit{#1 \\ \, \\}}

\usepackage{amsmath}
\usepackage{bm}
\usepackage{mathtools}

\usepackage{graphicx}
\usepackage{esvect}

\def\aap{Astro. Astrophys.}
\def\apjl{Astrophys. J. Let.}

\def\apjs{Astrophys. J., Supp.}

\def\deg{$^\circ$}

\shorttitle{Properties of Kepler Planets}
\shortauthors{Judkovsky et al.}

\begin{document}
\received{August 30, 2021}

\title{Physical Properties and Impact Parameter Variations of Kepler Planets from Analytic Light Curve Modeling}

\correspondingauthor{Yair Judkovsky}
\author[0000-0003-2295-8183]{Yair Judkovsky}
\affiliation{Weizmann Institute of Science, Rehovot 76100 Israel}
\email{yair.judkovsky@weizmann.ac.il}

\author[0000-0002-9152-5042]{Aviv Ofir}
\affiliation{Weizmann Institute of Science, Rehovot 76100 Israel}

\author[0000-0001-9930-2495]{Oded Aharonson}
\affiliation{Weizmann Institute of Science, Rehovot 76100 Israel}
\affiliation{Planetary Science Institute, Tucson, AZ, 85719-2395 USA}


\graphicspath{{./}{Figures/}}


\begin{abstract}
We apply \texttt{AnalyticLC}, an analytic model described in an accompanying paper, to interpret {Kepler} data of systems that contain two or three transiting planets.
We perform tests to verify that the obtained solutions agree with full N-body integrations, and that the number of model parameters is statistically justified. We probe non-coplanar interactions via impact parameter variations (TbVs), enabled by our analytic model. The subset of systems with a valid solution includes 54 systems composed of 140 planets, more than half of which without previously reported mass constraints. Overall we provide: (1) Estimates on physical and orbital properties for all systems analyzed. (2) 102 planets with mass detections significant to better than $3\,\sigma$, 43 of which are lighter than $5\,m_\oplus$. (3) 35 TbVs significant to better than $3\,\sigma$. We focus on select systems showing strong TbVs, which can result from either interaction among the known transiting planets, or with a non-transiting object, and provide: (4) a method to constrain the parameters of such unseen companions. These results are enabled by an accurate, 3D, photodynamical model, of a kind expected to become increasingly important for modeling multi-decade photometric and composite (RV, astrometry) data sets.
\end{abstract}


\section{Introduction} \label{sec:Introduction}
Orbital and physical properties of exoplanets are of great interest not only because of the richness of dynamical phenomena they embody, but also because of their relevance to theories of planetary formation, evolution, and habitability. 
While the first detections of exoplanets used the radial velocity (RV) method \citep{Latham1989,Mayor1995} or pulsar timing variations \citep{Wolszczan1992}, the {Kepler} mission \citep{BoruckiEtAl2010} provided a leap in the number of identified exoplanets, using the transit method. The  vast space-based photometric dataset started by {Kepler} is continuously expanded by further missions, including the current Transiting Exoplanet Survey Satellite (TESS) \citep{RickerEtAl2010} and the future PLAnetary Transits and Oscillations of stars (PLATO) \citep{Rauer2014} missions.

Systematic study of individual systems data provides a broader view of large-scale phenomena that sculpt the structure of planetary systems; a few examples include the radius gap \citep{Fulton2017}, which is related to photo-evaporation \citep{Owen2017} or to core-driven mass loss \citep{Ginzburg2018, Loyd2020}; the so-called "Neptune desert", which might be related to different evolution and migration paths of Jovian planets and super-Earths \citep{Mazeh2016}; the tendency of planet pairs to deviate from resonances due to dissipation \citep{LithwickWu2012a, Millholland2019}; and the "peas in a pod" pattern in orbital spacings and planetary sizes \citep{Weiss2018a}, which may be related to the similar environment in which those planets were formed. Let us review the methods that enabled inverting the observational data to planetary physical and orbital properties.

The most direct way to invert data for physical and orbital parameters is to run multiple N-body integrations guided by a non-linear search algorithm; this has been done for RV data \citep[e.g.][]{RiveraEtAl2005} and for photometric data  \citep[e.g.][]{MillsFabrycky2017, MillsEtAl2019, FreudenthalEtAl2018, GrimmEtAl2018}. This method is highly expensive in calculation time, making this a cumbersome, sometimes prohibitive, approach. 

Other techniques use analytic formulae for TTVs \citep{LithwickXieWu2012, HaddenLithwick2016, DeckAgol2015}, or modal decomposition that enables inversion to masses and eccentricities \citep{LinialGilbaumSari2018}. Fitting for TTV data was used to extract masses and eccentricities (or combinations thereof) for a large number of planets, e.g. in \citet{HaddenLithwick2016, HaddenLithwick2017, JontofHutter2021} (hereafter HL17, JH21 for the latter two), and many others.

While in many cases fitting times-of-mid-transit can yield adequate estimates of planetary properties, this method has drawbacks. 
Fitting transit times rather than flux does not fully exploit the information encoded in the light-curve. Many degrees of freedom are required in order to translate the light-curve to individual transit times, and the information is only partially preserved. Fitting transit times also induces a bias that favours large planets (of clear transits with well-defined mid-transit times) with strong TTV signals. \citet{OfirEtAl2018} highlights the benefits of a global fit that utilizes a full light-curve model; such an approach was used by \citet{Yoffe2021} to perform a global light-curve fit, using \texttt{TTVFaster} \citep{DeckAgol2015} as the dynamical model. This allowed placing new constraints on the masses and eccentricities of several previously known planetary systems.

Fitting transit times rather than flux reduces the light-curve data to mid-transit-times alone, erasing valuable information encoded in other types of transit variations, such as depth and/or duration. These are of great interest, as they can probe non-coplanar interactions within the system.  Detection of mutual inclination among planets is limited in both transit and RV, and hence in many cases statistical methods are used to make inferences on the {\it dispersion} of mutual inclinations rather than individual values, e.g. in \citet{FabryckyEtAl2014}, and \citet{XieEtAl2016}. If we would be able to probe mutual inclinations of individual systems, or at least identify systems that are likely to contain mutual inclination, this may shed light on the relations between smaller, inner, rocky planets, and external giants that can stir the structure of the inner system \cite{LaiPu2017}. The small number statistics of systems with both external and inner transiting planets show that mutual inclinations within such systems tends to be lower when the inner system contains multiple planets \citep{Masuda2020}; thus, additional examples are needed. A recent study by \citet{Millholland2021} established a link between the number of detected TDVs in the catalog of \citet{Shahaf2021} and the mutual inclination dispersion of planetary systems. This study showed that an AMD \citep[Angular Momentum Deficit,][]{Laskar2017} based, non-dichotomous model, is favored over a dichotomous bi-modal inclinations distribution, which was suggested in order to explain the singly-transiting-biased multiplicity distribution \citep[e.g.][]{Lissauer2011, BallardJohnson2016, XieEtAl2016}. These examples highlight the importance of extending the reach of TbVs and/or TDVs detection in order to advance understanding of the nature of mutual inclinations in planetary systems.

In a preceding paper \citep{Judkovsky2021a}, we described \texttt{AnalyticLC}, a analytic modeling tool we developed to enable accurate global light-curve modeling without the need for full N-body integration, while incorporating forces out of the plane. In this paper, we describe the usage of \texttt{AnalyticLC} in fitting {Kepler} photometric data. In \S\ref{sec:Methods} we describe the data reduction, the fitting process and the tests performed to check the physical sanity and statistical significance of the fitted models. In \S\ref{sec:Results}, we describe the overall results, with emphasis on planetary masses, densities and impact parameter variations. In \S\ref{sec:Individual}, we describe individual results for select systems, and compare them to former literature data. In \S\ref{sec:IndividualNoSolution} we describe notable systems for which the model did not succeed to fit a valid solution, but that deserve attention in future studies. In \S\ref{sec:Discussion} we summarize our findings and discuss future prospects, and in \S\ref{appendix} we provide all the results of the individual planets in tables, along with the literature stellar data used for this work. 

\section{Methods}\label{sec:Methods}

\subsection{Data Reduction}

A detrended light curve is obtained by applying a cosine filter based on \citet{Mazeh2010} to the latest version of the Presearch Data Conditioning (PDC) Maximum A Posteriori  (MAP) {Kepler} flux data \citep{Stumpe2014} \citep[which follows previous MAP versions:][]{Stumpe2012, Smith2012}. The cosine filter is modified relative to \citet{Mazeh2010} such that the shortest Fourier component is four times the duration of the transit as provided by NASA exoplanets archive\footnote{\url{https://exoplanetarchive.ipac.caltech.edu/}} (NExScI), and not twice the orbital period, due to different science cases (we are not interested in photometric variability on the scale of the orbital period, so we can filter more strongly). Outliers beyond $5\,\sigma$ are iteratively rejected from the cosine filter calculation.

Since the detrending and model fitting may affect each other, we perform an iterative detrending: after obtaining a best-fitting model on the detrended light-curve, we used that model to remove the transit signals from the raw flux and recompute the detrending step again. The fitting is then also repeated with an initial guess on the model parameters obtained from the parameter distribution in the previous iteration. This iterative process is taken to have converged when the improvement in $\chi^2$ score over the previous iteration is less than the 1$\,\sigma$ equivalent (the 68th percentile of the $\chi^2({\rm DOF})$ distribution, where DOF is the number of degrees of freedom in the fitted model). Visual inspection of the results from multiple systems shows this process 
is robust, and that convergence was achieved both from the point of view of light-curve detrending and  posterior distributions of the parameters. The method of fitting a model and generating a sample distribution is described below.

The fitting was performed on all available long and short cadence {Kepler} data. Binning was applied to the (instantaneous) model for the long cadence data points, where the number of points for binning was set by using the formulae given by \citet{Kipping2010a} and requiring that the error caused by binning would be at most ten times smaller than the typical data uncertainty. The iterative outliers rejection described above was performed separately on the long and short cadence data. 

For the calculation of the model flux we used \texttt{AnalyticLC}, which in turn applies the  formulae of \citet{MandelAgol2002} on the planet positions derived from dynamics, using the limb-darkening parameters from NExScI.

\subsection{Model Parameters}\label{sec:ModelParameters}
\begin{figure}[h]
    \includegraphics[width=1\linewidth]{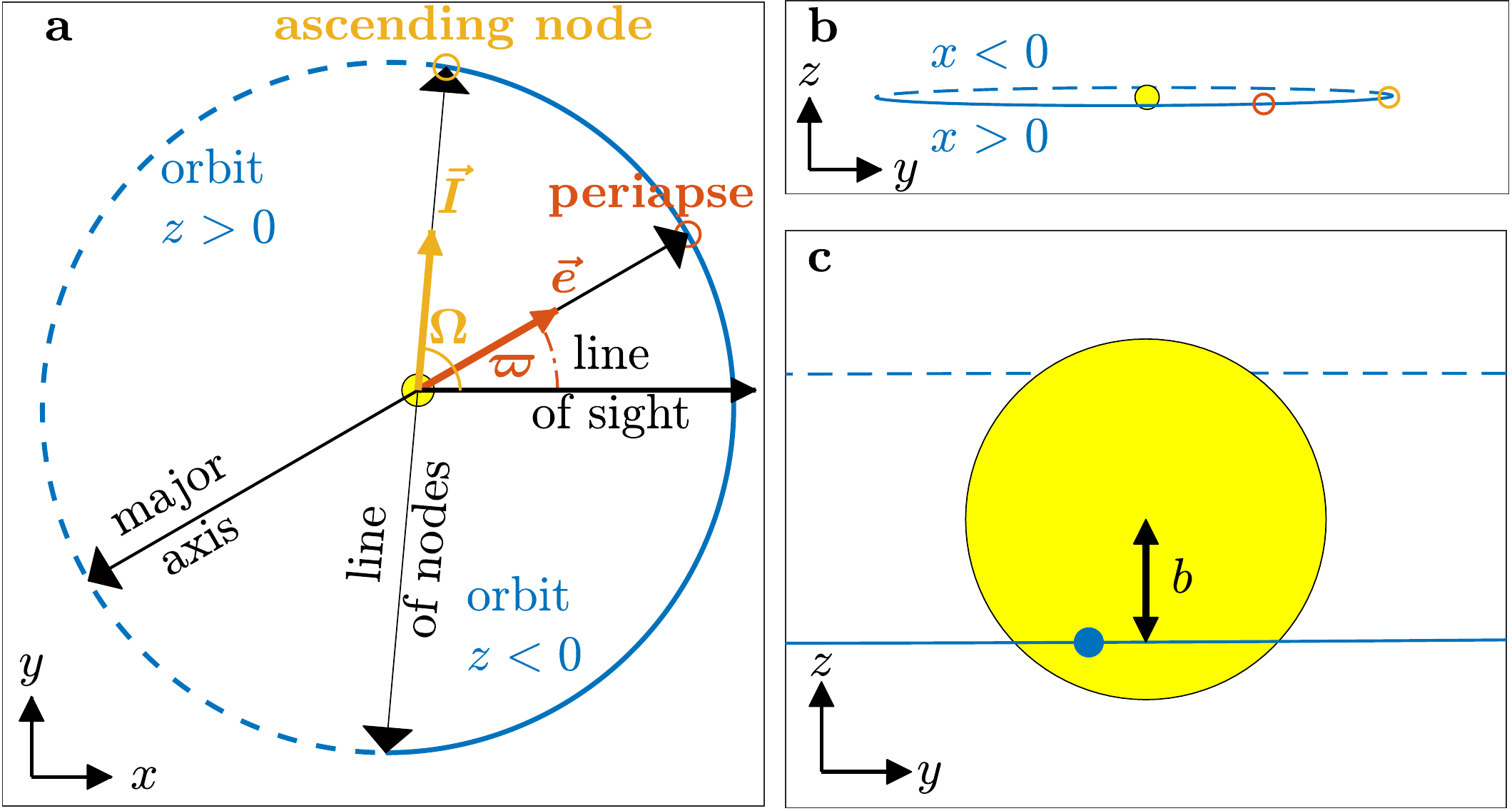}
    \caption{Illustration of some of the model parameters. (a) The vectors $\vec{e}, \vec{I}$ point at the periapse and at the longitude of ascending node, correspondingly. Their components $e_x, e_y, I_x, I_y$ are their projections on the $x$ and $y$ axes. The angle $\varpi$ is a compound angle - the sum of $\Omega$ and the argument of periapse $\omega$ which are not at the same plane. For small inclination values it is approximately the azimuth of the periapse on the $xy$ plane with respect to the line of sight. (b) The same orbit, as seen from the observer point of view. At the dashed part the orbit goes behind the star, at the solid part - before the star. The periapse and ascending node are shown as in (a). (c) is a zoom-in of (b), showing the transiting planet with the impact parameter $b$. \label{fig:OrbitalElementsIllustration}}
\end{figure}

The model includes eight parameters per planet: $P$ (orbital period), $T_{\rm mid0}$ (reference time of mid-transit), $R_{\rm p}/R_*$ (planet-to-star radius ratio), $\mu$ (planet-to-star mass ratio), $\Delta e_x=\Delta(e\cos{\varpi}), \Delta e_y=\Delta(e\sin{\varpi}), I_x=I\cos{\Omega},I_y=I\sin{\Omega}$, where $e$ is the orbital eccentricity, $\varpi$ is the longitude of periapse, $I$ is the orbital inclination and $\Omega$ is the longitude of ascending node. The $\Delta$ notation here indicates that the fitted parameter is not the value of each planet's eccentricity component, but rather the difference in eccentricity components relative to the previous planet inwards, towards the star. This choice was already used in \citet{Yoffe2021}, and is based on the recognition that that TTVs depend approximately on these differences, as opposed to the value of each eccentricity component \citep{LithwickXieWu2012}. For the innermost planet, $\Delta e_x=e_x$, and $\Delta e_y=e_y$. 
In our coordinate system, $x$ points towards the observer and $y,z$ form the sky plane, such that the orbit is inclined with respect to the $xy$ plane with a small inclination angle $I$ between its normal and the $z$ axis. The longitude of ascending node $\Omega$ is measured with respect to $x$. 
In Figure~\ref{fig:OrbitalElementsIllustration} some of the model parameters are graphically illustrated.

\subsection{Non-Linear Fitting Method} 

We fit a model to the light-curve and generate a posterior distribution using our implementation of DE-MCzs (Differential-Evolution Markov-Chain with Snooker update) \citep{BraakVrugt2008}, driving the model function \texttt{AnalyticLC}, which is described in \citet{Judkovsky2021a}. DE-MC is suitable for multi-dimensional parameters with correlations among them, a typical situation in our problem. 

The prior is uniform in all the model parameters specified above, with eccentricity components difference limited between -0.6 and 0.6, and inclination components to -50\deg{} to 50\deg{}; in practice, the solutions usually converged to eccentricity differences of a few per cent and inclinations of a few degrees. 

The initial states of the walkers in orbital periods, reference times of mid-transit and planet-to-star radius ratio and planet-to-star mass ratio are distributed based on values and errors from NExScI.  if no planetary mass is given, we translate the archive planetary radius to an initial guess on mass by using an empirical formula given by \citet[][Fig.~8]{Weiss2018}.

We ran the non-linear fitting procedure five times per system in order to check if it converges to the same minimum consistently, using a different realization of the initial states at each run. We empirically found that within five runs, for most cases the results are consistent. We also note that each run contains 48 walkers (for 2-planet systems) or 80 walkers (for 3-planet systems), and tens of thousands of generations, generating samples that span the parameters space.
We therefore consider five to be sufficient in the majority of cases. In \S~\ref{sec:Individual} we discuss the results system-by-system, and note cases with ambiguous solutions that we believe require further investigation.

For the innermost planet, $I_x$ was fixed to 0, as there is no preference for any direction on the sky plane, and we are free to fix this parameter. Similarly to \citet{OfirDreizler2013}, the innermost planet has another parameter, $a^{(1)}/R_*$, which is the semi-major axis in units of stellar radius; the semi-major axes of the other planets are obtained from their orbital periods using Kepler's law. The initial guess of $a^{(1)}/R_*$ was obtained from the literature stellar mass and planetary orbital period; whenever possible (for the decisive majority of planets) we used stellar data from \citet{Berger2020}. In a handful of cases, we used \cite{Fulton2018}, or NExScI.

The number of walkers is chosen to be close to $N\log{N}$ where $N$ is the number of fitted parameters. There is no general prescription for the optimal number of walkers; our choice is based on experimentation varying the number of walkers when running DE-MCzs on different types of non-linear problems. The runs were parallelized on model evaluations across the walkers; for parallelization efficiency on our specific cluster we rounded the number of walkers to 48 for two planets systems (16 parameters) and 80 for three planets systems (24 parameters). Convergence was declared upon reaching a Gelman-Rubin criterion \citep{Gelman1992,Brooks1998} of less than 1.2 for all parameters over 10,000 generations, excluding the orbital periods, which are known to a high degree of accuracy. After removal of burn-in we were left with a well-mixed sample, from which the statistical inference of the parameters was done. The parameters values described in the tables at \ref{appendix} are the median of the samples and the difference between the median and the 15.865 and the 84.135 percentiles, corresponding to $1 \,\sigma$ in a normal distribution.

As described above, the fitting process was iterated to improve the raw flux detrending.

\subsection{Verification vs. N-body Integration}\label{sec:Nbody}
\texttt{AnalyticLC} constructs a light curve based on variations that result from an approximate solution of Lagrange's equations (further detail in \citet{Judkovsky2021a}). The assumptions that the method is based on are appropriate for a significant portion of {Kepler} planetary systems (stable systems, out of resonance, with moderate eccentricities). After finding a best-fitting values and a posterior distribution, we wish to ensure the accuracy of the calculation in the parameter space around the solution. To do this, we perform an N-body integration using \texttt{Mercury6} \citep{Chambers1999} at the best fitting parameters set, evaluate the flux values at times when data was taken, and calculate the flux differences between the model generated by \texttt{AnalyticLC} and the model generated by the N-body integration. The sum of squares of these differences, normalized by the measurement errors of the data, provides a $\chi^2$-like estimate of the mismatch between the analytic and the N-body model at the best fit. We defined this quantity as

\begin{equation}
    \chi^2_{\rm N-body}=\sum_{t}\frac{(F_{\rm AnalyticLC}(t)-F_{\rm N-body}(t))^2}{dF(t)^2},
\end{equation}

where $F_{\rm AnalyticLC}(t)$ and $F_{\rm N-body}(t)$ are the flux values obtained from \texttt{AnalyticLC} and from the N-body integration at each data point respectively, and $dF(t)$ is the data uncertainty of each data point. $\chi^2_{\rm N-body}$ measures the ability to statistically discern between the two models given the data available.

We used the empirical CDF obtained from the posterior distribution to translate $\chi^2_{\rm N-body}$ value to CDF by interpolation, and then to equivalent number of standard deviations assuming a normal distribution.
We refer to this as $\sigma_{\rm N-body}$, which quantifies the systematic error of the model relative to the statistical error of the data.

\begin{figure}[h]
    \includegraphics[width=1\linewidth]{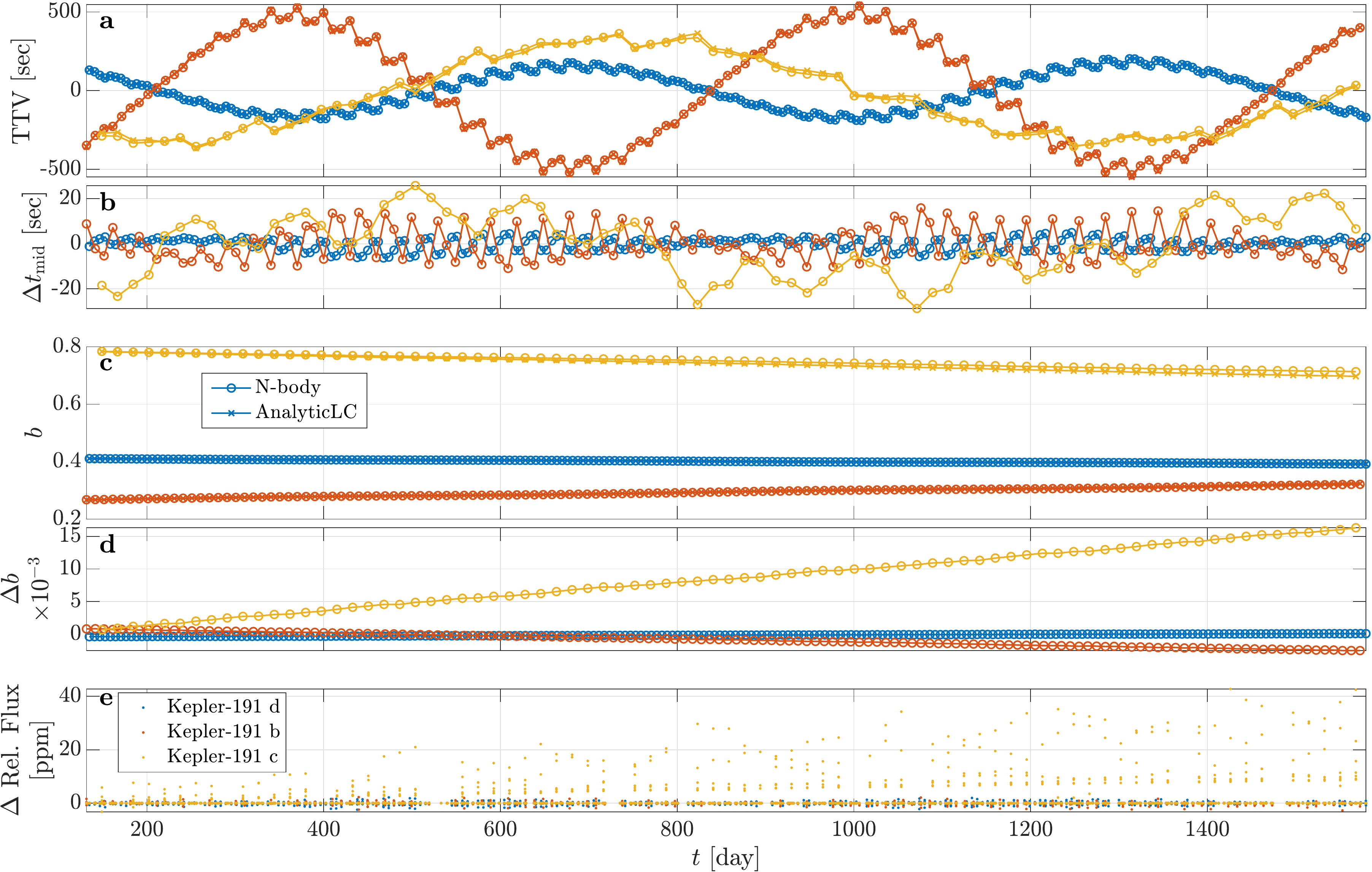}
    \caption{An example for the N-body matching test performed on the best fitting solution of the planetary system of Kepler-191 (KOI-582), demonstrating the ability of \texttt{AnalyticLC} to generate a model consistent with a full N-body integration. The mismatch is quantified as $\chi^2_{\rm N-body}=1.1389$,
  ($\sigma_{\rm N-body}=8.5003\times 10^{-5}$) - a systematic error much smaller than the statistical error arising from the data uncertainty. (a) TTV pattern of the N-body model (o) and the \texttt{AnalyticLC} model (x) for the three planets in this system (d in blue, b in red, c in yellow). The symbols are on top of each other at this scale. (b) "Residuals": Times-of-mid-transit mismatch between the N-body generated model and \texttt{AnalyticLC}, which is of order a few seconds for the two inner planets and order ten seconds for the outer planet. (c) Impact parameter evolution for the three planets from the N-body (o) and \texttt{AnalyticLC} (x). (d) "Residuals": Impact parameter evolution mismatch between the N-body and \texttt{AnalyticLC}, which is of order $10^{-3}$. (e) Manifestation of the mismatch to terms of relative flux, which is of order a few ppm for the two inner planets, and of order up to 30 ppm for the outer planet, much smaller than the typical {Kepler} long-cadence data uncertainty for this star ($\sim 240-300$ ppm).}
    \label{fig:NbodyTest}
\end{figure}

Figure~\ref{fig:NbodyTest} shows an example of the comparison between \texttt{AnalyticLC} and the results of a full N-body integration for the {Kepler}-191 system (KOI-582). The best fit solution of \texttt{AnalyticLC} for this system agrees well with an N-body integration ($\sigma_{\rm N-body}=8.5003\times 10^{-5}$), implying that the model error is four orders of magnitude smaller than the statistical error arising from the data uncertainty.

\begin{figure}[h]
    \includegraphics[width=1\linewidth]{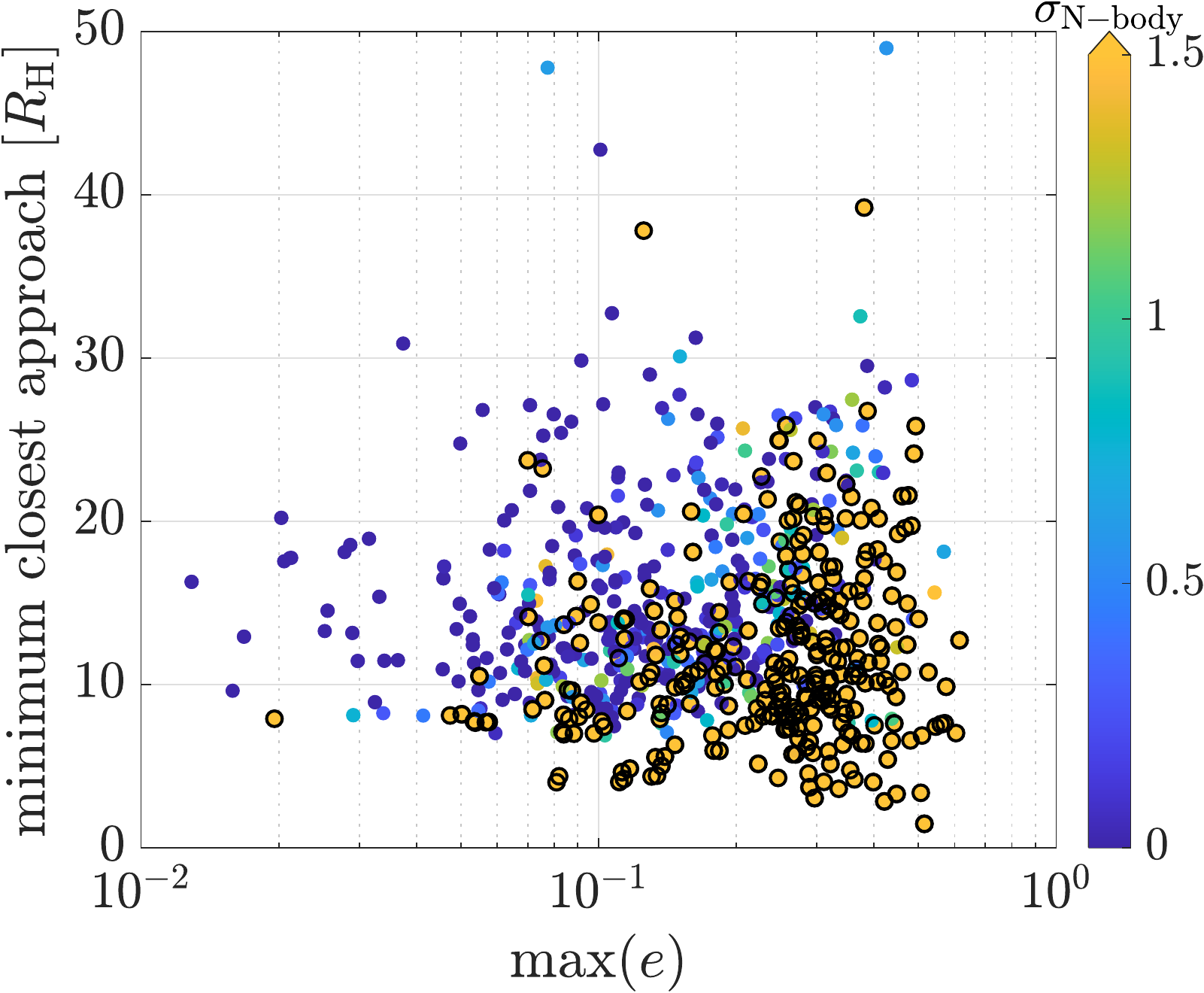}
    \caption{An overview of the N-body match with \texttt{AnalyticLC} for all our runs, showing the accuracy limits of \texttt{AnalyticLC}. The horizontal axis is the largest free eccentricity in the system at the best-fitting solution derived by \texttt{AnalyticLC}, the vertical axis is the minimal closest approach in the system in units of mutual Hill radii.  The color of the points represents the value of $\sigma_{\rm N-body}$, with points at which $\sigma_{\rm N-body}>1.5$ also marked  with black edges. In general, the light-curve generated by \texttt{AnalyticLC} matches the light-curve resulting from a full N-body integration up to the data uncertainty when the eccentricities are mild and when there are no close approaches much closer than $\approx 10$ mutual Hill radii.}
    \label{fig:NbodyMatchAllSystems}

\end{figure}

Inspecting cases where the model generated by \texttt{AnalyticLC} does not match the N-body-derived model sheds light on the limitations of our analytic approach. As might be expected, we observe that in cases with significant mismatch (empirically $>1.5\,\sigma$) there are large eccentricities and/or close approaches between planets in the system, which are a result of small separation in semi-major axes and/or large eccentricities. All $>$700 solutions where the closest approach in the system is above ten mutual Hill radii, and where the maximal eccentricity in the system is below 0.05, yielded a match better than $1.5\,\sigma$. This is illustrated in Figure~\ref{fig:NbodyMatchAllSystems}.

From the 54 systems with $\sigma_{\rm N-body}<1.5$, for 31 systems $\sigma^2_{\rm N-body}<0.1$, implying a very good match (such that when summing in quadrature the systematic and statistical errors, the systematic error contribution is an order of magnitude smaller than the statistical one). Only for seven systems $1<\sigma_{\rm N-body}<1.5$.

\subsection{Comparison to Strictly Periodic Circular Model}\label{sec:AlternativeModel}
Our model has eight parameters per planet, while a minimal model of a circular, strictly periodic model would involve only 5 for the innermost planet ($P, T_{\rm mid0}, R_{\rm p}/R_*, a^{(1)}/R_*, b$, where $b$ is the impact parameter) and only 4 for the other planets ($a/R_*$ can be deduced from the period ratios). The larger number of parameters in our model requires statistical justification. Therefore, we performed another fit on the detrended data with the minimal number of parameters, nulling the eccentricities, masses and sky-plane roll angle. Such a model does not generate any transit variations. In order to compare between two different models of a different number of parameters, an objective information criterion such as BIC (Bayesien Information Criterion) \citep{Schwartz1978} or AIC (Akaike Information Criterion) \citep{Akaike1974}. The BIC ensures that if one of the proposed models is the {\it true} model from which the data was derived, then for a sufficiently large number of data points BIC would favour the true model, while AIC can be useful if the candidate models are approximate descriptions of the full physical process from which the data originated \citep{Vrieze2012}. In our case, none of the models is the complete true model because both the dynamical model and the Mandel-Agol occultation model \cite{MandelAgol2002} include approximations. 
Therefore, we adopted the AIC as the information criterion to evaluate if the additional parameters are justified.

\begin{figure}[h]
    \includegraphics[width=1\linewidth]{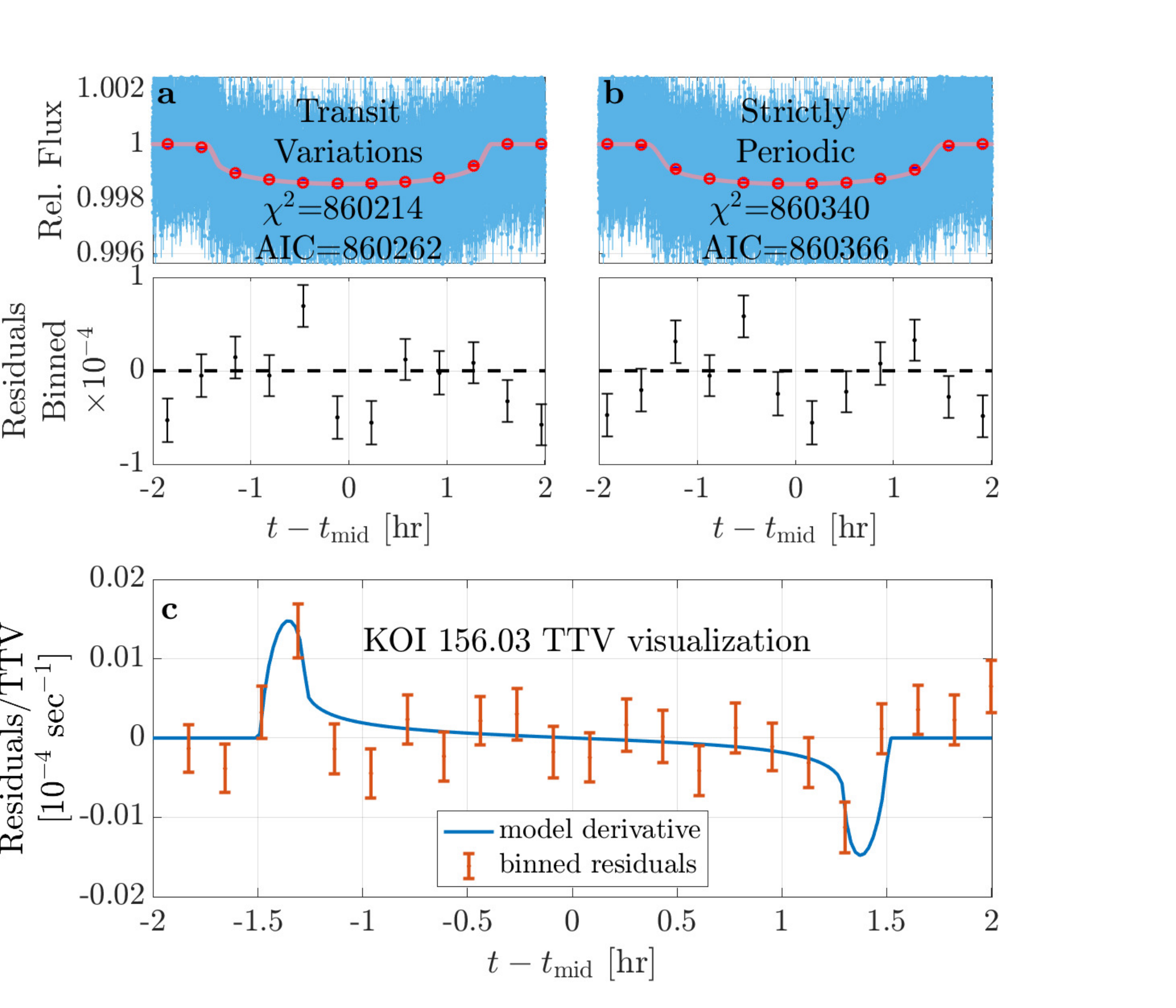}
    \caption{An example for the comparison between the best fitting model and a strictly periodic model. This figure visualizes the mismatch of a strictly periodic model, which might not be apparent from inspection of the binned residuals only. (a) Best fitting model including transit variations (TV) (b) Best fitting strictly periodic model, not including TV. The difference in fit quality is not easily seen in the residuals. (c) When dividing the strictly periodic model residuals by the TTV at each transit, the pattern of transit derivative with respect to time is clearly seen, as expected \citep{OfirEtAl2018}.}
    \label{fig:TTVDerivative}
\end{figure}

In Figure~\ref{fig:TTVDerivative} we show an example for the comparison between the best fitting \texttt{AnalyticLC} model and the best fitting strictly-periodic model. The difference in fit quality is expressed in the $\chi^2$ value, but even a difference of ($\Delta\chi^2=116$) is not visible in the binned residuals plot. When calculating the ratio between the residuals of the strictly periodic model and the TTV value, we obtain a clear pattern matching the transit model derivative with respect to transit time \citep{OfirEtAl2018}. This is done purely for visualization purposes to illustrate the effect of transit variations and how they are manifested in the data. This visualization works best for low-amplitude TTVs (smaller than ingress/egress time) which includes the bulk of the systems presented here, and it visually demonstrates the preference of the full model over the strictly periodic one as suggested by the AIC.

\subsection{Solution Consistency with Stellar parameters}
One of our model parameters is $a^{(1)}/R_*$, the ratio between the semi-major axis of the innermost planet and the stellar radius. This parameter can be obtained alternatively, from literature values for the stellar mass and radius \citep{Berger2020,Fulton2018}, and from the measured orbital period of the innermost planet. We checked the consistency of our converged solution with the literature-obtained value with respect to the literature error and our posterior distribution error on $a^{(1)}/R_*$. In the systems for which we provide a dynamical solutions, there are six out of 54 cases in which this difference is larger than $3\,\sigma$, where $\sigma$ is the root-sum-squares of the error estimates on $a^{(1)}/R_*$ from our posterior distribution and from literature. Such cases appear at KOIs with weak constraints on the planetary masses (e.g. KOI-279, KOI-1307) or multiple solutions (e.g. KOI-312, KOI-757, KOI-853), and not in KOIs for which there is a clear solution with significant planetary masses. Further detail on these individual systems is provided in \S~\ref{sec:Individual}.

\subsection{Consistency among Solutions}\label{sec:Consistencty}
\begin{figure}[h]
    \includegraphics[width=1\linewidth]{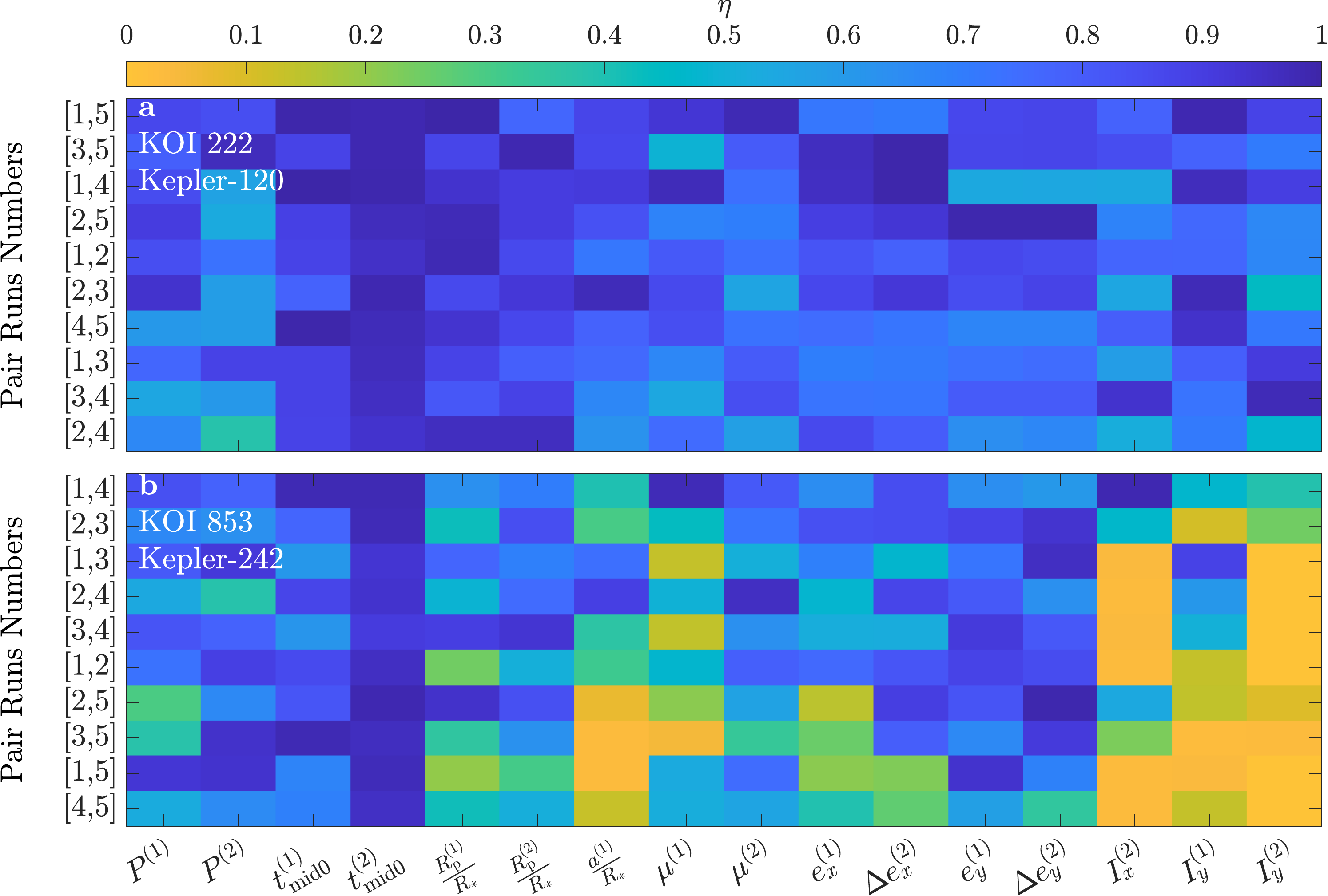}
    \caption{Consistency test example of two systems. This figure shows that the method is sufficiently robust to converge to the same global minimum (if one exists), and that in some of the cases there is no global minimum but a few local minima that match the data. (a) consistency map among different runs for KOI-222. All runs converged to the same minimum, as expressed in the high values of the overlapping parameter $\eta$. (b) consistency map for KOI-853. Different runs converged to different local minima, in which some runs agree on some of the parameters and some runs do not. For e.g., the first row suggests that runs 1 and 4 are in complete agreement on all parameters, but the second row suggests that runs 2 and 3 disagree on $I^{(1)}_y$}
    \label{fig:Consistency}
\end{figure}

We ran DE-MCzs five times for each planetary system, with different initial states of the walkers, as described in \S~\ref{sec:Methods}. We discard solutions for which our model is not compatible with a full N-body integration (see \S\ref{sec:Nbody}) and solutions for which the number of model parameters is statistically unjustified (see \S\ref{sec:AlternativeModel}). In addition, we discard solutions that imply unreasonably high planetary density - more than $2\,\sigma$ above $12\,{\rm g}\,{\rm cm}^{-3}$, the approximate density at the base of Earth's outer core \citep[][chapter 2]{SOROKHTIN201113}. 
This process leaves us with a subset of runs for each KOI. If all of them converged to the same maximum-likelihood region in parameters space, the solutions are consistent with each other and we report the obtained solution. If they converged to different regions in parameters space, we report all of them, and regard one of them as the "adopted" solution for this KOI, recognizing this solution may not be unique. Selection of the adopted solution is based on differences in fit quality ($\chi^2$), or if solutions are of similar quality, on physical reasoning (e.g. favoring solutions with plausible planetary densities and with small eccentricities for short-orbit planets). For 28 out of the 54 systems with valid dynamical solutions we report one solution, either because one of them was much better than the others (e.g. KOI-370), or because all solutions were consistent with each other (e.g. KOI-222, see Figure~\ref{fig:Consistency}) or because only one solution passed all our validity criteria. For 26 systems we report more than one solution (15 with two solutions, 11 with more). 

In order to quantify the test of solutions consistency, for each pair of runs we check the overlapping area between the Probability Density Function (PDF) of the posterior marginal distributions of each parameter. This overlapping area, denoted as $\eta$, is a measure of the similarity of two distributions \citep{Pastore2019}, defined for two PDFs of the parameter $x$, $f_1(x)$ and $f_2(x)$, as

\begin{equation}
    \eta = \int_{x~\rm support} \min{[f_1(x),f_2(x)]}dx.
\end{equation}

In Figure~\ref{fig:Consistency} we illustrate two cases of consistency among solutions. In one case (KOI-222) all our runs converged to the same maximum-likelihood region in parameters space, and in the second case different runs converged to different minima, resulting in more than one possible dynamical solution.

\subsection{Sample of Fitted Planetary Systems}
As the first usage of \texttt{AnalyticLC} on real data, we chose a set of 151 {Kepler} Objects of Interest (KOIs) that contain two or three confirmed or validated transiting planets. 
For each we performed an iterative detrending and fitting as described in \S~\ref{sec:Methods}, until reaching convergence. As explained above, we repeated this process 5 times per system, to verify convergence to a single minimum or to detect different local minima in the parameters space.
The process converged for 144 systems, 72 two-planets systems and 72 three-planets systems. Of those 144 systems, 54 systems passed all our three tests in at least one run: (i) N-body matching (illustrated in Figure~\ref{fig:NbodyTest}); (ii) AIC improvement over a strictly-periodic model (illustrated in Figure~\ref{fig:TTVDerivative}); and (iii) plausible planetary densities. Out of the 90 systems that did not pass our tests, the decisive majority failed due to criteria (i) and (ii) (about equally). Only a few systems failed due to high planetary densities. The 54 systems with valid solutions include 22 two-planets systems and 32 three-planets systems, for which we report dynamical solutions, {\it i.e.} orbital elements and planetary masses. 
The fit parameters of \texttt{AnalyticLC} are dimensionless; we translate them to absolute masses and radii by using stellar literature data from \citet{Berger2020}. If the specific star is not included in their catalog, we use the value from \cite{Fulton2018} (only KOI-582 is in this category), or from NExScI (only KOI-2711 is in this category).

We combine the error obtained from our posterior distribution and the literature error on stellar mass and radius in quadrature to obtain the error on the planetary absolute masses and radii.

\clearpage

\section{Results}\label{sec:Results}

\subsection{General}

\begin{figure}[h]
    \includegraphics[width=0.9\linewidth]{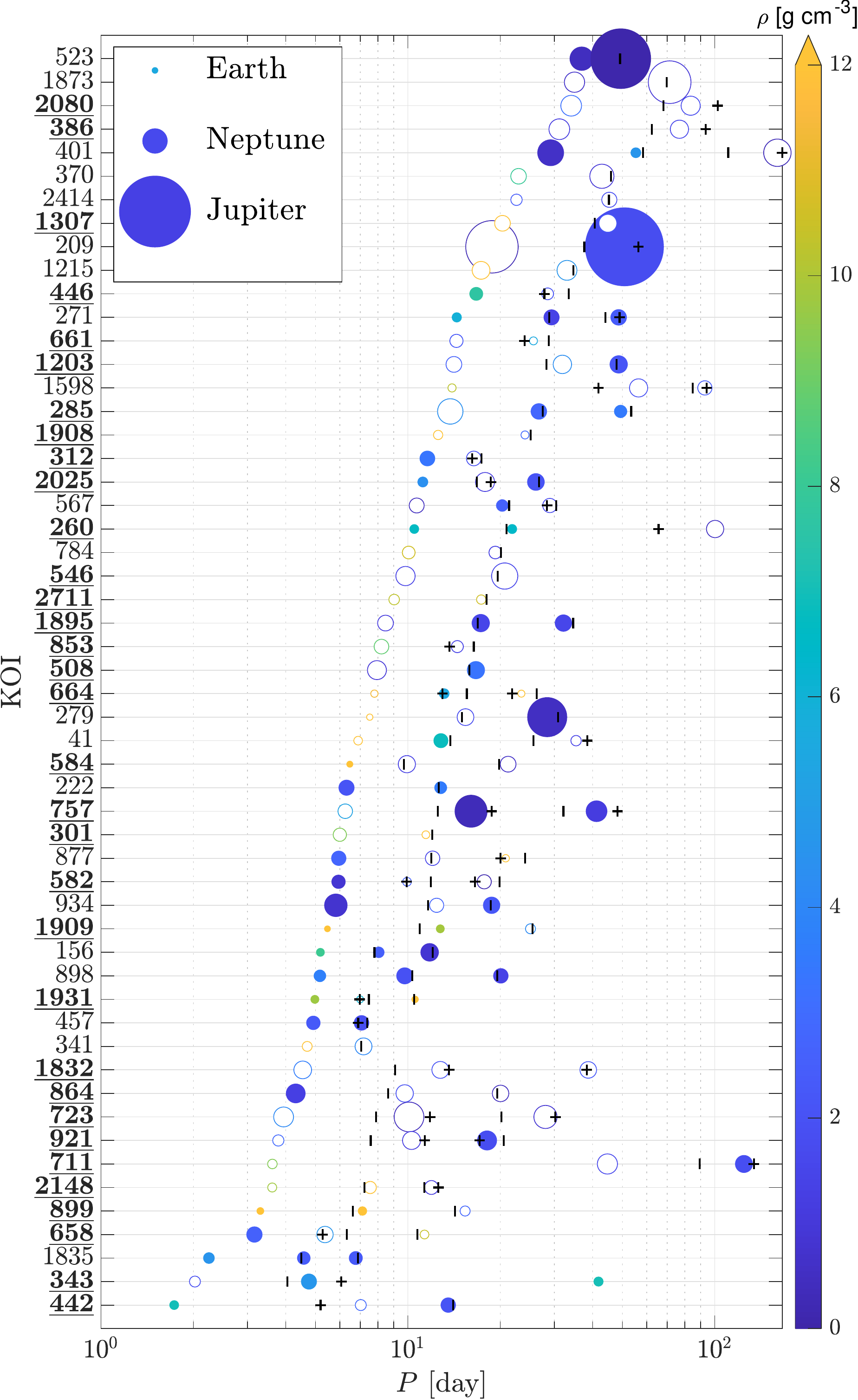}
    \caption{Orbital period, radii and densities of the planets with mass estimates from this work. KOI numbers for which at least one planet does not have a previously reported mass value or upper limit are shown in \underline{\textbf{underlined  bold}} text. The size of the circles represents absolute planetary size, and the color scale indicates the median density. Only planets with densities estimated to significance of more than $4\,\sigma$ are color-filled; others are empty. For reference, the legend shows the size and density of Earth, Jupiter and Neptune. The short vertical black lines indicate the locations of the closest first order MMRs to the observed period ratio of adjacent planets, and the black crosses similarly indicate the locations of second order MMRs - note that many planets are found close to these MMRs. Both first and second MMRs are indicated only if they are close to the observed period ratios. We do not show second order MMRs which are a multiplication of a first order MMR (e.g. 4:2 and 2:1).}
    \label{fig:FittedPlanetsOverview2}
\end{figure}

For the 54 systems for which we found a valid solution, 32 contain three transiting planets and 22 contain two transiting planets, summing to 140 planets in total. In Figure~\ref{fig:FittedPlanetsOverview2} we show a map of orbital periods, planetary radii and planetary densities obtained in this work, along with resonances locations. This map gives a brief overview of the planetary systems for which physical properties were obtained in this work. Many of the systems are near at least one first or second order resonance - not a surprising outcome, as proximity to resonances generates the large TTVs that enable mass estimation. For a large portion of the three planets systems, it is visually apparent that the spacing in $\log{P}$ is uniform (for example KOI-285, KOI-1832) and that they have similar radii (KOI-582, KOI-1895) \citep{Weiss2018a}. For a few three-planet systems this is also apprent in the densities (KOI-898, KOI-1835).

\subsection{New Constraints on Planetary Masses}

\begin{figure}[h]
    \includegraphics[width=1\linewidth]{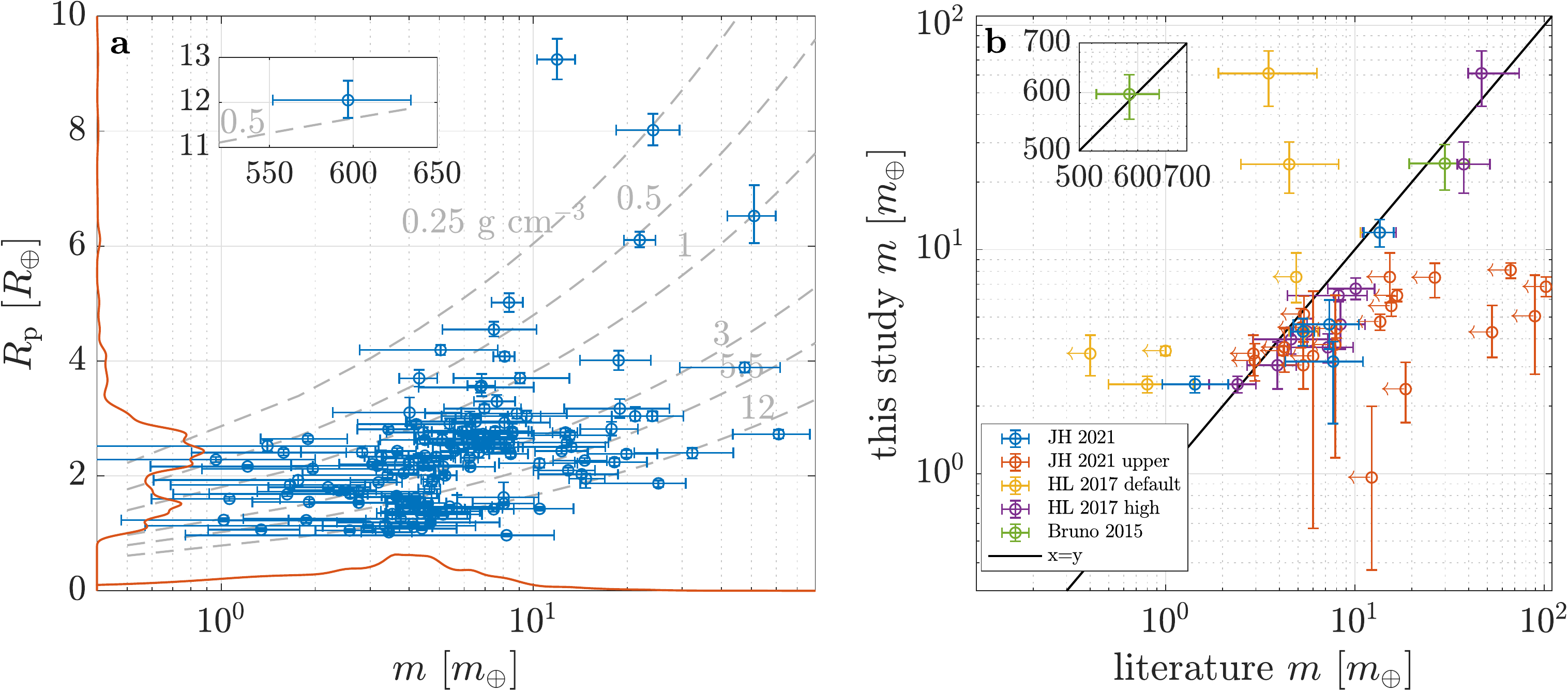}
    \caption{Planetary masses and radii obtained from this work. This figure shows the overall spread of masses and radii, and the good agreement of planet masses with former literature values. (a) Mass-radius diagram. Each blue error bar is related to a single planet. The red lines are weighted histograms of the masses and radii, obtained by summing up the PDFs of all points. As the sample is small and radii are well determined, the radii weighted histogram appeared more as a collection of discrete values, so it was smoothed using a Gaussian kernel of width $0.3\,R_\oplus$. The gray contours are constant-density curves, and the small insert above is for the outer planet of KOI-209 (KOI-209.01, Kepler-117 c). (b) Comparison of masses obtained in this work with literature values: JH21 values (blue) and upper limits (red); HL17 default mass prior, three of which show upper limits only (yellow) and high mass prior (purple); \citet{Bruno2015} (green). We omit planets for which literature values are within less than $2\,\sigma$ of zero. The black line shows the identity function.}
    \label{fig:FittedPlanetsOverview1}
\end{figure}

In Figure~\ref{fig:FittedPlanetsOverview1} we show the mass-radius spread of the planets in these 54 systems, with curves of constant density. Though this is a rather small sample of 140 planets, the so-called radius gap at $R_{\rm p} \sim 1.8 R_{\oplus}$ \citep{Fulton2017} is visually apparent. This gap is seen more clearly for planets of masses larger than $\sim 5 m_{\oplus}$. The sample is dominated by planets with densities of 1-3$\,{\rm g}\,{\rm cm}^{-3}$. Some planets have densities lower than 1$\,{\rm g}\,{\rm cm}^{-3}$ - these are planets massive enough to keep their large gas atmosphere. The smallest estimated density is that of Kepler-177 c/KOI-523.01: $0.0815^{+0.0157}_{-0.0141}\,{\rm g}\,{\rm cm}^{-3}$; a value similar to the one estimated by \citet{Vissapragada2020}, but not the lowest density planet found in literature: \citet{Masuda2014} estimated the densities of all the three planets in Kepler-51/KOI-620 to be about $0.03\,{\rm g}\,{\rm cm}^{-3}$. We chose the curves corresponding to 5.5$\,{\rm g}\,{\rm cm}^{-3}$ (approximate Earth's density) and 12$\,{\rm g}\,{\rm cm}^{-3}$ \citep[approximate density at the base of Earth's outer core,][chapter 2]{SOROKHTIN201113}.
In addition, we show a comparison of the masses obtained in this work with values of HL17 and JH21, which are studies that have a large number of common KOIs with our work, and with the masses of \citet{Bruno2015}, who analyzed the special case of high mass planets in KOI-209. 
Our results are in good agreement with the planets that have upper and lower error bars in JH21 and are also included in our sample, and agree with their upper bounds for planets that are also included in our sample.

 HL17 applied a default mass prior and a high mass prior and got two results for each planet; the masses we obtained are larger than their low-prior mass and are close to, or smaller than, most of their masses obtained from high-mass priors - a reasonable outcome, as our prior is uniform in mass, the same as their high-mass prior. We also add to the comparison the planetary masses of KOI-209 from \citet{Bruno2015}, which include a good example of the ability of \texttt{AnalyticLC} to fit for large planetary masses.

Overall, there is good agreement between the masses obtained in this work and previously reported planetary masses, thus giving confidence in the fitting process and on the reliability of the newly reported masses. The masses, radii and orbital elements of our adopted solution are tabulated in \S \ref{appendix}, along with machine-readable files.

\subsection{Out-of-Plane Forces}\label{subsec:OutOfPlaneForces}
In this section we focus on the specific issue of mutual inclinations within the system, as manifested in TbVs (see \S \ref{sec:Introduction}). 

The observational evidence for inclinations lies in impact parameter variations, manifested in the light curve shape (duration, ingress-egress) and in the transit depth, related to the impact parameter due to limb darkening \citep{Claret2000a}. The 3-dimensional motion of the orbit due to interactions among the system components can generate impact parameter variations, which are a sign of forces acting out of the plane. Let us devote a few words to describe the manifestation of orbital motion in impact parameter variations (TbVs). The impact parameter $b$ is determined by the planet-star separation at mid-transit $d/R_*$ and the sky-plane inclination $i$.

In our coordinates convention (see also \citet{Judkovsky2021a}), where $x$ points towards the observer,

\begin{equation}
    \cos{i} = \sin{I}\sin{\Omega}
\end{equation}

.
 
For small $I$, $\cos{i}\approx I\sin{\Omega}=I_y$. The component $I_x=I\cos{\Omega}$, which is related to the roll angle of the orbit on the sky plane, affects the {\it rate} of impact parameter variations mainly via the geometrical projection of the orbit on the sky plane as it moves, and to lesser extent via the rate of nodal regression (the correction to the nodal motion rate is only to second order in inclination, \citet{XuFabrycky2019}). For slow, gradual variations of $\Omega$ due to secular effects, the impact parameter variations can be approximated by a linear dependency on time, as the time scale of secular effects is proportional to the orbital period divided by the perturber-to-star mass ratio \citep{SSD1999}.

\begin{figure}[h]
    \includegraphics[width=1\linewidth]{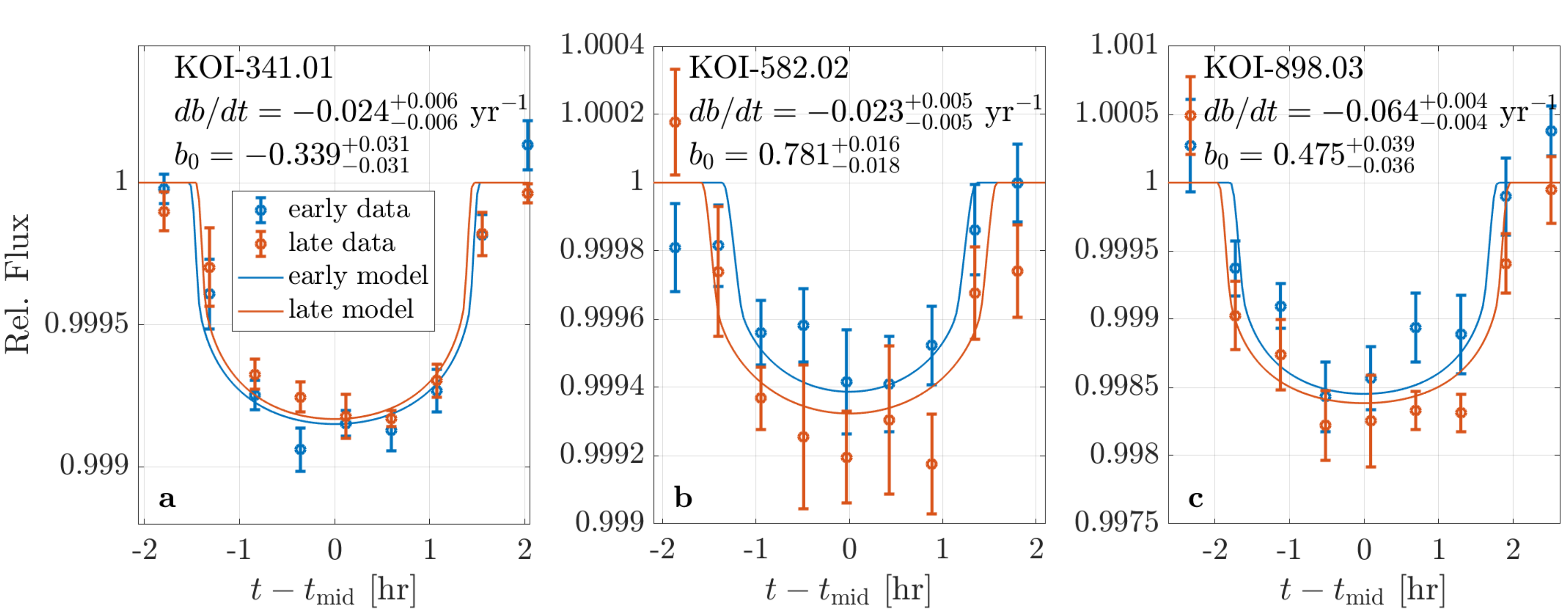}
    \caption{Visualization of impact parameter variations for three selected planets. This figure demonstrates the ability of the global fit approach to capture small TbVs. In each panel, the blue error bars indicate the binned first eight transit events in the long-cadence data, and the red error bars the binned last eight transit events. The solid lines indicate the binned model across these events densely sampled in time (blue for early events, red for late events). Printed are the impact parameter value $b_0$ (namely, at BKJD=0, which is January 1st 2009) and $db/dt$, the mean rate of change over the entire {Kepler} time span. 
    }
    \label{fig:bVariations}
\end{figure}

Along the fitting process, at each model evaluation we recorded the mean impact parameter variations rate for each planet by a linear regression, thereby receiving a posterior distribution not only of the orbital and physical properties, but also on $db/dt$. This enables us to detect planets that experience significant impact parameter variations regardless of the uniqueness of the best-fit solution. Because the changes in depth and duration are usually small \citep[for an interesting exception see ][]{Judkovsky2020}, such variations are usually not clearly visible. However, for some of these planets, a careful analysis of the data visualizes the TbVs.
Figure~\ref{fig:bVariations} illustrates the TbVs of three {Kepler} planets, all of which are the outermost transiting planet in either a doubly-transiting (KOI-341) or triply-transiting (KOI-582, KOI-898) systems. It is clearly visible that the earlier and later events differ in the transit depths and duration, and that the model captures these variations.
Long-term impact parameter variations are usually a manifestation of the combination of nodal regression along with some roll angle on the sky plane (in our model, this angle is $I_x$). For these three planets, the posterior distributions suggest significant $I_x$ values: $\sim 5^{\circ}.0^{+1.1}_{-1}$ for KOI-341.01; $\sim 10^{\circ}.0^{+2.1}_{-2.1}$ for KOI-582.02;  $\sim 30^{\circ}.4^{+3.3}_{-3.6}$ for KOI-898.03. Large mutual inclinations are required in order to explain these impact parameter variations using a model consisting only of the transiting planets, and it is not surprising that the strength of the $db/dt$ signal is increasing with the value of $I_x$. 
Large mutual inclinations for multi-transiting planets system is of low probability due to geometrical considerations \citep{RagozzineHolman2010} and from statistical analysis of impact parameter variations \citep{FabryckyEtAl2014} or durations distributions \citep{XieEtAl2016}. Therefore, it is entirely possible that these impact parameter variations are not a result of the interactions among the transiting planets, but a result of interaction with non-transiting external companion(s). In the latter case, the true $I_x$ can be smaller than obtained from the fit. In the following analysis we investigate the possible properties of such a hypothesized non-transiting planetary companion.

We stress that detection of statistically significant impact parameter variations does not mean that the dynamical model behind it is the true one. Different dynamical scenario can yield similar impact parameter variations, by an interplay among the controlling parameters. Let us take a simplified mathematical description of TbVs in order to make this clear. The impact parameter is given by \citep{Judkovsky2021a}:

\begin{equation}
    b=\frac{a(1-e^2)}{R_*(1+e\cos{\varpi})}\sin{I}\sin{\Omega},
\end{equation}

Let us assume that the long term TbVs arise solely due to the nodal motion of the orbit, and that its inclination magnitude with respect to the system's invariable plane is constant in time. In addition, we assume that the plane about which the precession occurs is close to the $xy$ plane as defined in Figure~\ref{fig:OrbitalElementsIllustration}. Under this assumption, we can write

\begin{equation}
    \frac{db}{dt}=b\frac{d\Omega}{dt}\cot{\Omega}.
    \label{e:dbdt}
\end{equation}

For the limiting case of $I_x=0$ (no roll angle on the sky with respect to the other planets),  $\cos{\Omega}=\cot{\Omega}=0$ and therefore $db/dt=0$.
Eq.~\ref{e:dbdt} demonstrates the notion  that given estimates of $b$ and $db/dt$ may correspond to multiple combinations of nodal motion rates and longitude-of-ascending-node values. If we obtain a solution that includes a large value of $I_x$, it is still possible that the true scenario actually includes an external, unseen companion that induces motion with a smaller $I_x$ but faster precession rate. Extending this idea further, for a system of transiting planets with known masses (for example, from the TTVs), with  one planet displaying impact parameter variations, we can constrain the properties of a possible non-transiting companion driving these variations, as follows. First, we calculate the nodal regression rate it experiences from all the transiting planets in the system. For each pair of planets, the nodal motion rate is approximately

\begin{equation}
    \dot{\Omega} = -\frac{1}{4}\alpha b_{3/2}^{(1)}(\alpha)(n^{\rm (in)}\mu^{\rm (out)}\alpha+n^{\rm (out)}\mu^{\rm (in)}),
\end{equation}

where $n$ is the mean motion, $\mu$ is the planet-to-star mass ratio, $\alpha$ the semi-major axis ratio and $b_{3/2}^{(1)}(\alpha)$ a Laplace coefficient, with the upper script relating each parameter to the inner planet or outer planet \citep{SSD1999, XuFabrycky2019, Millholland2019}. The hypothesised external unseen companion would generate {additional} precession, and for any value of the orbital period $P^{(\rm ext)}$ of the presumed perturber and any value of $\Omega$ of the perturbed, transiting planet, we can deduce the required perturber-to-star mass ratio $\mu^{(\rm ext)}$ to generate this additional precession rate. This creates a map of the three connected parameters $P^{(\rm ext)},\Omega$ and $\mu^{(\rm ext)}$.
We note that the assumption that the precession occurs about a plane close to $xy$ requires that the orbit of the external, non-transiting companion is only slightly inclined with respect to the $xy$, but it does not transit due to its large orbital radius.

\begin{figure}[h]
    \includegraphics[width=1\linewidth]{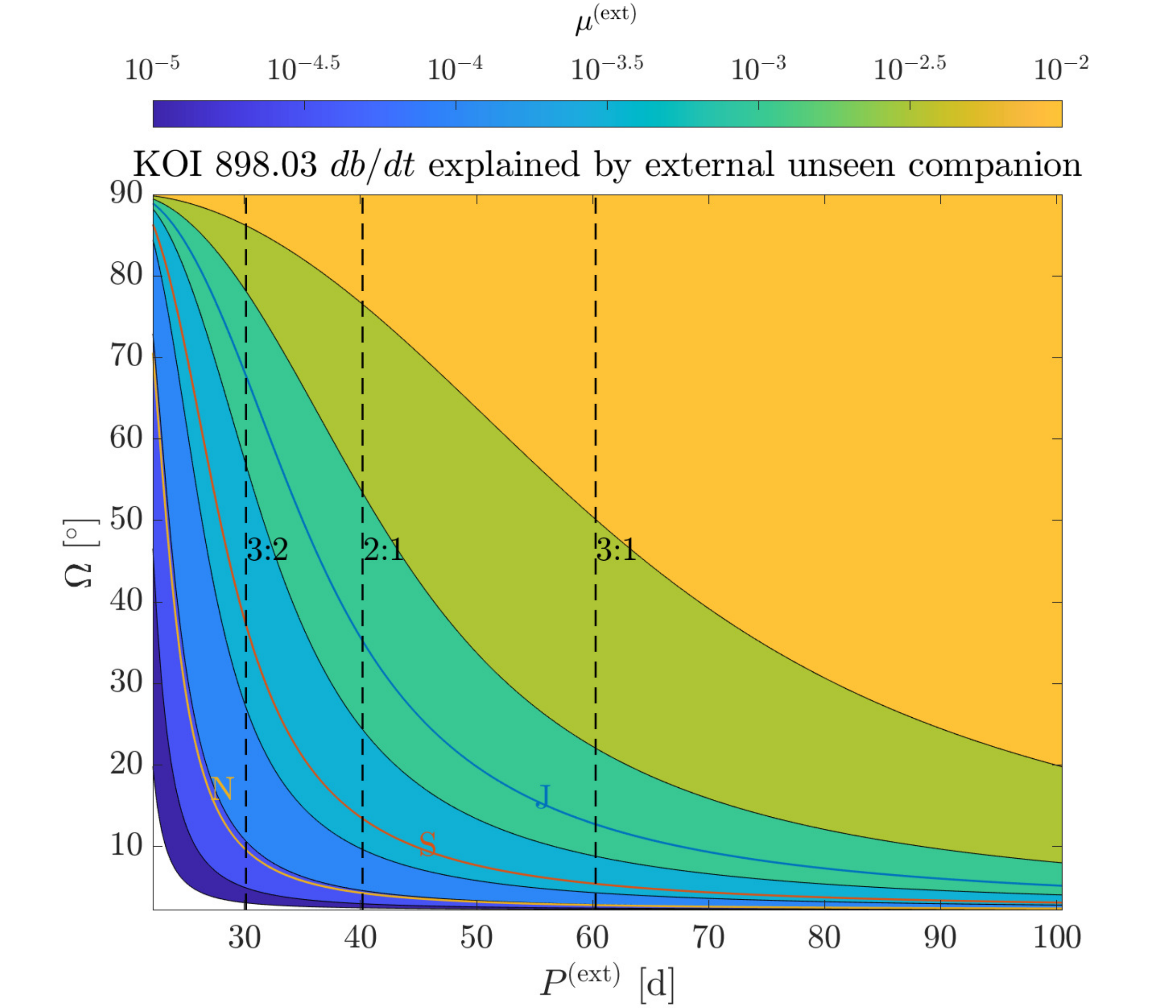}
    \caption{Characterization of a possible external non-transiting companion based on impact parameter variations of KOI-898.03. This figure shows that for a large portion of the parameters plane spanned by $\Omega$ (the longitude of ascending node of KOI-898.03) and $P^{(\rm ext)}$ (the orbital period of the external, hypothesized perturber), a planetary-mass arises, as seen in the map of $\mu^{(\rm ext)}$  (perturber-to-star mass ratio). Three contours that represent a perturber mass of Jupiter, Saturn and Neptune (denoted J, S, N) are overlaid on the color scheme. Some of the main resonances of KOI-898.03 are shown; because the TTVs seem to be well-explained by the transiting planets in the system it is likely that this external perturber, if it exists, is not located close to those resonances, as that would cause an additional strong TTV pattern. The RV signal arising from such an external non-transiting companion would be from a few m/s for Neptune-mass to up to 100 m/s for Jupiter-mass, detectable with the observational capabilities existing today.
    }
    \label{fig:ExternalUnseenPerturber1}
\end{figure}

In Figure~\ref{fig:ExternalUnseenPerturber1} we show such an analysis for KOI-898.03, that displays the strongest impact parameter variations of the three examples shown above. In addition to calculating the mass of the perturber as a function of its orbital period and as a function of $\Omega$ of KOI-898.03, we calculated the estimated radial velocity semi-amplitude, $K$, that would arise from such a non-transiting companion. The resulting RV signal semi-amplitude, of a few m/s to a few tens of m/s for most of the parameters' space, should be detectable with current instrumental capabilities.

Planets with significant impact parameter variations ($>2\,\sigma$ detection for the adopted solution, or $>2\,\sigma$ for all our 5 runs for systems without any valid solution) are listed in table \ref{tab:SignificantDbdt}. This table includes both systems with dynamical solutions, and systems without dynamical solutions. The study of mutual inclinations via transit variations in those systems would be very much benefit from future observations.

\newpage

\section{Individual Systems Description}\label{sec:Individual}

\begin{figure}[h]
    \includegraphics[width=1\linewidth]{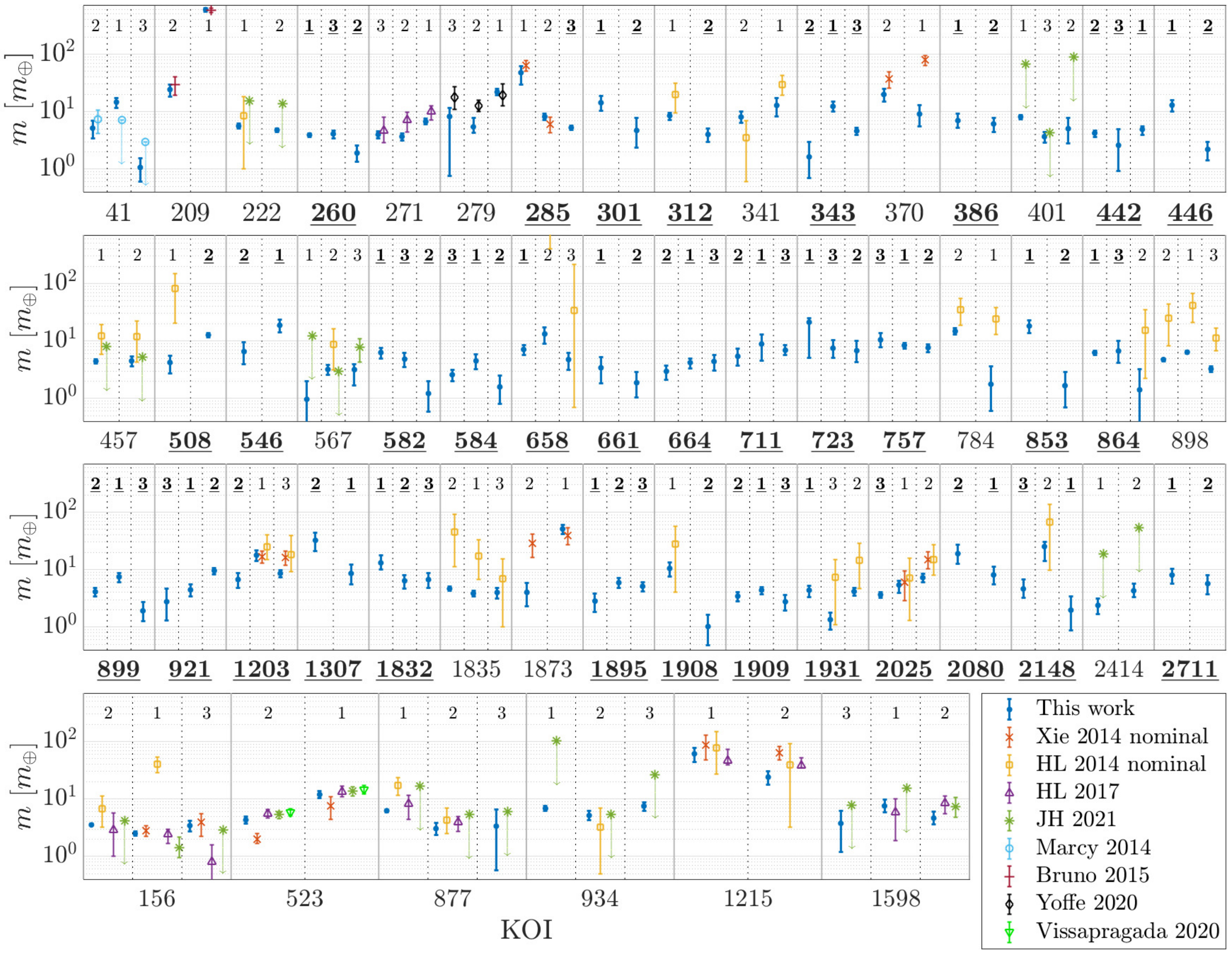}
    \caption{Comparison of the masses obtained from this work with past literature, by KOI. In each small panel, the planets in the system are sorted (left to right) by their orbital period, innermost planet at the left hand side. The small numbers above are the suffix of the planets' KOI numbers, for example in KOI-41 the number 2 stands for KOI-41.02. Dotted lines are plotted to distinguish between different planets. Bold underlined KOI numbers are used for systems with at least one planet without a previously reported mass value or upper limit, as in Figure~\ref{fig:FittedPlanetsOverview2}. Bold underlined KOI suffix stands for a specific planet without a previously reported mass. The legend shows the colors and markers representing different literature sources. Arrows pointing down represent upper limits on mass.
    \label{fig:MassVsLiterature1}
    }
\end{figure}

In this section we devote a few words for each one of the 54 systems with a dynamical solution. We relate to former literature planetary masses as well as to the dynamical features of the system. The systems appear in order of their KOI number.
The comparison to former literature masses is given in Figure~\ref{fig:MassVsLiterature1} for all the KOIs. We include both past estimates of the true planetary mass and estimates of planetary nominal mass \citep[][hereon HL14 and X14]{HaddenLithwick2014,Xie2014} which is, in many cases, an upper limit on the true mass. For the values of HL17, we use their high-mass prior (which is appropriate for comparing with our results, obtained from using the same prior as theirs).
For a several systems new significant mass constraints were found for planets without previously reported masses (e.g. KOI-260, KOI-582). For a number of  systems, the obtained mass constraints are substantially tighter than previous works (e.g. KOI-271). Out of the 140 planets for which we report masses, 102 have mass constraints significant to more than $3\,\sigma$, 43 of which are smaller than $5\,m_\oplus$. The two lowest mass planets with $>3\,\sigma$ detection are KOI-260.02/Kepler-126 d ($1.9^{+0.64}_{-0.56}\,m_\oplus$) and KOI-1931.03/Kepler-339 c ($1.34^{+0.44}_{-0.44}\,m_\oplus$).

\ksec{Kepler-100 (KOI-41)}
This system has been analyzed by \citet{Marcy2014}, who used both spectroscopic and imaging data to show that no companion star is evident. By using multi-Keplerian fit of the RV they obtained constraints on the mass of KOI-41.02, and upper limits on the masses of KOI-41.01 and KOI-41.03.
Our fitted mass for the innermost planet, Kepler-100 b (KOI-41.02), $5.14^{+1.7}_{-1.75}\,m_\oplus$, is within the error bar of \citet{Marcy2014}, $7.34^{+3.2}_{-3.2}\,m_\oplus$. \citet{Marcy2014} assign only upper limits for the masses of the two outer planets to the 1$\,\sigma$ level: $4.8\,m_\oplus$ and $1.7\,m_\oplus$ for the intermediate planet (Kepler-100 c, KOI-41.01) and outer planet (Kepler-100 d, KOI-41.03), respectively. Our solution (masses of $5.1^{+1.7}_{-1.8}\,m_\oplus$ for KOI-41.02, $14.6^{+2.6}_{-3}\,m_\oplus$ for KOI-41.01, $1.1^{+0.5}_{-0.5}\,m_\oplus$ for KOI-41.03) agrees with that limit for the outer planet, but not with the limit for the intermediate planet. We leave for the future a joint fitting of the RV and photometric data.


\ksec{Kepler-114 (KOI-156)}
HL14 have analyzed the TTV phases of collection of {Kepler} planet pairs  in a statistical manner to calculate the nominal mass of 156.01 (Kepler-114 c) and 156.02 (Kepler-114 b): $40.0^{+13.2}_{-11.4}\,m_\oplus$ and $6.8^{+4.3}_{-3.6}\,m_\oplus$ , respectively. The phases are not strongly constrained in this system (their table 1), and therefore the free eccentricity can take different values, meaning that the mass can be smaller or larger than the nominal mass. For Kepler-114 c, a mass of 40$\,m_\oplus$ would lead to high density, and hence the true mass should differ from the nominal one. Based on this high density, they relate both of these planets to the ``high-eccentricity" category.
X14 used the formalism of \citet{LithwickXieWu2012} to analyze this system (among others) by calculating the nominal mass, and by estimating maximal true masses by fitting masses and eccentricities to the TTVs, assuming co-planarity. He calculated a nominal mass of $2.8^{+0.6}_{-0.6}\,m_\oplus$ (KOI-156.01) and $3.9^{+1.7}_{-1.7}\,m_\oplus$ (KOI-156.03) with a 3-sigma maximal mass of $45.3\,m_\oplus$ and $22.6\,m_\oplus$ obtained from an MCMC-fit to the mass-eccentricity parameters.
We found five different solutions with similar fit quality, which means that the problem is under-constrained. All of them point at rocky planets with low eccentricities, with the innermost planet having a density similar to iron, the intermediate planet a density of $\sim 3\,{\rm g}\,{\rm cm}^{-3}$, and the outer planet a density below $1\,{\rm g}\,{\rm cm}^{-3}$. We  choose to report solution 1 as our adopted solution, as it points at zero eccentricity of the innermost planet, but this choice is rather arbitrary.
Our adopted solution agrees with the planets radii of JH21, and as for the masses, our estimated mass of KOI-156.01 lies between this of HL17 and the one of JH21, within their 95 per cent confidence interval. The mass of 156.02 is well below the upper limit of JH21, and the mass of 156.03 (outer planet) is near their upper limit, and consistent with it.



\ksec{Kepler-117 (KOI-209)}
This system was analyzed by using a combination of RV and TTV by \citet{Bruno2015}. Our masses ($24.2^{+5.2}_{-5.8}\,m_\oplus$ for KOI-209.02, $597^{+37}_{-45}\,m_\oplus$ for KOI-209.01) agree with their estimated values ($29.875^{+10.488}_{-10.488}\,m_\oplus$ and $584.78^{+57.21}_{-57.21}\,m_\oplus$, respectively), though we used photometry only without RV. Another piece of information we have from the global fit is the strong impact parameter variations of both planets, especially the inner one. These impact parameter variations are explained by our model by a mutual inclination of $2.^{\circ}05^{+0.11}_{-0.10}$ between the two planets in this system, causing nodal regression which is manifested in TbVs. A recent paper pointed at the orbital precession of the inner planet in this system via the detection of transit duration variations \citep{Millholland2021}.

We chose to include the solution for this system in this work even though it did not meet the $\sigma_{\rm N-body}<1.5$ criterion (here $\sigma_{\rm N-body}=1.92$), because it is a good example of the ability of \texttt{AnalyticLC} to extract planetary masses solely from photometry, and it may show that the $\sigma_{\rm N-body}<1.5$ criterion we set is conservative.



\ksec{Kepler-120 (KOI-222)}
HL14 calculated the nominal mass of KOI-222.01 (inner planet) to be $8.5^{+9.7}_{-7.5}\,m_\oplus$, based on their TTV analysis. In this system all our runs converged to consistent posterior distributions in all parameters. The 2:1 indirect term generates an asymmetry in the TTVs of the inner and outer planet, thus offering the possibility to break some of the inherent first-order MMR TTV degeneracies described by \citet{LithwickXieWu2012}.
Our solutions are consistent with a low-eccentricity, nearly co-planar structure, where the inner planet ($\approx 6$ days period) has a density just over half that of the outer planet ($\approx 12$ days period). The radii we obtained agree with the recent results of JH21, and our masses ($5.6^{+0.6}_{-0.6}\,m_\oplus$ for KOI-222.01, $4.8^{+0.4}_{-0.4}\,m_\oplus$ for KOI-222.02) are below their estimated upper limit. Interestingly, our solution implies that both planets have similar masses but different densities of $\sim~2$ and $3.5\,{\rm g}\,{\rm cm}^{-3}$, suggesting that surprisingly the inner planet held to its atmosphere, while the outer planet did not. This finding calls for further investigations; this system can be a target for studying models of atmosphere loss such as photoevaporation \citep{Owen2017} or core-driven mass loss \citep{Ginzburg2018, Loyd2020}. 



\ksec{Kepler-126 (KOI-260)}

We did not find literature masses for this 3-planet system. Our solution points at masses of $3.9^{+0.3}_{-0.3}\,m_\oplus$ for KOI-260.01, $4.1^{+0.6}_{-0.7}\,m_\oplus$ for KOI-260.03, $1.9^{+0.6}_{-0.6}\,m_\oplus$ for KOI-260.02, with earth-like densities for the two close-in planets and a low density of order $0.5\,{\rm g}\,{\rm cm}^{-3}$ for the outer, 100-days-orbit planet. Low-mass low-density planets are an unexpected, interesting and relatively rare sub-type of planets that is nevertheless found by different groups and methods \citep[e.g.][]{Ofir2014, Masuda2014, Vissapragada2020}. Therefore, this system deserves further observations to validate this proposed solution. We note the significant impact parameter variations of the intermediate and outer planets.



\ksec{Kepler-127 (KOI-271)}
This system has been analyzed by HL17, who highlighted the mass-eccentricity degeneracies of the near-2:1 inner pair and the near-5:3 outer pair.
We found different solutions, that all point at similar densities: an earth-like density for the inner, 14 days orbit planet, and densities of 1.5-2.5 ${\rm g~cm^{-1}}$ of the two outer planets (29 and 48 days orbit). Our adopted solution points at masses of $4^{+0.5}_{-0.6}\,m_\oplus$ for KOI-271.03, $3.7^{+0.5}_{-0.5}\,m_\oplus$ for KOI-271.02, $6.7^{+0.8}_{-0.7}\,m_\oplus$ for KOI-271.01. The two outer planets show significant impact parameter variations, that arise from mutual inclination of a few degrees in all solutions; an alternative explanation could be the existence of a fourth, external planet.



\ksec{Kepler-450 (KOI-279)}
The first mass constraints for this system were obtained by \citet{Yoffe2021}: $17.6^{+9.5}_{-6.7}\,m_\oplus$ for KOI-279.03 (planet d), $12.5^{+3.2}_{-2.6}\,m_\oplus$ for planet KOI-279.02 (planet c); and $19.4^{+11.1}_{-6.8}\,m_\oplus$ for KOI-279.01 (planet b). Our solution suggests masses of $8.2^{+3.5}_{-7.4}\,m_\oplus$ for KOI-279.03, $5.4^{+2.4}_{-1.2}\,m_\oplus$ for KOI-279.02, $21.9^{+2.7}_{-2.4}\,m_\oplus$ for KOI-279.01. We note the large uncertainty on the inner planet mass for both solutions; this large uncertainty, along with the mismatch of our solution $a^{(1)}/R_*$ with stellar literature data ($\sim 3.97\,\sigma$), suggest that further data is required in order to understand the properties of the planets in this system.

\ksec{Kepler-92 (KOI-285)}
X14 Analyzed the TTVs of the inner pair of this system, and found a nominal mass of $64.3^{+13.9}_{-13.9}\,m_\oplus$ and $6.1^{+1.8}_{-1.8}\,m_\oplus$, respectively. His MCMC analysis found a $3\,\sigma$ upper limit on the true mass of $51.4\,m_\oplus$ and $28.3\,m_\oplus$; The fact that the inner planet true mass should be smaller than the nominal mass implies that this pair possesses a non-negligible combined free eccentricity with respect to the normalized distance from resonance, $\Delta=-0.0282$.
We found 5 solutions, all of them point at eccentricity consistent with zero for the two inner planets and small eccentricity of up to a few per cent for the outer planet.



\ksec{Kepler-135 (KOI-301)}
We found no literature mass for the pair of planets in this system.
While the inner planet TTV magnitude is of order one minute, the outer planet TTV is an order of magnitude larger. Therefore, even though the two-planets model fits the data well, it is entirely possible that this TTV arises from an external unseen companion.



\ksec{Kepler-136 (KOI-312)}
HL14 calculated the nominal mass of the inner planet in this 2-planet system to be $19.8^{+11.7}_{-10.4}\,m_\oplus$ and listed it as having a small eccentricity.
We report two solutions of similar fit quality. Both of them agree on masses and eccentricities and suggest reasonable densities. The two solutions for $a^{(1)}/R_*$, combined with the orbital period, are not consistent with the literature stellar mass and radius; this could be a result of the degeneracy between $a$ and $b$ \citep{Carter2008, Kipping2010}. 



\ksec{Kepler-414 (KOI-341)}
HL14 calculated the nominal masses of the inner and outer planets to be $3.5^{+3.4}_{-2.9}\,m_\oplus$ and $29.9^{+12.4}_{-10.5}\,m_\oplus$, respectively. They classify this system as a ``high-eccentricity" one. As the orbital periods are roughly 4.7 and 7 days, we would expect that the absolute eccentricities would be small, but yet the ratio $Z_{\rm free}/\Delta$ should be order unity or larger.
We report three possible solutions that agree on the masses but not on the eccentricities. The density of the inner planet in all solutions is much larger than the density of the outer one. A possible explanation would be that their radii ($\approx1.5,2.6\,R_\oplus$) are below and above the radius gap \citep{Fulton2017}; however, interestingly the solution suggests that the outer planet may still be consistent with rocky bulk density. A better determination of the mass of the outer planet, e.g. via RV observations, could better pin down its true nature. We note the significant impact parameter variations of the outer planet KOI-341.01 ($\sim 3\,\sigma$), that might imply the existence of an external non-transiting planet.



\ksec{Kepler-142 (KOI-343)}
We found no literature masses for this 3-planets system. Because of the weak dynamical interaction the TTVs are rather small, and hence the AIC of a strictly periodic model is almost as good as that of the dynamical model. The solution is highly questionable as it involves large eccentricity of the two short-orbit planets.


\ksec{Kepler-145 (KOI-370)}
X14 calculated nominal masses for this system to be $37.1^{+11.6}_{-11.6}\,m_\oplus$ and $79.4^{+16.4}_{-16.4}\,m_\oplus$ for the inner (KOI-370.02) and outer (KOI-370.01) planets respectively, with 3-sigma upper limits of $68.3\,m_\oplus$ and $183.1\,m_\oplus$. All these are questionable given that the planets have radii of $\sim2.4\,R_\oplus$ and $\sim3.7\,R_\oplus$, respectively.
In our solution the masses are significantly smaller, and still the inner planet's density ($\sim8\,{\rm g}\,{\rm cm}^{-3}$) implies that it does not have a significant atmosphere, even though its radius is about 2.4$\,R_\oplus$. 


\ksec{Kepler-146 (KOI-386)}
We found no literature masses for this 2-planets system. The planets have similar radii ($\sim3.2\,R_\oplus$ and $\sim2.8\,R_\oplus$) and we were able to constrain their masses to $7.0^{+2.2}_{-1.8}\,m_\oplus$ and $6^{+1.7}_{-1.6}\,m_\oplus$, thus both likely have significant volatile envelopes. The best model TTV is dominated by the super-period of the near-5:2 MMR, which is about 1026 days. The solution has numerically moderate eccentricities - but large errors make them consistent with zero under $2\,\sigma$. The two planets seem to be significantly misaligned in both the $x$ and $y$ directions and indeed exhibit significant TbVs (see Table~\ref{tab:SignificantDbdt}).


\ksec{Kepler-149 (KOI-401)}
JH21 recently published estimated radii for these planets and upper limits for their masses; our solutions include significant constraints on the masses of the inner pair to $8.09^{+0.61}_{-0.65}\,m_\oplus$ and $3.66^{+0.79}_{-0.83}\,m_\oplus$, respectively, which agree with the upper limits above, and also seem reasonable for the radii of $\sim4.1\,R_\oplus$ and $\sim1.6\,R_\oplus$,  respectively. The third, outer planet's mass remains not well determined.
Our solutions were consistent with each other in all parameters except the mass of the outer planet. They all suggest small eccentricity of a few per cent of the innermost planet, and mild eccentricity of the intermediate planet.


\ksec{Kepler-157 (KOI-442)}
We did not find literature masses for this 3-planets system. We could constrain the masses of the innermost and outermost planet to be quite similar at $4.22^{+0.51}_{-0.55}\,m_\oplus$ and $4.89^{+0.65}_{-0.91}\,m_\oplus$, respectively. The different radii, though, suggest that the two planets lie on different sides of the radius gap: with a density of $\rho=7.0\pm1.1\,{\rm g}\,{\rm cm}^{-3}$ the inner planet does not have significant volatiles, while the outer planet, with a density of $\rho=2.1^{+0.37}_{-0.44}\,{\rm g}\,{\rm cm}^{-3}$, likely does.
All solutions point at eccentricities consistent with zero. It makes sense that the inner would be on a circular orbit, as it is on a 1.7-day orbit, and should have gone through circularization.


\ksec{Kepler-158 (KOI-446)}
We did not find literature masses for this 2-planets system.
All our solutions suggest that the inner planet's mass (which is determined to better than $4\,\sigma$) is of order $12-14\,m_\oplus$ with density of about $\rho\approx$7-8$\,{\rm g}\,{\rm cm}^{-3}$, while the outer planet's density is likely lower than Earth's but not well constrained. Our solutions are consistent in all parameters, except a degeneracy between $a^{(1)}/R_*$ and $I_y^{(1)}$ (the known $a-b$ degeneracy \citep{Carter2008, Kipping2010}.



\ksec{Kepler-161 (KOI-457)}
HL14 calculated the nominal mass of the two planets in this system to be $12.1^{+7.4}_{-6.3}\,m_\oplus$ and $11.8^{+10.5}_{-7.5}\,m_\oplus$ for the inner (KOI-457.01) and outer (KOI-457.02) planets respectively, and classify them as a ``high-eccentricity" pair.
JH21 estimated upper limits on the planets masses of $7.99\,m_\oplus$ and $5.19\,m_\oplus$ for the inner and outer planets, respectively, and indeed our adopted solution suggests masses of $4.4^{+0.4}_{-0.4}\,m_\oplus$ for KOI-457.01, and $4.5^{+1}_{-0.8}\,m_\oplus$ for KOI-457.02, both with sub-Earth density. The solutions are consistent with each other in most parameters. 


\ksec{Kepler-170 (KOI-508)}
HL14 calculated a nominal mass of the inner planet (508.01) of $82.6^{+67.5}_{-62.4}\,m_\oplus$. Out solution suggests sub-earth densities for both planets; however, the weak TTV signal of the best fit model (3.5 minutes of the inner planet, 2 minutes of the outer planet) implies that more data would be required in order to confidently constrain the planets masses.


\ksec{Kepler-177 (KOI-523)}
This interesting system is in the proximity of the 4:3 MMR ($\Delta \sim 0.0055$), with a super-period of 2258 days, about 1.5 times the time {Kepler} data spans.
The system has been analyzed by X14 (nominal masses of $2^{+0.5}_{-0.3}\,m_\oplus$ and $7.5^{+3.5}_{-3.1}\,m_\oplus$ for the inner and outer planets), HL17 (mass estimate of $5.6^{+0.9}_{-0.9}\,m_\oplus$ and $13.5^{+3.0}_{-2.5}\,m_\oplus$ for the inner and outer planets), \citet{Vissapragada2020} (mass estimates of $5.84^{+0.86}_{-0.82}\,m_\oplus$ and $14.7^{+2.7}_{-2.5}\,m_\oplus$), and JH21 (mass estimates of $5.32^{+0.78}_{-0.75}\,m_\oplus$ and $13.54 ^{+2.52}_{-2.39}\,m_\oplus$). Our adopted solution suggests masses of $4.3^{+0.6}_{-0.6}\,m_\oplus$ for KOI-523.02, $11.9^{+1.7}_{-1.6}\,m_\oplus$ for KOI-523.01, in agreement with those results up to the uncertainty (see Figure~\ref{fig:FittedPlanetsOverview1}).
We note the extremely low density of the outer planet (KOI-523.01, Kepler-177 c) of less than $0.1\,{\rm g}\,{\rm cm}^{-3}$; a surprising result, but not a first one: \citet{Masuda2014} estimated similar, and even lower, densities for all the planets in the Kepler-51 system (KOI-620).
We note that the outer planet experiences impact parameter variations of $2.2\,\sigma$ (significance of $0.00147^{+0.00066}_{-0.00062}$); though this is a slow rate, the manifestation in transit depth is strong due to the initial large impact parameter magnitude (about 0.86); hence for this object further observations are of high value.


\ksec{Kepler-182 (KOI-546)}
We found no literature masses for this 2-planets system. The planets lie at a distance of $\Delta=0.0525$ from the 2:1 MMR. Though the resulting TTV signal is weak, our best-fit model generates a TTV of the expected super-period and its harmonics. The estimated masses are $6.5^{+2.9}_{-2.6}\,m_\oplus$ for KOI-546.02 and $18.8^{+5}_{-4.6}\,m_\oplus$ for KOI-546.01, with an eccentricity for the inner planet of a few per cent, a borderline value for a 9.82 days orbit planet.


\ksec{Kepler-184 (KOI-567)}
HL14 calculated the nominal mass of the intermediate planet in this 3-planets system, estimating it to be $8.8^{+7.4}_{-5.7}\,m_\oplus$, based on the TTV arising from the interaction with the outer planet. The intermediate-outer pair is classified as having a low combined eccentricity. The phase of the TTV of the inner planet due to the intermediate planet, and the phase of the outer planet due to the intermediate, are weakly constrained.
JH21 estimated an upper limit of $12.27\,m_\oplus$ and $2.96\,m_\oplus$ for the two inner planets, and estimated a mass of $7.68^{+3.36}_{-3.39}\,m_\oplus$ for the outer planet.

We provide two different solutions, which differ significantly in the all the planets' masses.
It is interesting to note that this near-resonant chain of 3:2:1 has two close super-periods: 202.392 and 205.7121 days. This means that there will be a few-minutes-amplitude Super-Mean-Motion-Resonance (SMMR) TTV pattern (for further details, see \citet{Judkovsky2021a}) with a period of about 34 years.



\ksec{Kepler-191 (KOI-582)}
We did not find any literature masses for this 3-planets system, which have radii of about $2.1\,R_\oplus$, $1.3\,R_\oplus$ and $2.1\,R_\oplus$, from the innermost to the outermost planet.
All our solutions are consistent in planetary masses (with significant detections for the inner pair), and suggest sub-Earth densities to all three planets in the system. All the solutions point at small eccentricities of a few per cent, though they suggest different values.

Due to the strong impact parameter variations of the outer planet (Kepler-191 c, KOI-582.02), all the solutions suggest a mutual inclination of order 10 degrees between the outermost planet and the inner two planets, which, coupled with nodal regression, causes impact parameter variations. An alternative interpretation would be that the mutual inclination is lower, but the precession rate is faster - this would require an external, unseen, companion (see discussion at \S \ref{subsec:OutOfPlaneForces}).



\ksec{Kepler-192 (KOI-584)}
We found no literature masses for this 3-planets system, but were able to constrain the masses of the inner pair.
Our solution suggests a super-dense innermost planet, with a radius slightly larger than Earth's, and 2-3 times its mass. The intermediate planet is determined to have a mass of  $4.5^{+1.4}_{-1.3}\,m_\oplus$, i.e. it is above the radius gap. 
Note that the two super-periods in this system ($\sim 145.1, 153.8$ days) are close to each other, implying a 3-planets SMMR TTV component \citep{Judkovsky2021a} with period of about 7 years.



\ksec{Kepler-203 (KOI-658)}
HL14 estimated nominal masses which are not statistically significant but are numerically unreasonably high for the two inner $\sim2.5\,R_\oplus$ planets - as high as $\sim750\,m_\oplus$.
Our solution suggests masses of $7.1^{+1.2}_{-1.4}\,m_\oplus$ for KOI-658.01, $13.2^{+3.8}_{-4.3}\,m_\oplus$ for KOI-658.02 and $4.7^{+1.5}_{-1.6}\,m_\oplus$ for outermost KOI-658.03, all with small eccentricities, a physically reasonable scenario for such short-orbit planets.
The inner pair, with a period ratio of roughly 1.698, is mainly affected by the 5:3 MMR, causing a low amplitude $\sim1$ minute TTV of the inner planet with super-period of about 57 days - an example of the ability of the global fit method to capture low-amplitude TTVs. The outer pair, with a period ratio of 2.1, is dominated by the 2:1 resonance with a super-period of 103 days.


\ksec{Kepler-204 (KOI-661)}
We found no literature for the masses in this 2-planets system.
We provide two solutions; however we note that they are highly questionable, because they involve large eccentricities and mutual inclinations, which are not expected in such close orbits (14.4 and 25.66 days). We assume that the fit converged to such values in the attempt to explain the large TTV of the outer planet, which could be in fact a result of an external unseen companion (the two transiting planets are not close to any first order MMR, and hence are not expected to generate large TTVs; in fact our best model TTV is dominated by the proximity to the 9:5 MMR, which is manifested in 284-days oscillations).


\ksec{Kepler-206 (KOI-664)}
We found no literature for the masses in this 3-planets system.
Our solution suggests masses of $3^{+0.8}_{-0.8}\,m_\oplus$ for KOI-664.02, $4.2^{+0.8}_{-0.9}\,m_\oplus$ for KOI-664.01, and $4.4^{+1.3}_{-1.3}\,m_\oplus$ for KOI-664.03. This gives a high density inner planet, an earth-like density intermediate planet, and a high density outer planet. 


\ksec{Kepler-218 (KOI-711)}
We found no literature mass for the planets in this three-planet system. We were able to constrain the innermost planet to $m=5.4^{+2.0}_{-1.7}\,m_\oplus,\,R\approx1.46\,R_\oplus$ and the outermost planet to $m=6.9^{+1.5}_{-1.3}\,m_\oplus,\,R\approx2.7\,R_\oplus$, so on opposite sides of the radius gap.
The best model TTV of the outer pair is dominated by a 555 days super-period of the 3:1 MMR (period ratio of about 2.78).


\ksec{Kepler-222 (KOI-723)}
We found no literature masses for this 3-planets system. The inner pair is affected by the 2:1 and 5:2 MMRs; our solutions suggest masses of $21.2^{+3.7}_{-16.1}\,m_\oplus$ for KOI-723.01, $7.5^{+2.8}_{-2.4}\,m_\oplus$ for KOI-723.03, and $6.8^{+3.3}_{-2.5}\,m_\oplus$ for KOI-723.02. We note that the obtained solution is highly questionable due to the large eccentricity (of order 0.3) of the innermost 3.9-day orbit planet.


\ksec{Kepler-229 (KOI-757)}
We found no literature masses for this 3-planets system. We present 3 solutions, the adopted one was chosen on the basis of both fit quality and physical reasoning: the other solutions suggest either significant eccentricity of the innermost, 6.25-days orbit planet, or a large misalignemnt ($\sim$20\deg) of the outermost planet. The adopted solution suggests masses of $10.5^{+3.4}_{-2.7}\,m_\oplus$ for KOI-757.03, $8.4^{+0.9}_{-1.0}\,m_\oplus$ for KOI-757.01, $7.6^{+1.1}_{-1.2}\,m_\oplus$ for KOI-757.02, yielding a rocky density for the innermost, 6.25-days orbit, planet, and much lower densities for the two other planets.
The difference between our best-fit $a^{(1)}/R_*$ and the value obtained from literature stellar properties and the orbital period of the innermost planet is $3.09\,\sigma$, suggesting a possible degeneracy between $a^{(1)}/R_*$ and $b$ \citep{Carter2008, Kipping2010}.

\ksec{Kepler-231 (KOI-784)}
HL14 report nominal masses of $35.4^{+19.7}_{-16.6}\,m_\oplus$ and $24.1^{+13.5}_{-11.2}\,m_\oplus$, and classify this as a ``high eccentricity" system.
We were able to constrain the mass of the inner planet in this two-planet system to $14.7^{+2.2}_{-1.7}\,m_\oplus$ by leveraging  the outer planet's TTVs (of order 10 minutes). This mass, however, implies a high density of $10.6^{+4.1}_{-1.7}\,{\rm g}\,{\rm cm}^{-3}$ which is possible but rather high. An alternative explanation for the TTVs of the planet may be an unseen external companion - which is beyond the scope of the discussion here.


\ksec{Kepler-242 (KOI-853)}
We found no literature masses for this 2-planets system.
This system is interestingly affected by the 7:4, 5:3 and 2:1 resonances simultaneously (orbital period ratio of 1.767). Though there are apparently different solutions for the masses in the system - spanning a factor of $\sim4$ in absolute mass of both planets - they tightly agree on the mass ratio between the planets, and they are all statistically similar (closer than $3\,\sigma$ apart). The adopted solution is the lowest-$\chi^2$ one, which also has the highest masses.

In all our runs the outer planet showed $db/dt$ in significance of $2\,\sigma$; this finding along with its TTV might imply the possible existence of an external unseen companion.
We also note that the $a^{(1)}/R_*$ disagreement with literature stellar properties is $\sim 5\,\sigma$, further supporting the understanding that this system requires more observations for constraining its properties.


\ksec{Kepler-244 (KOI-864)}
In this 3-planets system the inner pair is rather far away from the 2:1 MMR ($\Delta\sim0.1326$), while the outer is close to the 2:1 MMR with ($\Delta\sim0.0264$) and a super-period of about 379 days. HL14 analyzed the TTV of the outer pair, and estimated the nominal mass of the outermost planet as $15.2^{+20}_{-13}\,m_\oplus$. They classify the pair in the ``low eccentricity" category.

We present two solutions of different mass estimates, both significantly detect the mass of the innermost planet to be about $5\,m_\oplus$ with a sub-Earth density. 


\ksec{Kepler-81 (KOI-877)}
HL14 calculated the nominal masses of the two inner planets of this 3-planets system to be $17.2^{+6.3}_{-5.5}\,m_\oplus$ and $4.3^{+2.7}_{-1.8}\,m_\oplus$. 
HL17 gave two mass estimates for the same planets, given different priors: they found masses that are both consistent with zero for the default mass prior, and masses of $8.2^{+3.4}_{-3.8}\,m_\oplus$ and $3.9^{+1.0}_{-1.2}\,m_\oplus$ for the high-mass prior. 
JH21 estimated upper limits of 16.7$\,m_\oplus$ for KOI-877.01 (planet b), 5.34$\,m_\oplus$ for KOI-877.02 (planet c), and 6.02$\,m_\oplus$  for KOI-877.03 (planet d). Our adopted solution suggests masses of $6.2^{+0.4}_{-0.4}\,m_\oplus$, $3^{+0.8}_{-0.7}\,m_\oplus$, and $3.3^{+3.2}_{-2.8}\,m_\oplus$ respectively, consistent with the high-mass prior of HL17, and in agreement with the limits set by JH21.


\ksec{Kepler-83 (KOI-898)}
The system was analysed by HL14 with no significant nominal mass constraints.
Our solution suggests sub-earth density for all three planets, with masses of (sorted by orbital period): $4.7^{+0.3}_{-0.3}\,m_\oplus$ for KOI-898.02, $6.4^{+0.4}_{-0.4}\,m_\oplus$ for KOI-898.01, and $3.3^{+0.4}_{-0.4}\,m_\oplus$ for KOI-898.03. Given the planets' radii this means monotonically decreasing density (i.e. increasing volatiles fraction) with orbital period.
The best-fit solution displays TTVs of order 1 minute for the innermost planet, 4 minutes for the intermediate planet and 10 minutes for the outermost planet. In addition, the outer planet displays significant $db/dt$, clearly seen in the long-cadence data (see Figure~\ref{fig:bVariations}). The solution involves large inclinations to explain these impact parameter variations; alternatively, the system should involve an external unseen companion (see Figure~\ref{fig:ExternalUnseenPerturber1}).




\ksec{Kepler-249 (KOI-899)}
We found no literature masses for this 3-planets system of small planets of radii between 1.2 and 1.5$\,R_\oplus$.
Our solution suggests masses of $4.1^{+0.7}_{-0.7}\,m_\oplus$ for KOI-899.02, $7.5^{+1.2}_{-1.4}\,m_\oplus$ for KOI-899.01, and $1.9^{+0.8}_{-0.7}\,m_\oplus$ for KOI-899.03. The solution is questionable as it involves rather high densities for the two inner planets, along with significant eccentricities of more than 0.1, which are not expected for 3.3 and 7.1 days orbit planets.


\ksec{Kepler-253 (KOI-921)}
We found no literature masses for this 3-planets system.
Our solution suggests masses of $2.8^{+2}_{-1.4}\,m_\oplus$ for KOI-921.03, $4.4^{+1.1}_{-1.1}\,m_\oplus$ for KOI-921.01, $9.5^{+1.0}_{-1.2}\,m_\oplus$ for KOI-921.02, implying sub-earth densities at co-planar orbits.


\ksec{Kepler-254 (KOI-934)}
Previous analyses (HL14, JH21) were unable to firmly detect any of the masses in the system.
Our two solutions have significant mass detections in the 5-10$\,\sigma$ range for all thee planets, which are consistent with each other in mass, agree with the upper limits of JH21, and differ in some of the eccentricity and inclination components. They imply sub-earth densities for all three transiting planets in this system, which is compatible with previous trends \citep{Rogers2015} given that they have radii in the range of 2.1-3.6$\,R_\oplus$. The two outer planets display impact parameter variations of $\sim 2\,\sigma$ significance.


\ksec{Kepler-276 (KOI-1203)}
HL14 did not provide firm nominal masses in this 3-planet system. X14 estimated nominal masses of $16.6^{+4.4}_{-3.6}\,m_\oplus$ for c and $16.3^{+5}_{-4.3}\,m_\oplus$ for d (the outer pair), which are already rather high masses as all planets in the system have radii of $\approx2.5\,R_\oplus$, with upper limits on the true masses higher still at $82.2\,m_\oplus$ and $74.9\,m_\oplus$.
Our solution suggests masses of $6.7^{+2}_{-1.9}\,m_\oplus$ for KOI-1203.02, $17.8^{+3.8}_{-3.7}\,m_\oplus$ for KOI-1203.01, and $8.6^{+1.3}_{-1.3}\,m_\oplus$ for KOI-1203.03, implying sub-earth densities for all three planets in this system.


\ksec{Kepler-277 (KOI-1215)}
HL14 and X14 could not put firm constraints on the nominal masses in the system. Later, HL17 highlighted the mass-eccentricity degeneracy inherent in this system, resulting in masses consistent with zero when using the default mass prior, and masses of $46.7^{+27.0}_{-7.0}\,m_\oplus$ and $37.6^{+14.1}_{-2.8}\,m_\oplus$ for planets b and c, respectively, when using a high mass prior.
This system is near the 2:1 from the inside, with $\Delta=-0.0474$, with the outer planet having a TTV about twice as large as the inner planet TTV ($\sigma_{\rm TTV}$ of about 5 and 10 minutes). Two solutions arise, with different masses. The second solution, with a high-density inner planet, is adopted based on fit quality. The eccentricities of both solutions are of order a few per cent.


\ksec{Kepler-287 (KOI-1307)}
We found no literature masses for this 2-planets systems. Because it is quite far from the 2:1 resonance ($\Delta \sim 0.1$) the TTV amplitude is small; our best fitting amplitude is of order 1 minute (inner planet) and 5 minutes (outer planet).
Our solution suggests masses of $32^{+11}_{-11}\,m_\oplus$ for KOI-1307.02, $8.6^{+3.6}_{-3}\,m_\oplus$ for KOI-1307.01, less than $3\,\sigma$ mass detection.
The small TTV magnitude, the low mass significance ($<3\,\sigma$) and the $a^{(1)}/R_*$ misagreement with literature stellar properties ($\sim 5.9\,\sigma$) all suggest that this system requires more observations in order to constrain its properties.


\ksec{Kepler-310 (KOI-1598)}
HL17 did not provide mass estimates better than $3\,\sigma$ significant. JH21 estimated an upper limit of mass to $15.3\,m_\oplus$ on KOI-1598.01, $7.34^{+3.16}_{-2.61}\,m_\oplus$ for KOI-1598.02 and upper limit of $7.88\,m_\oplus$ for KOI-1598.03. Our results (masses of $3.8^{+2.5}_{-2.6}\,m_\oplus$ for KOI-1598.03, $7.5^{+2.1}_{-1.7}\,m_\oplus$ for KOI-1598.01 and $4.6^{+1.3}_{-1.0\,m_\oplus}$ for KOI-1598.02) agree with their upper limits, and agree with the values of 1598.02 up to the uncertainty.


\ksec{Kepler-325 (KOI-1832)}
We found no literature mass for this 3-planets system.
We present three solutions, the third one rather eccentric. The adopted solution suggests small eccentricities, and sub-Earth densities for all three $\sim2.5\,R_\oplus$ planets in this system.


\ksec{Kepler-326 (KOI-1835)}
All three planets in the system lie just above the with/without gaseous envelope transition with radii between 1.8-2.1$\,R_\oplus$ \citep{Rogers2015}. Nominal masses of HL14 are all consistent with zero.
We obtained 5 different solutions, different in their eccentricity and inclination angles but all of them similar in quality and suggest planetary masses of up to $\sim 5\,m_\oplus$ and sub-Earth density.



\ksec{Kepler-328 (KOI-1873)}
Nominal masses from X14 are $28.5^{+12.9}_{-12.3}\,m_\oplus$   (inner planet, b) and    $39.4^{+13.6}_{-12.6}\,m_\oplus$   (outer planet, c).
Maximal mass estimate are 23.6 and 874.5.
The interaction of this pair is dominated by the proximity to the 2:1 resonance, with a super-period of about 1680 days, close to the span of the data taken by {Kepler}.
We present two solutions, both implying a density of roughly $1\,{\rm g}\,{\rm cm}^{-3}$ to the planets, with the outer planet mass a few times that of Neptune and with the inner planet mass a few times that of Earth, with eccentricities of a few per cent.


\ksec{Kepler-331 (KOI-1895)}
We found no literature mass for this 3-planets system.
We report two solutions, one of them preferred due to better fit quality. Both agree on the masses, of which the masses of the outer pair are statistically significant at $5.9^{+1.1}_{-1.2}m\oplus$ and  $5.11^{+0.85}_{-0.96}m\oplus$ for planets Kepler-331 c and Kepler-331 d, respectively. All planets have radii of 2.4-2.7$\,R_\oplus$, and thus low densities, i.e., they are above the radius gap \citep{Fulton2017}, and with small inclinations. There is no evidence for impact parameter variations.


\ksec{Kepler-333 (KOI-1908)}

We present two solutions, both suggest an inner mass of a several $m_\oplus$ for the inner planet with a radius of roughly $1.45\,R_\oplus$, implying a high density and small atmosphere. The outer planet mass is of low significance ($\sim 2\,\sigma$).

The large mass fitted for the inner planet stems from the need to explain the large TTVs ($>10$ minutes) of the outer planet; Alternatively, the TTVs may be explained by an unseen companion in an even larger orbit, potentially relaxing the almost unnaturally-high density of the innermost planet. 



\ksec{Kepler-334 (KOI-1909)}
We found no literature mass for the three planets in this system, which range from 1.0 to 1.5$\,R_\oplus$ in size. The solution we present provides statistically significant masses for all planets, but suggests a high density for the innermost and smallest planet; to make it physically plausible the true value must lie about 2-3 standard deviations below the median. The deduced densities of the other planets seem reasonable, but the system as a whole would benefit from additional observational data. There is no evidence for significant impact parameter variations.


\ksec{Kepler-339 (KOI-1931)}
HL14 did not provide firm nominal mass estimates for the planets in the system. The three planets are small, ranging from 1.1 to 1.3 $R_\oplus$ in size.
Our adopted solutions suggests masses of $4.4^{+0.9}_{-1}\,m_\oplus$ for KOI-1931.01, $1.3^{+0.4}_{-0.4}\,m_\oplus$ for KOI-1931.03, and $4.1^{+0.6}_{-0.6}\,m_\oplus$ for KOI-1931.02, and suggests a borderline density for the outermost planet.


\ksec{Kepler-350 (KOI-2025)}
X14 analyzed the TTV of the interaction between planets c and d and estimated nominal masses of $6.1^{+3.3}_{-3.2}\,m_\oplus$ for planet c and $14.9^{+5.3}_{-4.7}\,m_\oplus$ for planet d, with maximal masses of 33.6$\,m_\oplus$ and 1117.4$\,m_\oplus$.
HL14 analyzed the same system, and obtained $7.2^{+8.5}_{-5.9}\,m_\oplus$ for planet c and $14.8^{+12.3}_{-6.8}\,m_\oplus$ for planet d, classifying them as ``high eccentricity".
We were able to constrain all three masses: $3.6^{+0.4}_{-0.4}\,m_\oplus$ for KOI-2025.03, $5.4^{+1}_{-1.5}\,m_\oplus$ for KOI-2025.01, $7.2^{+1.3}_{-1.3}\,m_\oplus$ for KOI-2025.02. The density of the innermost, smallest transiting planet is significantly larger than the densities of the two outer and larger transiting planets, which suggests that the former ($R\approx1.6\,R_\oplus$) is below the radius gap and the latter ($R\approx2.7\,R_\oplus$ and $1.9\,R_\oplus$) are above it.


\ksec{Kepler-358 (KOI-2080)}
We found no literature masses for this 2-planets system.
This planet pair, with period ratio of 2.4512, is affected mainly by the 5:2 MMR (super-period of 855 days) and secondly by the 2:1 MMR ($\Delta\approx0.23$, super-period of 185 days). Both corresponding frequencies appear in the best model TTV. The masses are of only $\sim 3\,\sigma$ significance, and a strictly periodic model fits the data almost as good as the dynamical model. Adding the fact that the $a^{(1)}/R_*$ is not consistent with stellar literature data ($\sim 3.25\,\sigma$), we conclude that more data is required in order to shed light on the true masses of these planets.


\ksec{Kepler-363 (KOI-2148)}
HL14 analyzed the TTV of the interaction between planets c and d and classified it as ``high eccentricity". They got nominal masses only for planet c: $66.4^{+71.9}_{-56.6}\,m_\oplus$.
Our solution suggests masses of small significance ($\sim 2 \,\sigma$ for the two outer planets, and $\sim 3.5\,\sigma$ for the innermost planet). The solution involves significant eccentricity for the innermost, 3.6-days-orbit, planet, which is physically questionable. Further data would be required in order to shed light on this system. 


\ksec{Kepler-384 (KOI-2414)}
JH21 placed upper limits on the masses of KOI-2414.01 and KOI-2414.02: $18.55\,m_\oplus$ and $53.07\,m_\oplus$, correspondingly. Our resulting masses are $2.38^{+0.76}_{-0.69}\,m_\oplus$ and $4.28^{+1.35}_{-0.98}\,m_\oplus$ agree with their upper limits. 
A question arises if the system is locked in resonance (the period ratio is 2.0069). The N-body integration we performed in the best-fitting parameters set showed that the resonant angles circulate and not librate. Hence, if our best fitting solution is indeed correct, than the system is not locked in resonance (satisfying the validity requirement of \texttt{AnalyticLC} that the resonant angles circulate).


\ksec{Kepler-400 (KOI-2711)}
We found no literature mass for this 2-planets system.
Our solutions are consistent in most parameters, and imply two dense planets below the radius gap.

\newpage


\section{Notable Systems with no Dynamical Solution}\label{sec:IndividualNoSolution}

In this section we devote a few words to some of the systems for which we did not get a valid solution based on \texttt{AnalyticLC}, but that deserve attention due to special dynamical features or due to significant impact parameter variations. The reason for the invalidity of the solutions obtained by fitting \texttt{AnalyticLC} is given in the table \ref{tab:Reason}.

\ksec{Kepler-113 (KOI-153)}
This system has mass estimates of \citet{Marcy2014}, with $11.7^{+4.2}_             {-4.2}$ for the inner planet (153.02) and an upper limit of 8.7 for the outer planet (153.01). The system is at a distance of $\sim 0.061$ from the 2:1 MMR.
We did not get a solution that passed our tests for this system; however we note the strong impact parameter variations of both planets, that might imply the possible existence of an external companion.

\ksec{Kepler-9 (KOI-377)}
This system was dynamically analyzed in many studies \citep[e.g.][]{Holman2010, FreudenthalEtAl2018}. It was highlighted as having significant transit duration variations by \citet{Shahaf2021}. Our solutions did not pass the N-body test, though we highlight it as having significant impact paramater variations of both planets.

\ksec{Kepler-28 (KOI-870)}
This 2-planets system was first analyzed by \citet{Steffen2012}, who showed that the anti-correlated TTVs, arising from the proximity ($\Delta \sim 0.0132$ to the 3:2 MMR, is likely to be due to the interaction of two planets in one planetary system. HL14 classified this as a ``low eccentricity" system, reporting the nominal masses to be $8.8^{+3.8}_{-3.1}$ and $10.9^{+6.1}_{-4.5}$. HL17 made mass estimates using low and high mass priors.
The proximity of the planets pair to the 3:2 resonance ($\Delta \sim 0.0132$) creates a TTV signal of a few minutes to both planets. 
Though our solutions did not pass the N-body test, we report $db/dt$ at the $3\,\sigma$ level for both planets, when the magnitude of the impact parameter decreases for both planets; this can be explained by their cross-interaction if their impact parameters are of opposite signs, or by an external non-transiting companion if their impact parameters are of the same sign. 




\ksec{Kepler-29 (KOI-738)}
This system was analyzed in terms of TTV by \citet{Fabrycky2012}, \citet{JontofHutter2016}, and \citet{Vissapragada2020}. The planets are close to both the 9:7 MMR and the 5:4 MMR. Though we do not provide a dynamical solution, we report strong impact parameter variations of both planets in this system (see table \ref{tab:SignificantDbdt}).

\ksec{Kepler-51 (KOI-620)}
\citet{Masuda2014} analyzed this system using TTV fitting based on N-body integrations, and showed that the existing planets can explain their TTVs. He also concentrated on a planet-planet eclipse event at BKJD=1513 days, and used it to constrain the mutual inclination by estimating the roll angle on the sky of these planets to be $-25.3^{+6.2}_{-6.8}$\deg; a significant value. 
We did not get a dynamical solution based on \texttt{AnalyticLC}; however, we report strong impact parameter variations, which are, along with the finding of \citet{Masuda2014}, another sign of mutual inclination. A further analysis including full light curve fitting of this system based on an N-body model and that accounts for mutual transits could possibly better constrain the mutual inclination in this system. this is beyond the scope of this study and is left for future work.


\ksec{Kepler-18 (KOI-137)}
This system has been analyzed by joint fitting its TTVs and RV \citep{Cochran2011}, and by its TTVs \citep{HaddenLithwick2014, HaddenLithwick2017}.
Though we did not get a dynamical solution based on \texttt{AnalyticLC}, we report significant impact parameter variations of KOI-137.02 (Kepler-18 d) - the outermost transiting planet in this system.

\ksec{Kepler-261 (KOI-988)}
We did not find literature masses for this 2-planet system.
We did not obtain a dynamical solution based on \texttt{AnalyticLC}; all our runs converged to large eccentricity values which led to solutions inconsistent with a full N-body integration. We report strong impact parameter variations of the outer transiting planet, that might imply the existence of a non-transiting external companion. Such a non-transiting companion might also explain the strong TTV signal of the outer transiting planet, which \texttt{AnalyticLC} was unable to explain with a two-planet model.


\ksec{Kepler-138 (KOI-314)}
This M-dwarf host system has been analyzed in a few studies  \citep{Kipping2014, HaddenLithwick2014, JontofHutter2015, Almenara2018}. It is of high interest due to the small planetary masses it possesses.
\texttt{AnalyticLC} failed to fit a model consistent with a full N-body integration. We found significant impact parameter variations of planet b in all our runs (see table \ref{tab:SignificantDbdt}). In most of our runs, also planets c and d experienced significant impact parameter variations. Such impact parameter variations are possibly a result of mutual inclinations, for which a few different configurations were found to fit the data using a global light-curve fitting based on full N-body modeling \citep{Almenara2018}.


\ksec{Kepler-278 (KOI-1221)}
We found no literature masses for this two-planets system, which resides near the 5:3 MMR.
We did not find a solution compatible with a full N-body model; however we report strong impact parameter variations of the outer planet. This, along with the strong TTV of the outer planet \citep{HolczerEtAl2016}, suggests the possible existence of an external non-transiting companion.



\ksec{Kepler-367 (KOI-2173)}
We found no literature mass for this two-planets system. The period ratio of 1.4168 implies a contribution of both the 7:5 and the 10:7 MMRs. None of our runs yielded a solution that passes all our tests; however we report significant impact parameter variations of the inner transiting planet (see table \ref{tab:SignificantDbdt}).





\ksec{Kepler-337 (KOI-1929)}
We found no literture mass for this two-planets system. Not all the runs converged to a solution with significant impact parameter variations; however four out of five did, and they imply impact parameter variations of both planets, the values of which are given in table \ref{tab:Observable}.


\ksec{Kepler-190 (KOI-579)}
We found no literature mass for this tight system (orbital periods of 2.02 and 3.76 days). The outer planet presents significant impact parameter variations (table \ref{tab:SignificantDbdt}) which could be a result of an external non-transiting companion.



\newpage

\section{Summary and Future Prospects}\label{sec:Discussion}

In this work we have fitted a photometric model to an ensemble of {Kepler} two- and three- planet systems. The model is based on the analytic global light-curve modeling method \texttt{AnalyticLC}, which was described in \citet{Judkovsky2021a}. The posterior distribution obtained from the DE-MC process, combined with literature stellar data, resulted in estimates of the planetary physical and orbital properties. Out of 144 analyzed systems, 54 passed all our validation tests. These include 32 three-transiting-planets systems, and 22 two-transiting-planets systems, summing to 140 planets. For 72 of those we did not find a previously reported literature mass value or upper limit, and out of these we were able to provide the first mass constraints formally better than $3\,\sigma$ for 50 planets.
In addition to planetary masses, we emphasized the role of impact parameter variations (TbVs), which if interpreted as a manifestation of secular orbital nodal motion, are a sign of mutual inclination in the system. Because secular motion occurs on a time scale longer than the observation span, we sought linear impact parameter variations.

Along our fitting process, we recorded the impact parameter variations rate for any fitted planet, thus yielding a posterior distribution of the $db/dt$ even in the case where the provided solution did not pass all our validity tests. We list the planets with the most significant impact parameter variations in Table \ref{tab:SignificantDbdt}. In a recent study, \citet{Shahaf2021} used duration measurements from the catalog of \citet{HolczerEtAl2016}, and listed 15 KOIs with significant TDVs, that might be a result of orbital precession. This work, which only included systems with either two or three planets, did not analyze most of their reported significant-TDV KOIs; for the three KOIs which are included in our analysis (137.02, 209.02, 377.01, 377.02) we did also find impact parameter variations.
Table \ref{tab:SignificantDbdt} lists the KOIs for which we detected significant TbVs; for some of them a dynamical solution is available. For three of those planets we show the illustration of the TbVs in Figure~\ref{fig:bVariations}. Note that Table \ref{tab:SignificantDbdt} includes 59 objects culled only from the two- and three- planet systems, while \citet{Shahaf2021} detected 15 KOIs with a significant TDVs in the entire Kepler dataset. We interpret this difference as stemming from the low duration-information content in the individual Kepler long cadence transits, versus the global model used here.

The model used here includes only the transiting planets, therefore it can explain their TbVs only by the interaction between them; however it is entirely possible that some of these TbVs arise from a non-transiting companion. We therefore performed an analysis which attempts to characterize the possible existence of a non-transiting planet that causes the TbVs; these are illustrated in Figure~\ref{fig:ExternalUnseenPerturber1}. In future work (Ofir et al., in prep.) we will provide a more complete catalog of TbVs, akin to the one given in \citet{OfirEtAl2018}.

The results of this work are first examples for the usage of \texttt{AnalyticLC}. The validation process, which includes full N-body integrations, along with the good agreement with previously reported masses (Figure~\ref{fig:FittedPlanetsOverview1}) give confidence in the reliability of the model, thus giving confidence in the new reported masses and TbVs. We note that the algorithm is robust, and the exact same method was used for all KOIs without special treatment to specific objects. The results are also in good agreement with previous mass determinations and trends (see Figure \ref{fig:FittedPlanetsOverview1}) 

The analytic approach of \texttt{AnalyticLC}, described in \citet{Judkovsky2021a} and used in this work, can be further used for analyzing combined data sets and specifically for ones that span long time epochs for which N-body integrations become increasingly time consuming. Examples for such joint analyses are: Kepler photometry and other photometry (e.g., ground-based, TESS, PLATO), or photometry and RV, or photometry and GAIA astrometry \citep{2016A&A...595A...1G, GAIACollaborationEtAl2018}, or any combination of the above components.
Another research avenue enabled by \texttt{AnalyticLC} is the large-scale search for unseen companions; there are many KOIs for which a significant TTV was detected but no transiting companion was observed \citep{HolczerEtAl2016, OfirEtAl2018}. Understanding the nature of systems with non-transiting companions, and particularly systems with a singly-transiting planet, are of high interest due to the interplay between observed multiplicity and mutual inclination distribution \citep[e.g.][]{XieEtAl2016, He2020}. Further analyses (such as the one exemplified in Figure \ref{fig:ExternalUnseenPerturber1}) of each system is needed.

\acknowledgments
This study was supported by the Helen Kimmel Center for Planetary Sciences and the Minerva Center for Life Under Extreme Planetary Conditions \#13599 at the Weizmann Institute of Science.

This paper includes data collected by the Kepler mission and obtained from the MAST data archive at the Space Telescope Science Institute (STScI). Funding for the Kepler mission is provided by the NASA Science Mission Directorate. STScI is operated by the Association of Universities for Research in Astronomy, Inc., under NASA contract NAS 5-26555.

\newpage



\appendix
\renewcommand\thesection{Appendix \Alph{section}}

\section{Tables}\label{appendix}

\renewcommand\thesubsection{\Alph{section}.\arabic{subsection}}
\subsection{Orbital and Physical Parameters}


\begin{table}
\caption{Physical and orbital elements of our solutions that passes all tests, including N-body matching, AIC improvement over a strictly-periodic model, and physically plausible planetary density. If multiple solutions were found they are all presented; the adopted solution is shown in bold text. For a detailed description of the meaning of these parameters, see \S\ref{sec:ModelParameters}. Only a part of the table is given here for illustration; the full table is given in the machine-readable file MRT1.txt.\label{tab:PhysicalOrbital}}
\begin{turn}{90}
\begin{tabular}{c c c c c c c c c c c c}
 \hline
id &  Kepler- & KOI & Period [d] & $\mu~[10^{-6}]$  & $m~[m_{\oplus}]$  & $\rho_{\rm p}\,[{\rm g\,cm^{-3}}]$ & $R_{\rm p}~[R_{\oplus}]$  & $\Delta e_x$   & $\Delta e_y$   & $I_x~[^{\circ}]$   & $I_y~[^{\circ}]$ \\
[0.5ex] 
\hline\hline
\bf 1 & \bf 100 b & \bf 41.02 & \bf 6.88707 & \bf $\mathbf{13.55^{+4.39}_{-4.5}}$ & \bf $\mathbf{5.14^{+1.7}_{-1.75}}$ & \bf $\mathbf{12.01^{+4.12}_{-4.23}}$ & \bf $\mathbf{1.329^{+0.027}_{-0.026}}$ & \bf $\mathbf{-0.046^{+0.026}_{-0.029}}$ & \bf $\mathbf{0.026^{+0.015}_{-0.016}}$ & \bf $\mathbf{0}$ & \bf $\mathbf{3.06^{+0.11}_{-0.13}}$  \\ 
\bf  & \bf 100 c & \bf 41.01 & \bf 12.8159 & \bf $\mathbf{38.5^{+6.28}_{-7.46}}$ & \bf $\mathbf{14.61^{+2.58}_{-3.04}}$ & \bf $\mathbf{6.87^{+1.3}_{-1.52}}$ & \bf $\mathbf{2.268^{+0.043}_{-0.042}}$ & \bf $\mathbf{0.041^{+0.024}_{-0.022}}$ & \bf $\mathbf{0.007^{+0.021}_{-0.019}}$ & \bf $\mathbf{-10.46^{+3.63}_{-3.21}}$ & \bf $\mathbf{-0.03^{+0.23}_{-0.2}}$  \\ 
\bf  & \bf 100 d & \bf 41.03 & \bf 35.3332 & \bf $\mathbf{2.8^{+1.25}_{-1.21}}$ & \bf $\mathbf{1.06^{+0.48}_{-0.46}}$ & \bf $\mathbf{1.42^{+0.64}_{-0.62}}$ & \bf $\mathbf{1.601^{+0.036}_{-0.036}}$ & \bf $\mathbf{0.016^{+0.031}_{-0.03}}$ & \bf $\mathbf{0.099^{+0.041}_{-0.039}}$ & \bf $\mathbf{-23.2^{+6.21}_{-5.14}}$ & \bf $\mathbf{-1.453^{+0.048}_{-0.051}}$  \\ 
[1ex]
\hline
\bf 1 & \bf 114 b & \bf 156.02 & \bf 5.18856 & \bf $\mathbf{14.51^{+0.42}_{-0.46}}$ & \bf $\mathbf{3.53^{+0.17}_{-0.15}}$ & \bf $\mathbf{8.07^{+0.57}_{-0.45}}$ & \bf $\mathbf{1.337^{+0.022}_{-0.024}}$ & \bf $\mathbf{0.0114^{+0.0042}_{-0.0043}}$ & \bf $\mathbf{0.0052^{+0.0082}_{-0.0081}}$ & \bf $\mathbf{0}$ & \bf $\mathbf{1.15^{+0.15}_{-0.16}}$  \\ 
\bf  & \bf 114 c & \bf 156.01 & \bf 8.04135 & \bf $\mathbf{10.3^{+0.77}_{-0.82}}$ & \bf $\mathbf{2.51^{+0.21}_{-0.21}}$ & \bf $\mathbf{2.55^{+0.24}_{-0.23}}$ & \bf $\mathbf{1.751^{+0.029}_{-0.031}}$ & \bf $\mathbf{0.001^{+0.006}_{-0.006}}$ & \bf $\mathbf{0.0037^{+0.0041}_{-0.0045}}$ & \bf $\mathbf{2.44^{+0.94}_{-0.98}}$ & \bf $\mathbf{1.22^{+0.11}_{-0.14}}$  \\ 
\bf  & \bf 114 d & \bf 156.03 & \bf 11.7761 & \bf $\mathbf{14.11^{+2.9}_{-2.85}}$ & \bf $\mathbf{3.43^{+0.72}_{-0.7}}$ & \bf $\mathbf{0.84^{+0.18}_{-0.17}}$ & \bf $\mathbf{2.812^{+0.045}_{-0.048}}$ & \bf $\mathbf{0.0012^{+0.0031}_{-0.003}}$ & \bf $\mathbf{-0.0017^{+0.0028}_{-0.0025}}$ & \bf $\mathbf{1^{+0.89}_{-0.82}}$ & \bf $\mathbf{1.098^{+0.056}_{-0.072}}$  \\ 
2 & 114 b & 156.02 & 5.18856 & $14.43^{+0.48}_{-0.45}$ & $3.51^{+0.18}_{-0.15}$ & $8.28^{+0.6}_{-0.47}$ & $1.323^{+0.022}_{-0.023}$ & $0.021^{+0.011}_{-0.011}$ & $-0.052^{+0.021}_{-0.023}$ & $0$ & $0.74^{+0.18}_{-0.18}$  \\ 
 & 114 c & 156.01 & 8.04135 & $11.11^{+0.89}_{-0.96}$ & $2.7^{+0.24}_{-0.25}$ & $2.81^{+0.31}_{-0.3}$ & $1.74^{+0.028}_{-0.031}$ & $-0.0002^{+0.0081}_{-0.0075}$ & $0.0148^{+0.0076}_{-0.0077}$ & $-2.73^{+2.81}_{-2.71}$ & $-1.04^{+0.11}_{-0.1}$  \\ 
 & 114 d & 156.03 & 11.7761 & $7.77^{+2.77}_{-2.57}$ & $1.89^{+0.68}_{-0.63}$ & $0.48^{+0.17}_{-0.16}$ & $2.789^{+0.043}_{-0.047}$ & $0.0006^{+0.0034}_{-0.0034}$ & $0.0064^{+0.0039}_{-0.0037}$ & $-1.14^{+2.72}_{-2.62}$ & $-0.985^{+0.043}_{-0.047}$  \\ 
3 & 114 b & 156.02 & 5.18856 & $16.37^{+0.32}_{-0.31}$ & $3.98^{+0.18}_{-0.14}$ & $9.05^{+0.65}_{-0.5}$ & $1.34^{+0.021}_{-0.023}$ & $0.005^{+0.016}_{-0.011}$ & $0.001^{+0.016}_{-0.019}$ & $0$ & $1.09^{+0.097}_{-0.129}$  \\ 
 & 114 c & 156.01 & 8.04135 & $11.54^{+1.44}_{-1.67}$ & $2.81^{+0.37}_{-0.41}$ & $2.86^{+0.4}_{-0.44}$ & $1.754^{+0.028}_{-0.03}$ & $0.0141^{+0.0075}_{-0.0093}$ & $0.005^{+0.0093}_{-0.0088}$ & $2.45^{+1.18}_{-1.21}$ & $1.137^{+0.086}_{-0.092}$  \\ 
 & 114 d & 156.03 & 11.7761 & $13.26^{+4.11}_{-4.73}$ & $3.23^{+1.01}_{-1.16}$ & $0.8^{+0.25}_{-0.29}$ & $2.802^{+0.044}_{-0.048}$ & $-0.0011^{+0.0047}_{-0.0036}$ & $-0^{+0.0032}_{-0.0026}$ & $2.04^{+0.89}_{-1.1}$ & $-1.064^{+0.031}_{-0.028}$  \\ 
4 & 114 b & 156.02 & 5.18856 & $17.26^{+0.65}_{-0.78}$ & $4.2^{+0.23}_{-0.23}$ & $9.85^{+0.83}_{-0.73}$ & $1.325^{+0.023}_{-0.024}$ & $0.0592^{+0.0065}_{-0.0074}$ & $-0.037^{+0.02}_{-0.02}$ & $0$ & $0.43^{+0.16}_{-0.16}$  \\ 
 & 114 c & 156.01 & 8.04135 & $12.48^{+0.56}_{-0.68}$ & $3.04^{+0.18}_{-0.19}$ & $3.2^{+0.25}_{-0.22}$ & $1.732^{+0.028}_{-0.031}$ & $0.0005^{+0.0113}_{-0.0079}$ & $0.0118^{+0.009}_{-0.0095}$ & $7.66^{+4.63}_{-3.22}$ & $1.078^{+0.08}_{-0.091}$  \\ 
 & 114 d & 156.03 & 11.7761 & $18.01^{+2.09}_{-2.41}$ & $4.38^{+0.54}_{-0.6}$ & $1.1^{+0.14}_{-0.15}$ & $2.791^{+0.043}_{-0.048}$ & $-0.0102^{+0.0023}_{-0.0021}$ & $0.005^{+0.0032}_{-0.0032}$ & $6.96^{+3.89}_{-2.8}$ & $-1.053^{+0.049}_{-0.038}$  \\ 
5 & 114 b & 156.02 & 5.18856 & $15.95^{+0.5}_{-0.65}$ & $3.88^{+0.2}_{-0.2}$ & $9.18^{+0.7}_{-0.64}$ & $1.322^{+0.021}_{-0.022}$ & $0.043^{+0.022}_{-0.015}$ & $-0.111^{+0.067}_{-0.042}$ & $0$ & $0.07^{+0.114}_{-0.051}$  \\ 
 & 114 c & 156.01 & 8.04135 & $9.86^{+0.76}_{-0.7}$ & $2.4^{+0.21}_{-0.19}$ & $2.58^{+0.26}_{-0.23}$ & $1.719^{+0.027}_{-0.029}$ & $0.0013^{+0.0067}_{-0.005}$ & $0.0262^{+0.0085}_{-0.0131}$ & $-1.75^{+2.33}_{-1.76}$ & $-0.867^{+0.093}_{-0.075}$  \\ 
 & 114 d & 156.03 & 11.7761 & $7.26^{+2.48}_{-2.04}$ & $1.77^{+0.61}_{-0.5}$ & $0.45^{+0.16}_{-0.13}$ & $2.777^{+0.043}_{-0.046}$ & $0.0007^{+0.0042}_{-0.0061}$ & $0.015^{+0.0059}_{-0.0095}$ & $0.24^{+1.88}_{-1.33}$ & $-0.95^{+0.028}_{-0.038}$  \\ 
[1ex]
\hline
\bf 1 & \bf 117 b & \bf 209.02 & \bf 18.7959 & \bf $\mathbf{57.9^{+12.1}_{-13.3}}$ & \bf $\mathbf{24.2^{+5.19}_{-5.76}}$ & \bf $\mathbf{0.254^{+0.061}_{-0.065}}$ & \bf $\mathbf{8.02^{+0.29}_{-0.26}}$ & \bf $\mathbf{-0.0516^{+0.0046}_{-0.0044}}$ & \bf $\mathbf{0.0172^{+0.0024}_{-0.0022}}$ & \bf $\mathbf{0}$ & \bf $\mathbf{1.294^{+0.032}_{-0.035}}$  \\ 
\bf  & \bf 117 c & \bf 209.01 & \bf 50.7903 & \bf $\mathbf{1428.7^{+57.9}_{-58.9}}$ & \bf $\mathbf{596.8^{+37.4}_{-44.9}}$ & \bf $\mathbf{1.84^{+0.23}_{-0.23}}$ & \bf $\mathbf{12.05^{+0.43}_{-0.39}}$ & \bf $\mathbf{0.0286^{+0.0025}_{-0.0026}}$ & \bf $\mathbf{-0.0336^{+0.0018}_{-0.002}}$ & \bf $\mathbf{-1.27^{+0.17}_{-0.18}}$ & \bf $\mathbf{-0.316^{+0.039}_{-0.034}}$  \\ 
[1ex]
\hline
\bf 1 & \bf 120 b & \bf 222.01 & \bf 6.31251 & \bf $\mathbf{24.15^{+2.33}_{-2.31}}$ & \bf $\mathbf{5.6^{+0.59}_{-0.56}}$ & \bf $\mathbf{2.12^{+0.24}_{-0.23}}$ & \bf $\mathbf{2.437^{+0.047}_{-0.046}}$ & \bf $\mathbf{-0.0063^{+0.0047}_{-0.0045}}$ & \bf $\mathbf{-0.0033^{+0.0076}_{-0.0086}}$ & \bf $\mathbf{0}$ & \bf $\mathbf{0.39^{+0.18}_{-0.18}}$  \\ 
\bf  & \bf 120 c & \bf 222.02 & \bf 12.7945 & \bf $\mathbf{20.54^{+1.42}_{-1.57}}$ & \bf $\mathbf{4.76^{+0.38}_{-0.39}}$ & \bf $\mathbf{3.57^{+0.33}_{-0.32}}$ & \bf $\mathbf{1.942^{+0.039}_{-0.04}}$ & \bf $\mathbf{0.0075^{+0.0077}_{-0.0078}}$ & \bf $\mathbf{0.009^{+0.013}_{-0.015}}$ & \bf $\mathbf{0.04^{+1.48}_{-1.69}}$ & \bf $\mathbf{-0.07^{+0.41}_{-0.32}}$  \\ 
[1ex]
\hline
\end{tabular}
\end{turn}
\end{table}

\clearpage

\subsection{Light-Curve Transit and MMR-Proximity Parameters}
\begin{table}
\caption{Light curve transit and MMR-proximity parameters of all analyzed planets. A part of the table is shown here for visualization; the full table is given in a machine-readable file MRT2.txt. The parameters of the planets whose system have a valid dynamical solution appear first, as in Table \ref{tab:PhysicalOrbital}, with the adopted solutions marked using the ``Adopted" flag in the file and shown in bold text in the lines printed below. The parameters of the systems with no valid solutions appear afterwards. \label{tab:Observable}}
\begin{turn}{90}
\begin{tabular}{c c c c c c c c c c c c }\hline
id & Kepler- & KOI & Period [d] & $\sigma_{\rm {TTV}}$ [min] & $J$ & $\Delta$ & S.P. [day]& $T~[\rm hr]$ & $\tau~[\rm min]$ & $b$ & $db/dt~[\rm yr^{-1}]$  \\
[0.5ex] 
\hline\hline
\bf 1 & \bf 100 b & \bf 41.02 & \bf 6.88707 & \bf 3.21 & \bf  & \bf  & \bf  & \bf $\mathbf{4.333^{+0.046}_{-0.049}}$ & \bf $\mathbf{3.13^{+0.18}_{-0.16}}$ & \bf $\mathbf{0.581^{+0.033}_{-0.034}}$ & \bf $\mathbf{-0.0166^{+0.0071}_{-0.0069}}$  \\ 
\bf  & \bf 100 c & \bf 41.01 & \bf 12.8159 & \bf 0.52 & \bf 2 & \bf -0.07 & \bf 92.1& $\mathbf{5.98^{+0.15}_{-0.17}}$ & \bf $\mathbf{4.9^{+0.13}_{-0.14}}$ & \bf $\mathbf{-0.009^{+0.06}_{-0.053}}$ & \bf $\mathbf{0.006^{+0.0032}_{-0.0028}}$  \\ 
\bf  & \bf 100 d & \bf 41.03 & \bf 35.3332 & \bf 1.17 & \bf 2 & \bf 0.38 & \bf 46.7& $\mathbf{5.71^{+0.17}_{-0.11}}$ & \bf $\mathbf{7.38^{+0.54}_{-0.55}}$ & \bf $\mathbf{-0.743^{+0.028}_{-0.023}}$ & \bf $\mathbf{0.0064^{+0.0027}_{-0.0026}}$  \\ 
[1ex]
\hline
\bf 1 & \bf 114 b & \bf 156.02 & \bf 5.18856 & \bf 0.28 & \bf  & \bf  & \bf  & \bf $\mathbf{2.361^{+0.012}_{-0.012}}$ & \bf $\mathbf{2.659^{+0.083}_{-0.082}}$ & \bf $\mathbf{0.314^{+0.035}_{-0.04}}$ & \bf $\mathbf{0.005^{+0.0019}_{-0.0021}}$  \\ 
\bf  & \bf 114 c & \bf 156.01 & \bf 8.04135 & \bf 1.47 & \bf 3 & \bf 0.033 & \bf 80.7& $\mathbf{2.594^{+0.025}_{-0.028}}$ & \bf $\mathbf{4.34^{+0.16}_{-0.18}}$ & \bf $\mathbf{0.449^{+0.035}_{-0.046}}$ & \bf $\mathbf{-0.0113^{+0.0043}_{-0.0038}}$  \\ 
\bf  & \bf 114 d & \bf 156.03 & \bf 11.7761 & \bf 2.05 & \bf 3 & \bf -0.024 & \bf 166& $\mathbf{2.812^{+0.015}_{-0.015}}$ & \bf $\mathbf{8.27^{+0.29}_{-0.29}}$ & \bf $\mathbf{0.522^{+0.02}_{-0.028}}$ & \bf $\mathbf{0.0027^{+0.002}_{-0.0023}}$  \\ 
2 & 114 b & 156.02 & 5.18856 & 0.24 &  &  & & $2.35^{+0.013}_{-0.014}$ & $2.464^{+0.061}_{-0.049}$ & $0.204^{+0.05}_{-0.049}$ & $-0.0058^{+0.0063}_{-0.0063}$  \\ 
 & 114 c & 156.01 & 8.04135 & 1.02 & 3 & 0.033 & 80.7& $2.614^{+0.047}_{-0.045}$ & $4.08^{+0.14}_{-0.12}$ & $-0.394^{+0.045}_{-0.039}$ & $0.0109^{+0.0085}_{-0.0085}$  \\ 
 & 114 d & 156.03 & 11.7761 & 1.99 & 3 & -0.024 & 166& $2.831^{+0.034}_{-0.031}$ & $7.78^{+0.17}_{-0.15}$ & $-0.479^{+0.017}_{-0.018}$ & $-0.0035^{+0.0029}_{-0.0029}$  \\ 
3 & 114 b & 156.02 & 5.18856 & 0.3 &  &  & & $2.364^{+0.013}_{-0.013}$ & $2.646^{+0.049}_{-0.061}$ & $0.301^{+0.029}_{-0.039}$ & $0.006^{+0.0031}_{-0.0032}$  \\ 
 & 114 c & 156.01 & 8.04135 & 1.22 & 3 & 0.033 & 80.7& $2.582^{+0.045}_{-0.035}$ & $4.157^{+0.119}_{-0.09}$ & $0.416^{+0.033}_{-0.034}$ & $-0.0095^{+0.0054}_{-0.0052}$  \\ 
 & 114 d & 156.03 & 11.7761 & 2.05 & 3 & -0.024 & 166& $2.819^{+0.031}_{-0.028}$ & $8.107^{+0.096}_{-0.094}$ & $-0.511^{+0.014}_{-0.011}$ & $-0.0003^{+0.0023}_{-0.002}$  \\ 
4 & 114 b & 156.02 & 5.18856 & 0.43 &  &  & & $2.367^{+0.012}_{-0.012}$ & $2.413^{+0.036}_{-0.032}$ & $0.111^{+0.04}_{-0.04}$ & $0.0205^{+0.0111}_{-0.0083}$  \\ 
 & 114 c & 156.01 & 8.04135 & 1.8 & 3 & 0.033 & 80.7& $2.688^{+0.029}_{-0.031}$ & $4.18^{+0.11}_{-0.14}$ & $0.393^{+0.029}_{-0.034}$ & $-0.028^{+0.013}_{-0.018}$  \\ 
 & 114 d & 156.03 & 11.7761 & 2.26 & 3 & -0.024 & 166& $2.908^{+0.031}_{-0.025}$ & $8.22^{+0.15}_{-0.18}$ & $-0.501^{+0.021}_{-0.016}$ & $-0.0034^{+0.0029}_{-0.0029}$  \\ 
5 & 114 b & 156.02 & 5.18856 & 0.33 &  &  & & $2.3555^{+0.0078}_{-0.0093}$ & $2.365^{+0.015}_{-0.011}$ & $0.018^{+0.029}_{-0.013}$ & $-0.0029^{+0.0044}_{-0.0032}$  \\ 
 & 114 c & 156.01 & 8.04135 & 1.23 & 3 & 0.033 & 80.7& $2.709^{+0.062}_{-0.05}$ & $3.946^{+0.067}_{-0.062}$ & $-0.323^{+0.038}_{-0.031}$ & $0.0087^{+0.0066}_{-0.008}$  \\ 
 & 114 d & 156.03 & 11.7761 & 2.19 & 3 & -0.024 & 166& $2.891^{+0.051}_{-0.033}$ & $7.71^{+0.23}_{-0.15}$ & $-0.458^{+0.012}_{-0.014}$ & $-0.0048^{+0.0022}_{-0.0022}$  \\ 
[1ex]
\hline
\bf 1 & \bf 117 b & \bf 209.02 & \bf 18.7959 & \bf 8.48 & \bf  & \bf  & \bf  & \bf $\mathbf{6.767^{+0.016}_{-0.016}}$ & \bf $\mathbf{24.02^{+0.32}_{-0.34}}$ & \bf $\mathbf{0.469^{+0.011}_{-0.011}}$ & \bf $\mathbf{-0.0156^{+0.0018}_{-0.002}}$  \\ 
\bf  & \bf 117 c & \bf 209.01 & \bf 50.7903 & \bf 0.69 & \bf 2 & \bf 0.35 & \bf 72.3& $\mathbf{9.63^{+0.041}_{-0.041}}$ & \bf $\mathbf{41.83^{+0.45}_{-0.47}}$ & \bf $\mathbf{-0.205^{+0.024}_{-0.021}}$ & \bf $\mathbf{0.00087^{+0.0002}_{-0.0002}}$  \\ 
[1ex]
\hline
\bf 1 & \bf 120 b & \bf 222.01 & \bf 6.31251 & \bf 1.56 & \bf  & \bf  & \bf  & \bf $\mathbf{2.638^{+0.011}_{-0.011}}$ & \bf $\mathbf{5.151^{+0.099}_{-0.07}}$ & \bf $\mathbf{0.125^{+0.056}_{-0.056}}$ & \bf $\mathbf{0.0001^{+0.0016}_{-0.0018}}$  \\ 
\bf  & \bf 120 c & \bf 222.02 & \bf 12.7945 & \bf 2.99 & \bf 2 & \bf 0.013 & \bf 476& $\mathbf{3.281^{+0.032}_{-0.041}}$ & \bf $\mathbf{5.13^{+0.139}_{-0.088}}$ & \bf $\mathbf{-0.03^{+0.21}_{-0.16}}$ & \bf $\mathbf{-0.0001^{+0.0028}_{-0.0023}}$  \\ 
[1ex]
\hline
\end{tabular}
\end{turn}
\end{table}


\clearpage

\subsection{Stellar Parameters}

\begin{center}
\begin{longtable}{c c c c c c c c } 
\caption{Stellar parameters of the systems for which valid dynamical solutions were found (systems that appear in table \ref{tab:PhysicalOrbital}). The full table is given in the machine-readable file MRT3.txt.} \\
\label{tab:Stellar} \\
 \hline
KOI & Star & KIC & $m~[m_{\odot}]$ & $R~[R_{\odot}]$ & $m,R$ source & $u_1$ & $u_2$  \\
[0.5ex] 
\hline\hline
41 & Kepler-100 & 6521045 & $1.14^{+0.078}_{-0.087}$ & $1.526^{+0.029}_{-0.028}$ & Berger et al. 2020 & 0.4046 & 0.2638  \\ 
[1ex]
\hline
156 & Kepler-114 & 10925104 & $0.731^{+0.029}_{-0.022}$ & $0.725^{+0.011}_{-0.012}$ & Berger et al. 2020 & 0.4639 & 0.2589  \\ 
[1ex]
\hline
209 & Kepler-117 & 10723750 & $1.255^{+0.06}_{-0.079}$ & $1.592^{+0.057}_{-0.052}$ & Berger et al. 2020 & 0.3166 & 0.303  \\ 
[1ex]
\hline
222 & Kepler-120 & 4249725 & $0.697^{+0.029}_{-0.022}$ & $0.698^{+0.013}_{-0.013}$ & Berger et al. 2020 & 0.4639 & 0.2589  \\ 
[1ex]
\hline
\end{longtable}
\end{center}

\clearpage

\subsection{Planets with Significant Impact Parameter Variations}

\begin{center}
\begin{longtable}{c c c c c c c } 
\caption{Planets with significant impact parameter variations. First to be included in the table are systems with dynamical solutions, i.e., that also appear in Table \ref{tab:PhysicalOrbital}, with a $db/dt$ significant to $>2\,\sigma$. We show the values of $db/dt$ for the adopted solution only. The next group to be included in the table are systems without a dynamical solution, but with a $db/dt$ significant to $>2\,\sigma$ for all our runs. The sign of $b_0$ is always positive for the innermost planet in the system. A negative value corresponds to a solution in which the transit occurs on the opposite hemisphere of the stellar disk than the innermost planet. The table is given in the machine-readable file MRT4.txt.} \\
\label{tab:SignificantDbdt} \\
 \hline
Planet & KOI & Period [d] & $N_{\rm pl}$ & Position & $b_0$ & $db/dt~[\rm yr^{-1}]$  \\
[0.5ex] 
\hline\hline
\bf Kepler-100 c & \bf 41.01 & \bf 12.8159 & \bf 3 & \bf 2 & \bf $\mathbf{-0.009^{+0.06}_{-0.053}}$ & \bf $\mathbf{0.006^{+0.0032}_{-0.0028}}$  \\ 
[1ex]
\hline
\bf Kepler-100 b & \bf 41.02 & \bf 6.88707 & \bf 3 & \bf 1 & \bf $\mathbf{0.581^{+0.033}_{-0.034}}$ & \bf $\mathbf{-0.0166^{+0.0071}_{-0.0069}}$  \\ 
[1ex]
\hline
\bf Kepler-100 d & \bf 41.03 & \bf 35.3332 & \bf 3 & \bf 3 & \bf $\mathbf{-0.743^{+0.028}_{-0.023}}$ & \bf $\mathbf{0.0064^{+0.0027}_{-0.0026}}$  \\ 
[1ex]
\hline
\bf Kepler-114 c & \bf 156.01 & \bf 8.04135 & \bf 3 & \bf 2 & \bf $\mathbf{0.449^{+0.035}_{-0.046}}$ & \bf $\mathbf{-0.0113^{+0.0043}_{-0.0038}}$  \\ 
[1ex]
\hline
\bf Kepler-114 b & \bf 156.02 & \bf 5.18856 & \bf 3 & \bf 1 & \bf $\mathbf{0.314^{+0.035}_{-0.04}}$ & \bf $\mathbf{0.005^{+0.0019}_{-0.0021}}$  \\ 
[1ex]
\hline
\bf Kepler-117 c & \bf 209.01 & \bf 50.7903 & \bf 2 & \bf 2 & \bf $\mathbf{-0.205^{+0.024}_{-0.021}}$ & \bf $\mathbf{0.00087^{+0.0002}_{-0.0002}}$  \\ 
[1ex]
\hline
\bf Kepler-117 b & \bf 209.02 & \bf 18.7959 & \bf 2 & \bf 1 & \bf $\mathbf{0.469^{+0.011}_{-0.011}}$ & \bf $\mathbf{-0.0156^{+0.0018}_{-0.002}}$  \\ 
[1ex]
\hline
\bf Kepler-126 b & \bf 260.01 & \bf 10.4957 & \bf 3 & \bf 1 & \bf $\mathbf{0.365^{+0.03}_{-0.031}}$ & \bf $\mathbf{-0.00407^{+0.00093}_{-0.00087}}$  \\ 
[1ex]
\hline
\bf Kepler-126 d & \bf 260.02 & \bf 100.283 & \bf 3 & \bf 3 & \bf $\mathbf{0.061^{+0.041}_{-0.045}}$ & \bf $\mathbf{-0.0008^{+0.00024}_{-0.00024}}$  \\ 
[1ex]
\hline
\bf Kepler-126 c & \bf 260.03 & \bf 21.8697 & \bf 3 & \bf 2 & \bf $\mathbf{0.278^{+0.044}_{-0.041}}$ & \bf $\mathbf{0.00437^{+0.00064}_{-0.00072}}$  \\ 
[1ex]
\hline
\bf Kepler-127 d & \bf 271.01 & \bf 48.6304 & \bf 3 & \bf 3 & \bf $\mathbf{0.5186^{+0.0088}_{-0.0079}}$ & \bf $\mathbf{-0.00236^{+0.0006}_{-0.00062}}$  \\ 
[1ex]
\hline
\bf Kepler-127 c & \bf 271.02 & \bf 29.3934 & \bf 3 & \bf 2 & \bf $\mathbf{0.078^{+0.05}_{-0.058}}$ & \bf $\mathbf{0.0033^{+0.001}_{-0.001}}$  \\ 
[1ex]
\hline
\bf Kepler-450 b & \bf 279.01 & \bf 28.4549 & \bf 3 & \bf 3 & \bf $\mathbf{-0.196^{+0.03}_{-0.018}}$ & \bf $\mathbf{0.0096^{+0.0061}_{-0.0031}}$  \\ 
[1ex]
\hline
\bf Kepler-450 c & \bf 279.02 & \bf 15.4131 & \bf 3 & \bf 2 & \bf $\mathbf{0.052^{+0.056}_{-0.069}}$ & \bf $\mathbf{-0.026^{+0.012}_{-0.014}}$  \\ 
[1ex]
\hline
\bf Kepler-92 b & \bf 285.01 & \bf 13.7488 & \bf 3 & \bf 1 & \bf $\mathbf{0.607^{+0.015}_{-0.019}}$ & \bf $\mathbf{-0.00158^{+0.00073}_{-0.00101}}$  \\ 
[1ex]
\hline
\bf Kepler-92 c & \bf 285.02 & \bf 26.7232 & \bf 3 & \bf 2 & \bf $\mathbf{-0.427^{+0.043}_{-0.035}}$ & \bf $\mathbf{0.0122^{+0.0052}_{-0.0051}}$  \\ 
[1ex]
\hline
\bf Kepler-414 c & \bf 341.01 & \bf 7.17071 & \bf 2 & \bf 2 & \bf $\mathbf{-0.339^{+0.031}_{-0.031}}$ & \bf $\mathbf{-0.0241^{+0.0063}_{-0.0058}}$  \\ 
[1ex]
\hline
\bf Kepler-414 b & \bf 341.02 & \bf 4.69965 & \bf 2 & \bf 1 & \bf $\mathbf{0.106^{+0.062}_{-0.059}}$ & \bf $\mathbf{0.032^{+0.011}_{-0.012}}$  \\ 
[1ex]
\hline
\bf Kepler-146 b & \bf 386.01 & \bf 31.1588 & \bf 2 & \bf 1 & \bf $\mathbf{0.267^{+0.092}_{-0.11}}$ & \bf $\mathbf{0.0087^{+0.0028}_{-0.0033}}$  \\ 
[1ex]
\hline
\bf Kepler-146 c & \bf 386.02 & \bf 76.7325 & \bf 2 & \bf 2 & \bf $\mathbf{0.523^{+0.046}_{-0.051}}$ & \bf $\mathbf{-0.0133^{+0.0052}_{-0.0049}}$  \\ 
[1ex]
\hline
\bf Kepler-170 b & \bf 508.01 & \bf 7.93058 & \bf 2 & \bf 1 & \bf $\mathbf{0.1^{+0.053}_{-0.046}}$ & \bf $\mathbf{0.0153^{+0.0057}_{-0.0054}}$  \\ 
[1ex]
\hline
\bf Kepler-170 c & \bf 508.02 & \bf 16.6659 & \bf 2 & \bf 2 & \bf $\mathbf{0.372^{+0.041}_{-0.041}}$ & \bf $\mathbf{-0.0059^{+0.0026}_{-0.0027}}$  \\ 
[1ex]
\hline
\bf Kepler-177 c & \bf 523.01 & \bf 49.4112 & \bf 2 & \bf 2 & \bf $\mathbf{-0.859^{+0.016}_{-0.012}}$ & \bf $\mathbf{0.00147^{+0.00066}_{-0.00062}}$  \\ 
[1ex]
\hline
\bf Kepler-182 c & \bf 546.01 & \bf 20.6842 & \bf 2 & \bf 2 & \bf $\mathbf{0.199^{+0.087}_{-0.102}}$ & \bf $\mathbf{0.013^{+0.006}_{-0.0053}}$  \\ 
[1ex]
\hline
\bf Kepler-182 b & \bf 546.02 & \bf 9.82578 & \bf 2 & \bf 1 & \bf $\mathbf{0.587^{+0.051}_{-0.064}}$ & \bf $\mathbf{-0.0341^{+0.0091}_{-0.0097}}$  \\ 
[1ex]
\hline
\bf Kepler-191 c & \bf 582.02 & \bf 17.7384 & \bf 3 & \bf 3 & \bf $\mathbf{0.781^{+0.016}_{-0.018}}$ & \bf $\mathbf{-0.0231^{+0.0052}_{-0.005}}$  \\ 
[1ex]
\hline
\bf Kepler-242 c & \bf 853.02 & \bf 14.4965 & \bf 2 & \bf 2 & \bf $\mathbf{-0.766^{+0.052}_{-0.047}}$ & \bf $\mathbf{0.034^{+0.011}_{-0.011}}$  \\ 
[1ex]
\hline
\bf Kepler-83 d & \bf 898.02 & \bf 5.1698 & \bf 3 & \bf 1 & \bf $\mathbf{0.064^{+0.033}_{-0.027}}$ & \bf $\mathbf{0.0274^{+0.0086}_{-0.0103}}$  \\ 
[1ex]
\hline
\bf Kepler-83 c & \bf 898.03 & \bf 20.0902 & \bf 3 & \bf 3 & \bf $\mathbf{0.475^{+0.039}_{-0.036}}$ & \bf $\mathbf{-0.064^{+0.0038}_{-0.0039}}$  \\ 
[1ex]
\hline
\bf Kepler-277 b & \bf 1215.01 & \bf 17.3242 & \bf 2 & \bf 1 & \bf $\mathbf{0.12^{+0.039}_{-0.036}}$ & \bf $\mathbf{0.0034^{+0.002}_{-0.0016}}$  \\ 
[1ex]
\hline
\bf Kepler-277 c & \bf 1215.02 & \bf 33.0063 & \bf 2 & \bf 2 & \bf $\mathbf{0.709^{+0.016}_{-0.017}}$ & \bf $\mathbf{-0.0106^{+0.0048}_{-0.0047}}$  \\ 
[1ex]
\hline
\bf Kepler-310 c & \bf 1598.01 & \bf 56.4758 & \bf 3 & \bf 2 & \bf $\mathbf{0.339^{+0.054}_{-0.06}}$ & \bf $\mathbf{0.0123^{+0.0023}_{-0.0024}}$  \\ 
[1ex]
\hline
\bf Kepler-310 d & \bf 1598.02 & \bf 92.8747 & \bf 3 & \bf 3 & \bf $\mathbf{-0.39^{+0.038}_{-0.036}}$ & \bf $\mathbf{-0.0212^{+0.0058}_{-0.0057}}$  \\ 
[1ex]
\hline
\bf Kepler-310 b & \bf 1598.03 & \bf 13.9307 & \bf 3 & \bf 1 & \bf $\mathbf{0.77^{+0.024}_{-0.029}}$ & \bf $\mathbf{-0.00225^{+0.00052}_{-0.00057}}$  \\ 
[1ex]
\hline
\bf Kepler-350 c & \bf 2025.01 & \bf 17.8485 & \bf 3 & \bf 2 & \bf $\mathbf{0.344^{+0.039}_{-0.051}}$ & \bf $\mathbf{-0.0179^{+0.0073}_{-0.0076}}$  \\ 
[1ex]
\hline
\bf Kepler-350 d & \bf 2025.02 & \bf 26.1363 & \bf 3 & \bf 3 & \bf $\mathbf{-0.361^{+0.037}_{-0.03}}$ & \bf $\mathbf{0.0102^{+0.005}_{-0.0047}}$  \\ 
[1ex]
\hline
 \\ 
[1ex]
\hline
Kepler-18 d & 137.02 & 14.8589 & 3 & 3 & $0.635^{+0.0086}_{-0.0123}$ & $-0.00737^{+0.0006}_{-0.00069}$  \\ 
[1ex]
\hline
Kepler-113 c & 153.01 & 8.92508 & 2 & 2 & $-0.492^{+0.016}_{-0.016}$ & $-0.023^{+0.0024}_{-0.0025}$  \\ 
[1ex]
\hline
Kepler-113 b & 153.02 & 4.754 & 2 & 1 & $0.025^{+0.024}_{-0.016}$ & $0.0395^{+0.0041}_{-0.0049}$  \\ 
[1ex]
\hline
Kepler-138 b & 314.03 & 10.3128 & 3 & 1 & $0.7287^{+0.009}_{-0.0078}$ & $0.0112^{+0.0012}_{-0.0013}$  \\ 
[1ex]
\hline
Kepler-9 b & 377.01 & 19.2708 & 3 & 2 & $0.3875^{+0.002}_{-0.0023}$ & $0.05795^{+0.00016}_{-0.00016}$  \\ 
[1ex]
\hline
Kepler-9 c & 377.02 & 38.9079 & 3 & 3 & $-0.98078^{+0.00052}_{-0.00055}$ & $-0.05903^{+0.00025}_{-0.00027}$  \\ 
[1ex]
\hline
Kepler-9 d & 377.03 & 1.59296 & 3 & 1 & $0.9182^{+0.0017}_{-0.0017}$ & $-0.003979^{+3.9e-05}_{-3.8e-05}$  \\ 
[1ex]
\hline
Kepler-190 b & 579.01 & 2.02 & 2 & 1 & $0.118^{+0.081}_{-0.063}$ & $-0.0227^{+0.0092}_{-0.011}$  \\ 
[1ex]
\hline
Kepler-190 c & 579.02 & 3.76304 & 2 & 2 & $-0.687^{+0.039}_{-0.043}$ & $0.021^{+0.0074}_{-0.007}$  \\ 
[1ex]
\hline
Kepler-51 b & 620.01 & 45.1554 & 3 & 1 & $0.071^{+0.015}_{-0.015}$ & $0.00992^{+0.00061}_{-0.00058}$  \\ 
[1ex]
\hline
Kepler-51 d & 620.02 & 130.178 & 3 & 3 & $0.18^{+0.009}_{-0.0093}$ & $-0.01599^{+0.00073}_{-0.00078}$  \\ 
[1ex]
\hline
Kepler-29 b & 738.01 & 10.3392 & 2 & 1 & $0.619^{+0.014}_{-0.012}$ & $0.01027^{+0.0009}_{-0.00101}$  \\ 
[1ex]
\hline
Kepler-29 c & 738.02 & 13.2869 & 2 & 2 & $-0.679^{+0.014}_{-0.016}$ & $-0.01317^{+0.00105}_{-0.00099}$  \\ 
[1ex]
\hline
Kepler-28 b & 870.01 & 5.91227 & 2 & 1 & $0.417^{+0.027}_{-0.028}$ & $-0.0289^{+0.0071}_{-0.0072}$  \\ 
[1ex]
\hline
Kepler-28 c & 870.02 & 8.98582 & 2 & 2 & $-0.279^{+0.045}_{-0.04}$ & $0.0365^{+0.0087}_{-0.0086}$  \\ 
[1ex]
\hline
Kepler-247 d & 884.02 & 20.4775 & 3 & 3 & $0.378^{+0.064}_{-0.083}$ & $-0.76^{+0.048}_{-0.037}$  \\ 
[1ex]
\hline
Kepler-252 b & 912.02 & 6.66833 & 2 & 1 & $0.542^{+0.05}_{-0.049}$ & $0.034^{+0.013}_{-0.012}$  \\ 
[1ex]
\hline
Kepler-261 c & 988.02 & 24.5709 & 2 & 2 & $0.997^{+0.018}_{-0.021}$ & $-0.0859^{+0.0042}_{-0.004}$  \\ 
[1ex]
\hline
Kepler-278 c & 1221.02 & 51.0771 & 2 & 2 & $0.397^{+0.123}_{-0.094}$ & $-0.0114^{+0.0019}_{-0.0013}$  \\ 
[1ex]
\hline
Kepler-367 b & 2173.01 & 37.8154 & 2 & 1 & $0.394^{+0.03}_{-0.037}$ & $-0.0151^{+0.0042}_{-0.0041}$  \\ 
[1ex]
\hline
Kepler-367 c & 2173.02 & 53.5783 & 2 & 2 & $0.069^{+0.06}_{-0.071}$ & $0.0033^{+0.0013}_{-0.0011}$  \\ 
[1ex]
\hline
Kepler-396 c & 2672.01 & 88.5069 & 2 & 2 & $0.388^{+0.028}_{-0.024}$ & $0.00394^{+0.00026}_{-0.00024}$  \\ 
[1ex]
\hline
Kepler-396 b & 2672.02 & 42.9937 & 2 & 1 & $0.633^{+0.012}_{-0.011}$ & $-0.00324^{+0.00018}_{-0.00019}$  \\ 
[1ex]
\hline
\end{longtable}
\end{center}
\newpage

\subsection{Systems with no Dynamical Solution}

\begin{center}
\begin{longtable}{c c c c } 
\caption{Systems that did not pass our tests and hence are not associated with a dynamical solution. For each system we provide the rejection reason. ``AIC" means that the strictly periodic model attains a better AIC \citep{Akaike1974} than the dynamical model (see \S\ref{sec:AlternativeModel}). ``N-body mismatch" means that the best fitting solution does not match an N-body integration to the level of $\sigma_{\rm N-body}<1.5$ (see \S\ref{sec:Nbody}). ``High density" means that the median density is larger than $12\,{\rm g}\,{\rm cm}^{-3}$ by more than two error bars (see \S\ref{sec:Consistencty}).} \\
\label{tab:Reason} \\
 \hline
Planet & KOI & $N_{\rm pl}$ & Reason  \\
[0.5ex] 
\hline\hline
Kepler-110 & 124 & 2 & AIC, N-body mismatch  \\ 
[1ex]
\hline
Kepler-18 & 137 & 3 & High density of innermost planet  \\ 
[1ex]
\hline
Kepler-113 & 153 & 2 & N-body mismatch  \\ 
[1ex]
\hline
Kepler-116 & 171 & 2 & AIC, N-body mismatch  \\ 
[1ex]
\hline
Kepler-124 & 241 & 3 & N-body mismatch  \\ 
[1ex]
\hline
Kepler-449 & 270 & 2 & AIC  \\ 
[1ex]
\hline
Kepler-128 & 274 & 2 & N-body mismatch  \\ 
[1ex]
\hline
Kepler-130 & 282 & 3 & High density of inner planet  \\ 
[1ex]
\hline
Kepler-134 & 295 & 2 & AIC  \\ 
[1ex]
\hline
Kepler-137 & 313 & 2 & N-body mismatch  \\ 
[1ex]
\hline
Kepler-138 & 314 & 3 & N-body mismatch  \\ 
[1ex]
\hline
Kepler-144 & 369 & 2 & AIC, N-body mismatch  \\ 
[1ex]
\hline
Kepler-9 & 377 & 3 & N-body mismatch  \\ 
[1ex]
\hline
Kepler-148 & 398 & 3 & AIC  \\ 
[1ex]
\hline
Kepler-151 & 413 & 2 & AIC  \\ 
[1ex]
\hline
Kepler-153 & 431 & 2 & N-body mismatch  \\ 
[1ex]
\hline
Kepler-165 & 475 & 2 & N-body mismatch  \\ 
[1ex]
\hline
Kepler-166 & 481 & 3 & AIC, N-body mismatch  \\ 
[1ex]
\hline
Kepler-171 & 509 & 3 & N-body mismatch  \\ 
[1ex]
\hline
Kepler-174 & 518 & 3 & AIC  \\ 
[1ex]
\hline
Kepler-175 & 519 & 2 & AIC  \\ 
[1ex]
\hline
Kepler-179 & 534 & 2 & AIC  \\ 
[1ex]
\hline
Kepler-180 & 542 & 2 & AIC  \\ 
[1ex]
\hline
Kepler-183 & 551 & 2 & AIC  \\ 
[1ex]
\hline
Kepler-190 & 579 & 2 & N-body mismatch  \\ 
[1ex]
\hline
Kepler-194 & 597 & 3 & AIC  \\ 
[1ex]
\hline
Kepler-51 & 620 & 3 & N-body mismatch  \\ 
[1ex]
\hline
Kepler-198 & 624 & 3 & AIC  \\ 
[1ex]
\hline
Kepler-207 & 665 & 3 & AIC  \\ 
[1ex]
\hline
Kepler-211 & 678 & 2 & AIC, N-body mismatch  \\ 
[1ex]
\hline
Kepler-214 & 693 & 2 & N-body mismatch  \\ 
[1ex]
\hline
Kepler-216 & 708 & 2 & AIC  \\ 
[1ex]
\hline
Kepler-217 & 710 & 3 & N-body mismatch  \\ 
[1ex]
\hline
Kepler-219 & 718 & 3 & N-body mismatch  \\ 
[1ex]
\hline
Kepler-29 & 738 & 2 & N-body mismatch  \\ 
[1ex]
\hline
Kepler-226 & 749 & 3 & AIC  \\ 
[1ex]
\hline
Kepler-228 & 756 & 3 & N-body mismatch  \\ 
[1ex]
\hline
Kepler-230 & 759 & 2 & High density of outer planet  \\ 
[1ex]
\hline
Kepler-234 & 800 & 2 & N-body mismatch  \\ 
[1ex]
\hline
Kepler-241 & 842 & 2 & AIC  \\ 
[1ex]
\hline
Kepler-28 & 870 & 2 & N-body mismatch  \\ 
[1ex]
\hline
Kepler-246 & 874 & 2 & N-body mismatch  \\ 
[1ex]
\hline
Kepler-247 & 884 & 3 & N-body mismatch  \\ 
[1ex]
\hline
Kepler-250 & 906 & 3 & AIC  \\ 
[1ex]
\hline
Kepler-252 & 912 & 2 & AIC, N-body mismatch  \\ 
[1ex]
\hline
Kepler-255 & 938 & 3 & AIC  \\ 
[1ex]
\hline
Kepler-257 & 941 & 3 & AIC  \\ 
[1ex]
\hline
Kepler-258 & 951 & 2 & AIC  \\ 
[1ex]
\hline
Kepler-261 & 988 & 2 & N-body mismatch  \\ 
[1ex]
\hline
Kepler-263 & 999 & 2 & N-body mismatch  \\ 
[1ex]
\hline
Kepler-755 & 1050 & 2 & N-body mismatch  \\ 
[1ex]
\hline
Kepler-267 & 1078 & 3 & N-body mismatch  \\ 
[1ex]
\hline
Kepler-770 & 1108 & 3 & AIC  \\ 
[1ex]
\hline
Kepler-270 & 1148 & 2 & AIC  \\ 
[1ex]
\hline
Kepler-272 & 1161 & 3 & AIC  \\ 
[1ex]
\hline
Kepler-278 & 1221 & 2 & N-body mismatch  \\ 
[1ex]
\hline
Kepler-280 & 1240 & 2 & AIC  \\ 
[1ex]
\hline
Kepler-284 & 1301 & 2 & AIC, N-body mismatch  \\ 
[1ex]
\hline
Kepler-288 & 1332 & 3 & N-body mismatch  \\ 
[1ex]
\hline
Kepler-293 & 1366 & 2 & N-body mismatch  \\ 
[1ex]
\hline
Kepler-295 & 1413 & 3 & AIC  \\ 
[1ex]
\hline
Kepler-298 & 1430 & 3 & AIC  \\ 
[1ex]
\hline
Kepler-301 & 1436 & 3 & AIC  \\ 
[1ex]
\hline
Kepler-315 & 1707 & 2 & N-body mismatch  \\ 
[1ex]
\hline
Kepler-318 & 1779 & 2 & AIC  \\ 
[1ex]
\hline
Kepler-319 & 1805 & 3 & AIC  \\ 
[1ex]
\hline
Kepler-321 & 1809 & 2 & AIC  \\ 
[1ex]
\hline
Kepler-323 & 1824 & 2 & AIC  \\ 
[1ex]
\hline
Kepler-327 & 1867 & 3 & N-body mismatch  \\ 
[1ex]
\hline
Kepler-332 & 1905 & 3 & AIC  \\ 
[1ex]
\hline
Kepler-336 & 1916 & 3 & AIC, N-body mismatch  \\ 
[1ex]
\hline
Kepler-337 & 1929 & 2 & N-body mismatch  \\ 
[1ex]
\hline
Kepler-343 & 1960 & 2 & AIC  \\ 
[1ex]
\hline
Kepler-348 & 2011 & 2 & AIC  \\ 
[1ex]
\hline
Kepler-349 & 2022 & 2 & AIC  \\ 
[1ex]
\hline
Kepler-351 & 2028 & 3 & AIC  \\ 
[1ex]
\hline
Kepler-354 & 2045 & 3 & AIC  \\ 
[1ex]
\hline
Kepler-356 & 2053 & 2 & AIC  \\ 
[1ex]
\hline
Kepler-357 & 2073 & 3 & AIC  \\ 
[1ex]
\hline
Kepler-417 & 2113 & 2 & AIC  \\ 
[1ex]
\hline
Kepler-364 & 2153 & 2 & High density of inner planet  \\ 
[1ex]
\hline
Kepler-367 & 2173 & 2 & N-body mismatch  \\ 
[1ex]
\hline
Kepler-368 & 2175 & 2 & N-body mismatch  \\ 
[1ex]
\hline
Kepler-379 & 2289 & 2 & AIC  \\ 
[1ex]
\hline
Kepler-396 & 2672 & 2 & N-body mismatch  \\ 
[1ex]
\hline
Kepler-398 & 2693 & 3 & AIC  \\ 
[1ex]
\hline
Kepler-445 & 2704 & 3 & AIC  \\ 
[1ex]
\hline
Kepler-399 & 2707 & 3 & AIC  \\ 
[1ex]
\hline
Kepler-401 & 2714 & 3 & AIC  \\ 
[1ex]
\hline
Kepler-446 & 2842 & 3 & AIC  \\ 
[1ex]
\hline
\end{longtable}
\end{center}


\bibliography{main.bib}{}

\begin{thebibliography}{}
\expandafter\ifx\csname natexlab\endcsname\relax\def\natexlab#1{#1}\fi
\providecommand{\url}[1]{\href{#1}{#1}}
\providecommand{\dodoi}[1]{doi:~\href{http://doi.org/#1}{\nolinkurl{#1}}}
\providecommand{\doeprint}[1]{\href{http://ascl.net/#1}{\nolinkurl{http://ascl.net/#1}}}
\providecommand{\doarXiv}[1]{\href{https://arxiv.org/abs/#1}{\nolinkurl{https://arxiv.org/abs/#1}}}

\bibitem[{Akaike(1974)}]{Akaike1974}
Akaike, H. 1974, IEEE Transactions on Automatic Control, 19, 716,
  \dodoi{10.1109/TAC.1974.1100705}

\bibitem[{{Almenara} {et~al.}(2018){Almenara}, {D{\'\i}az}, {Dorn}, {Bonfils},
  \& {Udry}}]{Almenara2018}
{Almenara}, J.~M., {D{\'\i}az}, R.~F., {Dorn}, C., {Bonfils}, X., \& {Udry}, S.
  2018, \mnras, 478, 460, \dodoi{10.1093/mnras/sty1050}

\bibitem[{{Ballard} \& {Johnson}(2016)}]{BallardJohnson2016}
{Ballard}, S., \& {Johnson}, J.~A. 2016, \apj, 816, 66,
  \dodoi{10.3847/0004-637X/816/2/66}

\bibitem[{{Berger} {et~al.}(2020){Berger}, {Huber}, {van Saders}, {Gaidos},
  {Tayar}, \& {Kraus}}]{Berger2020}
{Berger}, T.~A., {Huber}, D., {van Saders}, J.~L., {et~al.} 2020, \aj, 159,
  280, \dodoi{10.3847/1538-3881/159/6/280}

\bibitem[{{Borucki} {et~al.}(2010){Borucki}, {Koch}, {Basri}, {Batalha},
  {Brown}, {Caldwell}, {Caldwell}, {Christensen-Dalsgaard}, {Cochran},
  {DeVore}, {Dunham}, {Dupree}, {Gautier}, {Geary}, {Gilliland}, {Gould},
  {Howell}, {Jenkins}, {Kondo}, {Latham}, {Marcy}, {Meibom}, {Kjeldsen},
  {Lissauer}, {Monet}, {Morrison}, {Sasselov}, {Tarter}, {Boss}, {Brownlee},
  {Owen}, {Buzasi}, {Charbonneau}, {Doyle}, {Fortney}, {Ford}, {Holman},
  {Seager}, {Steffen}, {Welsh}, {Rowe}, {Anderson}, {Buchhave}, {Ciardi},
  {Walkowicz}, {Sherry}, {Horch}, {Isaacson}, {Everett}, {Fischer}, {Torres},
  {Johnson}, {Endl}, {MacQueen}, {Bryson}, {Dotson}, {Haas}, {Kolodziejczak},
  {Van Cleve}, {Chandrasekaran}, {Twicken}, {Quintana}, {Clarke}, {Allen},
  {Li}, {Wu}, {Tenenbaum}, {Verner}, {Bruhweiler}, {Barnes}, \&
  {Prsa}}]{BoruckiEtAl2010}
{Borucki}, W.~J., {Koch}, D., {Basri}, G., {et~al.} 2010, Science, 327, 977,
  \dodoi{10.1126/science.1185402}

\bibitem[{Brooks \& Gelman(1998)}]{Brooks1998}
Brooks, S.~P., \& Gelman, A. 1998, Journal of Computational and Graphical
  Statistics, 7, 434, \dodoi{10.1080/10618600.1998.10474787}

\bibitem[{{Bruno} {et~al.}(2015){Bruno}, {Almenara}, {Barros}, {Santerne},
  {Diaz}, {Deleuil}, {Damiani}, {Bonomo}, {Boisse}, {Bouchy}, {H{\'e}brard}, \&
  {Montagnier}}]{Bruno2015}
{Bruno}, G., {Almenara}, J.~M., {Barros}, S.~C.~C., {et~al.} 2015, \aap, 573,
  A124, \dodoi{10.1051/0004-6361/201424591}

\bibitem[{{Carter} {et~al.}(2008){Carter}, {Yee}, {Eastman}, {Gaudi}, \&
  {Winn}}]{Carter2008}
{Carter}, J.~A., {Yee}, J.~C., {Eastman}, J., {Gaudi}, B.~S., \& {Winn}, J.~N.
  2008, \apj, 689, 499, \dodoi{10.1086/592321}

\bibitem[{{Chambers}(1999)}]{Chambers1999}
{Chambers}, J.~E. 1999, \mnras, 304, 793,
  \dodoi{10.1046/j.1365-8711.1999.02379.x}

\bibitem[{{Claret}(2000)}]{Claret2000a}
{Claret}, A. 2000, \aap, 363, 1081

\bibitem[{{Cochran} {et~al.}(2011){Cochran}, {Fabrycky}, {Torres}, {Fressin},
  {D{\'e}sert}, {Ragozzine}, {Sasselov}, {Fortney}, {Rowe}, {Brugamyer},
  {Bryson}, {Carter}, {Ciardi}, {Howell}, {Steffen}, {Borucki}, {Koch}, {Winn},
  {Welsh}, {Uddin}, {Tenenbaum}, {Still}, {Seager}, {Quinn}, {Mullally},
  {Miller}, {Marcy}, {MacQueen}, {Lucas}, {Lissauer}, {Latham}, {Knutson},
  {Kinemuchi}, {Johnson}, {Jenkins}, {Isaacson}, {Howard}, {Horch}, {Holman},
  {Henze}, {Haas}, {Gilliland}, {Gautier}, {Ford}, {Fischer}, {Everett},
  {Endl}, {Demory}, {Deming}, {Charbonneau}, {Caldwell}, {Buchhave}, {Brown},
  \& {Batalha}}]{Cochran2011}
{Cochran}, W.~D., {Fabrycky}, D.~C., {Torres}, G., {et~al.} 2011, \apjs, 197,
  7, \dodoi{10.1088/0067-0049/197/1/7}

\bibitem[{{Deck} \& {Agol}(2015)}]{DeckAgol2015}
{Deck}, K.~M., \& {Agol}, E. 2015, \apj, 802, 116,
  \dodoi{10.1088/0004-637X/802/2/116}

\bibitem[{{Fabrycky} {et~al.}(2012){Fabrycky}, {Ford}, {Steffen}, {Rowe},
  {Carter}, {Moorhead}, {Batalha}, {Borucki}, {Bryson}, {Buchhave},
  {Christiansen}, {Ciardi}, {Cochran}, {Endl}, {Fanelli}, {Fischer}, {Fressin},
  {Geary}, {Haas}, {Hall}, {Holman}, {Jenkins}, {Koch}, {Latham}, {Li},
  {Lissauer}, {Lucas}, {Marcy}, {Mazeh}, {McCauliff}, {Quinn}, {Ragozzine},
  {Sasselov}, \& {Shporer}}]{Fabrycky2012}
{Fabrycky}, D.~C., {Ford}, E.~B., {Steffen}, J.~H., {et~al.} 2012, \apj, 750,
  114, \dodoi{10.1088/0004-637X/750/2/114}

\bibitem[{{Fabrycky} {et~al.}(2014){Fabrycky}, {Lissauer}, {Ragozzine}, {Rowe},
  {Steffen}, {Agol}, {Barclay}, {Batalha}, {Borucki}, {Ciardi}, {Ford},
  {Gautier}, {Geary}, {Holman}, {Jenkins}, {Li}, {Morehead}, {Morris},
  {Shporer}, {Smith}, {Still}, \& {Van Cleve}}]{FabryckyEtAl2014}
{Fabrycky}, D.~C., {Lissauer}, J.~J., {Ragozzine}, D., {et~al.} 2014, \apj,
  790, 146, \dodoi{10.1088/0004-637X/790/2/146}

\bibitem[{{Freudenthal} {et~al.}(2018){Freudenthal}, {von Essen}, {Dreizler},
  {Wedemeyer}, {Agol}, {Morris}, {Becker}, {Mallonn}, {Hoyer}, {Ofir}, {Tal-
  Or}, {Deeg}, {Herrero}, {Ribas}, {Khalafinejad}, {Hern{\'a}ndez}, \&
  {Rodr{\'\i}guez S.}}]{FreudenthalEtAl2018}
{Freudenthal}, J., {von Essen}, C., {Dreizler}, S., {et~al.} 2018, \aap, 618,
  A41, \dodoi{10.1051/0004-6361/201833436}

\bibitem[{{Fulton} \& {Petigura}(2018)}]{Fulton2018}
{Fulton}, B.~J., \& {Petigura}, E.~A. 2018, \aj, 156, 264,
  \dodoi{10.3847/1538-3881/aae828}

\bibitem[{{Fulton} {et~al.}(2017){Fulton}, {Petigura}, {Howard}, {Isaacson},
  {Marcy}, {Cargile}, {Hebb}, {Weiss}, {Johnson}, {Morton}, {Sinukoff},
  {Crossfield}, \& {Hirsch}}]{Fulton2017}
{Fulton}, B.~J., {Petigura}, E.~A., {Howard}, A.~W., {et~al.} 2017, \aj, 154,
  109, \dodoi{10.3847/1538-3881/aa80eb}

\bibitem[{{Gaia Collaboration} {et~al.}(2018){Gaia Collaboration}, {Brown},
  {Vallenari}, {Prusti}, {de Bruijne}, {Babusiaux}, \&
  {Bailer-Jones}}]{GAIACollaborationEtAl2018}
{Gaia Collaboration}, {Brown}, A.~G.~A., {Vallenari}, A., {et~al.} 2018, ArXiv
  e-prints.
\newblock \doarXiv{1804.09365}

\bibitem[{{Gaia Collaboration} {et~al.}(2016){Gaia Collaboration}, {Prusti},
  {de Bruijne}, {Brown}, {Vallenari}, {Babusiaux}, {Bailer-Jones}, {Bastian},
  {Biermann}, {Evans}, {Eyer}, {Jansen}, {Jordi}, {Klioner}, {Lammers},
  {Lindegren}, {Luri}, {Mignard}, {Milligan}, {Panem}, {Poinsignon},
  {Pourbaix}, {Randich}, {Sarri}, {Sartoretti}, {Siddiqui}, {Soubiran},
  {Valette}, {van Leeuwen}, {Walton}, {Aerts}, {Arenou}, {Cropper}, {Drimmel},
  {H{\o}g}, {Katz}, {Lattanzi}, {O'Mullane}, {Grebel}, {Holland}, {Huc},
  {Passot}, {Bramante}, {Cacciari}, {Casta{\~n}eda}, {Chaoul}, {Cheek}, {De
  Angeli}, {Fabricius}, {Guerra}, {Hern{\'a}ndez}, {Jean-Antoine-Piccolo},
  {Masana}, {Messineo}, {Mowlavi}, {Nienartowicz}, {Ord{\'o}{\~n}ez-Blanco},
  {Panuzzo}, {Portell}, {Richards}, {Riello}, {Seabroke}, {Tanga},
  {Th{\'e}venin}, {Torra}, {Els}, {Gracia-Abril}, {Comoretto},
  {Garcia-Reinaldos}, {Lock}, {Mercier}, {Altmann}, {Andrae}, {Astraatmadja},
  {Bellas-Velidis}, {Benson}, {Berthier}, {Blomme}, {Busso}, {Carry},
  {Cellino}, {Clementini}, {Cowell}, {Creevey}, {Cuypers}, {Davidson}, {De
  Ridder}, {de Torres}, {Delchambre}, {Dell'Oro}, {Ducourant}, {Fr{\'e}mat},
  {Garc{\'\i}a-Torres}, {Gosset}, {Halbwachs}, {Hambly}, {Harrison}, {Hauser},
  {Hestroffer}, {Hodgkin}, {Huckle}, {Hutton}, {Jasniewicz}, {Jordan},
  {Kontizas}, {Korn}, {Lanzafame}, {Manteiga}, {Moitinho}, {Muinonen},
  {Osinde}, {Pancino}, {Pauwels}, {Petit}, {Recio-Blanco}, {Robin}, {Sarro},
  {Siopis}, {Smith}, {Smith}, {Sozzetti}, {Thuillot}, {van Reeven}, {Viala},
  {Abbas}, {Abreu Aramburu}, {Accart}, {Aguado}, {Allan}, {Allasia},
  {Altavilla}, {{\'A}lvarez}, {Alves}, {Anderson}, {Andrei}, {Anglada Varela},
  {Antiche}, {Antoja}, {Ant{\'o}n}, {Arcay}, {Atzei}, {Ayache}, {Bach},
  {Baker}, {Balaguer-N{\'u}{\~n}ez}, {Barache}, {Barata}, {Barbier}, {Barblan},
  {Baroni}, {Barrado y Navascu{\'e}s}, {Barros}, {Barstow}, {Becciani},
  {Bellazzini}, {Bellei}, {Bello Garc{\'\i}a}, {Belokurov}, {Bendjoya},
  {Berihuete}, {Bianchi}, {Bienaym{\'e}}, {Billebaud}, {Blagorodnova},
  {Blanco-Cuaresma}, {Boch}, {Bombrun}, {Borrachero}, {Bouquillon}, {Bourda},
  {Bouy}, {Bragaglia}, {Breddels}, {Brouillet}, {Br{\"u}semeister},
  {Bucciarelli}, {Budnik}, {Burgess}, {Burgon}, {Burlacu}, {Busonero}, {Buzzi},
  {Caffau}, {Cambras}, {Campbell}, {Cancelliere}, {Cantat-Gaudin}, {Carlucci},
  {Carrasco}, {Castellani}, {Charlot}, {Charnas}, {Charvet}, {Chassat},
  {Chiavassa}, {Clotet}, {Cocozza}, {Collins}, {Collins}, {Costigan}, {Crifo},
  {Cross}, {Crosta}, {Crowley}, {Dafonte}, {Damerdji}, {Dapergolas}, {David},
  {David}, {De Cat}, {de Felice}, {de Laverny}, {De Luise}, {De March}, {de
  Martino}, {de Souza}, {Debosscher}, {del Pozo}, {Delbo}, {Delgado},
  {Delgado}, {di Marco}, {Di Matteo}, {Diakite}, {Distefano}, {Dolding}, {Dos
  Anjos}, {Drazinos}, {Dur{\'a}n}, {Dzigan}, {Ecale}, {Edvardsson}, {Enke},
  {Erdmann}, {Escolar}, {Espina}, {Evans}, {Eynard Bontemps}, {Fabre},
  {Fabrizio}, {Faigler}, {Falc{\~a}o}, {Farr{\`a}s Casas}, {Faye}, {Federici},
  {Fedorets}, {Fern{\'a}ndez-Hern{\'a}ndez}, {Fernique}, {Fienga}, {Figueras},
  {Filippi}, {Findeisen}, {Fonti}, {Fouesneau}, {Fraile}, {Fraser}, {Fuchs},
  {Furnell}, {Gai}, {Galleti}, {Galluccio}, {Garabato}, {Garc{\'\i}a-Sedano},
  {Gar{\'e}}, {Garofalo}, {Garralda}, {Gavras}, {Gerssen}, {Geyer}, {Gilmore},
  {Girona}, {Giuffrida}, {Gomes}, {Gonz{\'a}lez-Marcos},
  {Gonz{\'a}lez-N{\'u}{\~n}ez}, {Gonz{\'a}lez-Vidal}, {Granvik}, {Guerrier},
  {Guillout}, {Guiraud}, {G{\'u}rpide}, {Guti{\'e}rrez-S{\'a}nchez}, {Guy},
  {Haigron}, {Hatzidimitriou}, {Haywood}, {Heiter}, {Helmi}, {Hobbs},
  {Hofmann}, {Holl}, {Holland}, {Hunt}, {Hypki}, {Icardi}, {Irwin}, {Jevardat
  de Fombelle}, {Jofr{\'e}}, {Jonker}, {Jorissen}, {Julbe}, {Karampelas},
  {Kochoska}, {Kohley}, {Kolenberg}, {Kontizas}, {Koposov}, {Kordopatis},
  {Koubsky}, {Kowalczyk}, {Krone-Martins}, {Kudryashova}, {Kull}, {Bachchan},
  {Lacoste-Seris}, {Lanza}, {Lavigne}, {Le Poncin-Lafitte}, {Lebreton},
  {Lebzelter}, {Leccia}, {Leclerc}, {Lecoeur-Taibi}, {Lemaitre}, {Lenhardt},
  {Leroux}, {Liao}, {Licata}, {Lindstr{\o}m}, {Lister}, {Livanou}, {Lobel},
  {L{\"o}ffler}, {L{\'o}pez}, {Lopez-Lozano}, {Lorenz}, {Loureiro},
  {MacDonald}, {Magalh{\~a}es Fernandes}, {Managau}, {Mann}, {Mantelet},
  {Marchal}, {Marchant}, {Marconi}, {Marie}, {Marinoni}, {Marrese},
  {Marschalk{\'o}}, {Marshall}, {Mart{\'\i}n-Fleitas}, {Martino}, {Mary},
  {Matijevi{\v{c}}}, {Mazeh}, {McMillan}, {Messina}, {Mestre}, {Michalik},
  {Millar}, {Miranda}, {Molina}, {Molinaro}, {Molinaro}, {Moln{\'a}r},
  {Moniez}, {Montegriffo}, {Monteiro}, {Mor}, {Mora}, {Morbidelli}, {Morel},
  {Morgenthaler}, {Morley}, {Morris}, {Mulone}, {Muraveva}, {Musella},
  {Narbonne}, {Nelemans}, {Nicastro}, {Noval}, {Ord{\'e}novic},
  {Ordieres-Mer{\'e}}, {Osborne}, {Pagani}, {Pagano}, {Pailler}, {Palacin},
  {Palaversa}, {Parsons}, {Paulsen}, {Pecoraro}, {Pedrosa}, {Pentik{\"a}inen},
  {Pereira}, {Pichon}, {Piersimoni}, {Pineau}, {Plachy}, {Plum}, {Poujoulet},
  {Pr{\v{s}}a}, {Pulone}, {Ragaini}, {Rago}, {Rambaux}, {Ramos-Lerate},
  {Ranalli}, {Rauw}, {Read}, {Regibo}, {Renk}, {Reyl{\'e}}, {Ribeiro},
  {Rimoldini}, {Ripepi}, {Riva}, {Rixon}, {Roelens}, {Romero-G{\'o}mez},
  {Rowell}, {Royer}, {Rudolph}, {Ruiz-Dern}, {Sadowski}, {Sagrist{\`a}
  Sell{\'e}s}, {Sahlmann}, {Salgado}, {Salguero}, {Sarasso}, {Savietto},
  {Schnorhk}, {Schultheis}, {Sciacca}, {Segol}, {Segovia}, {Segransan},
  {Serpell}, {Shih}, {Smareglia}, {Smart}, {Smith}, {Solano}, {Solitro},
  {Sordo}, {Soria Nieto}, {Souchay}, {Spagna}, {Spoto}, {Stampa}, {Steele},
  {Steidelm{\"u}ller}, {Stephenson}, {Stoev}, {Suess}, {S{\"u}veges}, {Surdej},
  {Szabados}, {Szegedi-Elek}, {Tapiador}, {Taris}, {Tauran}, {Taylor},
  {Teixeira}, {Terrett}, {Tingley}, {Trager}, {Turon}, {Ulla}, {Utrilla},
  {Valentini}, {van Elteren}, {Van Hemelryck}, {van Leeuwen}, {Varadi},
  {Vecchiato}, {Veljanoski}, {Via}, {Vicente}, {Vogt}, {Voss}, {Votruba},
  {Voutsinas}, {Walmsley}, {Weiler}, {Weingrill}, {Werner}, {Wevers},
  {Whitehead}, {Wyrzykowski}, {Yoldas}, {{\v{Z}}erjal}, {Zucker}, {Zurbach},
  {Zwitter}, {Alecu}, {Allen}, {Allende Prieto}, {Amorim},
  {Anglada-Escud{\'e}}, {Arsenijevic}, {Azaz}, {Balm}, {Beck}, {Bernstein},
  {Bigot}, {Bijaoui}, {Blasco}, {Bonfigli}, {Bono}, {Boudreault}, {Bressan},
  {Brown}, {Brunet}, {Bunclark}, {Buonanno}, {Butkevich}, {Carret}, {Carrion},
  {Chemin}, {Ch{\'e}reau}, {Corcione}, {Darmigny}, {de Boer}, {de Teodoro}, {de
  Zeeuw}, {Delle Luche}, {Domingues}, {Dubath}, {Fodor}, {Fr{\'e}zouls},
  {Fries}, {Fustes}, {Fyfe}, {Gallardo}, {Gallegos}, {Gardiol}, {Gebran},
  {Gomboc}, {G{\'o}mez}, {Grux}, {Gueguen}, {Heyrovsky}, {Hoar}, {Iannicola},
  {Isasi Parache}, {Janotto}, {Joliet}, {Jonckheere}, {Keil}, {Kim},
  {Klagyivik}, {Klar}, {Knude}, {Kochukhov}, {Kolka}, {Kos}, {Kutka}, {Lainey},
  {LeBouquin}, {Liu}, {Loreggia}, {Makarov}, {Marseille}, {Martayan},
  {Martinez-Rubi}, {Massart}, {Meynadier}, {Mignot}, {Munari}, {Nguyen},
  {Nordlander}, {Ocvirk}, {O'Flaherty}, {Olias Sanz}, {Ortiz}, {Osorio},
  {Oszkiewicz}, {Ouzounis}, {Palmer}, {Park}, {Pasquato}, {Peltzer}, {Peralta},
  {P{\'e}turaud}, {Pieniluoma}, {Pigozzi}, {Poels}, {Prat}, {Prod'homme},
  {Raison}, {Rebordao}, {Risquez}, {Rocca-Volmerange}, {Rosen}, {Ruiz-Fuertes},
  {Russo}, {Sembay}, {Serraller Vizcaino}, {Short}, {Siebert}, {Silva},
  {Sinachopoulos}, {Slezak}, {Soffel}, {Sosnowska}, {Strai{\v{z}}ys}, {ter
  Linden}, {Terrell}, {Theil}, {Tiede}, {Troisi}, {Tsalmantza}, {Tur},
  {Vaccari}, {Vachier}, {Valles}, {Van Hamme}, {Veltz}, {Virtanen}, {Wallut},
  {Wichmann}, {Wilkinson}, {Ziaeepour}, \& {Zschocke}}]{2016A&A...595A...1G}
{Gaia Collaboration}, {Prusti}, T., {de Bruijne}, J.~H.~J., {et~al.} 2016,
  \aap, 595, A1, \dodoi{10.1051/0004-6361/201629272}

\bibitem[{Gelman \& Rubin(1992)}]{Gelman1992}
Gelman, A., \& Rubin, D.~B. 1992, Statist. Sci., 7, 457,
  \dodoi{10.1214/ss/1177011136}

\bibitem[{{Ginzburg} {et~al.}(2018){Ginzburg}, {Schlichting}, \&
  {Sari}}]{Ginzburg2018}
{Ginzburg}, S., {Schlichting}, H.~E., \& {Sari}, R. 2018, \mnras, 476, 759,
  \dodoi{10.1093/mnras/sty290}

\bibitem[{{Grimm} {et~al.}(2018){Grimm}, {Demory}, {Gillon}, {Dorn}, {Agol},
  {Burdanov}, {Delrez}, {Sestovic}, {Triaud}, {Turbet}, {Bolmont}, {Caldas},
  {Wit}, {Jehin}, {Leconte}, {Raymond}, {Grootel}, {Burgasser}, {Carey},
  {Fabrycky}, {Heng}, {Hernandez}, {Ingalls}, {Lederer}, {Selsis}, \&
  {Queloz}}]{GrimmEtAl2018}
{Grimm}, S.~L., {Demory}, B.-O., {Gillon}, M., {et~al.} 2018, \aap, 613, A68,
  \dodoi{10.1051/0004-6361/201732233}

\bibitem[{{Hadden} \& {Lithwick}(2014)}]{HaddenLithwick2014}
{Hadden}, S., \& {Lithwick}, Y. 2014, \apj, 787, 80,
  \dodoi{10.1088/0004-637X/787/1/80}

\bibitem[{{Hadden} \& {Lithwick}(2016)}]{HaddenLithwick2016}
---. 2016, \apj, 828, 44, \dodoi{10.3847/0004-637X/828/1/44}

\bibitem[{{Hadden} \& {Lithwick}(2017)}]{HaddenLithwick2017}
---. 2017, \aj, 154, 5, \dodoi{10.3847/1538-3881/aa71ef}

\bibitem[{{He} {et~al.}(2020){He}, {Ford}, {Ragozzine}, \& {Carrera}}]{He2020}
{He}, M.~Y., {Ford}, E.~B., {Ragozzine}, D., \& {Carrera}, D. 2020, \aj, 160,
  276, \dodoi{10.3847/1538-3881/abba18}

\bibitem[{{Holczer} {et~al.}(2016){Holczer}, {Mazeh}, {Nachmani}, {Jontof-
  Hutter}, {Ford}, {Fabrycky}, {Ragozzine}, {Kane}, \&
  {Steffen}}]{HolczerEtAl2016}
{Holczer}, T., {Mazeh}, T., {Nachmani}, G., {et~al.} 2016, The Astrophysical
  Journal Supplement Series, 225, 9, \dodoi{10.3847/0067-0049/225/1/9}

\bibitem[{{Holman} {et~al.}(2010){Holman}, {Fabrycky}, {Ragozzine}, {Ford},
  {Steffen}, {Welsh}, {Lissauer}, {Latham}, {Marcy}, {Walkowicz}, {Batalha},
  {Jenkins}, {Rowe}, {Cochran}, {Fressin}, {Torres}, {Buchhave}, {Sasselov},
  {Borucki}, {Koch}, {Basri}, {Brown}, {Caldwell}, {Charbonneau}, {Dunham},
  {Gautier}, {Geary}, {Gilliland}, {Haas}, {Howell}, {Ciardi}, {Endl},
  {Fischer}, {F{\"u}r{\'e}sz}, {Hartman}, {Isaacson}, {Johnson}, {MacQueen},
  {Moorhead}, {Morehead}, \& {Orosz}}]{Holman2010}
{Holman}, M.~J., {Fabrycky}, D.~C., {Ragozzine}, D., {et~al.} 2010, Science,
  330, 51, \dodoi{10.1126/science.1195778}

\bibitem[{{Jontof-Hutter} {et~al.}(2015){Jontof-Hutter}, {Rowe}, {Lissauer},
  {Fabrycky}, \& {Ford}}]{JontofHutter2015}
{Jontof-Hutter}, D., {Rowe}, J.~F., {Lissauer}, J.~J., {Fabrycky}, D.~C., \&
  {Ford}, E.~B. 2015, \nat, 522, 321, \dodoi{10.1038/nature14494}

\bibitem[{{Jontof-Hutter} {et~al.}(2021){Jontof-Hutter}, {Wolfgang}, {Ford},
  {Lissauer}, {Fabrycky}, \& {Rowe}}]{JontofHutter2021}
{Jontof-Hutter}, D., {Wolfgang}, A., {Ford}, E.~B., {et~al.} 2021, \aj, 161,
  246, \dodoi{10.3847/1538-3881/abd93f}

\bibitem[{{Jontof-Hutter} {et~al.}(2016){Jontof-Hutter}, {Ford}, {Rowe},
  {Lissauer}, {Fabrycky}, {Van Laerhoven}, {Agol}, {Deck}, {Holczer}, \&
  {Mazeh}}]{JontofHutter2016}
{Jontof-Hutter}, D., {Ford}, E.~B., {Rowe}, J.~F., {et~al.} 2016, \apj, 820,
  39, \dodoi{10.3847/0004-637X/820/1/39}

\bibitem[{{Judkovsky} {et~al.}(2020){Judkovsky}, {Ofir}, \&
  {Aharonson}}]{Judkovsky2020}
{Judkovsky}, Y., {Ofir}, A., \& {Aharonson}, O. 2020, \aj, 160, 195,
  \dodoi{10.3847/1538-3881/abb406}

\bibitem[{{Judkovsky} {et~al.}(submitted){Judkovsky}, {Ofir}, \&
  {Aharonson}}]{Judkovsky2021a}
---. submitted, \aj

\bibitem[{{Kipping}(2010{\natexlab{a}})}]{Kipping2010a}
{Kipping}, D.~M. 2010{\natexlab{a}}, \mnras, 408, 1758,
  \dodoi{10.1111/j.1365-2966.2010.17242.x}

\bibitem[{{Kipping}(2010{\natexlab{b}})}]{Kipping2010}
---. 2010{\natexlab{b}}, \mnras, 407, 301,
  \dodoi{10.1111/j.1365-2966.2010.16894.x}

\bibitem[{{Kipping} {et~al.}(2014){Kipping}, {Nesvorn{\'y}}, {Buchhave},
  {Hartman}, {Bakos}, \& {Schmitt}}]{Kipping2014}
{Kipping}, D.~M., {Nesvorn{\'y}}, D., {Buchhave}, L.~A., {et~al.} 2014, \apj,
  784, 28, \dodoi{10.1088/0004-637X/784/1/28}

\bibitem[{{Lai} \& {Pu}(2017)}]{LaiPu2017}
{Lai}, D., \& {Pu}, B. 2017, \aj, 153, 42, \dodoi{10.3847/1538-3881/153/1/42}

\bibitem[{{Laskar} \& {Petit}(2017)}]{Laskar2017}
{Laskar}, J., \& {Petit}, A.~C. 2017, \aap, 605, A72,
  \dodoi{10.1051/0004-6361/201630022}

\bibitem[{{Latham} {et~al.}(1989){Latham}, {Mazeh}, {Stefanik}, {Mayor}, \&
  {Burki}}]{Latham1989}
{Latham}, D.~W., {Mazeh}, T., {Stefanik}, R.~P., {Mayor}, M., \& {Burki}, G.
  1989, \nat, 339, 38, \dodoi{10.1038/339038a0}

\bibitem[{{Linial} {et~al.}(2018){Linial}, {Gilbaum}, \&
  {Sari}}]{LinialGilbaumSari2018}
{Linial}, I., {Gilbaum}, S., \& {Sari}, R. 2018, \apj, 860, 16,
  \dodoi{10.3847/1538-4357/aac21b}

\bibitem[{{Lissauer} {et~al.}(2011){Lissauer}, {Ragozzine}, {Fabrycky},
  {Steffen}, {Ford}, {Jenkins}, {Shporer}, {Holman}, {Rowe}, {Quintana},
  {Batalha}, {Borucki}, {Bryson}, {Caldwell}, {Carter}, {Ciardi}, {Dunham},
  {Fortney}, {Gautier}, {Howell}, {Koch}, {Latham}, {Marcy}, {Morehead}, \&
  {Sasselov}}]{Lissauer2011}
{Lissauer}, J.~J., {Ragozzine}, D., {Fabrycky}, D.~C., {et~al.} 2011, The
  Astrophysical Journal Supplement Series, 197, 8,
  \dodoi{10.1088/0067-0049/197/1/8}

\bibitem[{{Lithwick} \& {Wu}(2012)}]{LithwickWu2012a}
{Lithwick}, Y., \& {Wu}, Y. 2012, \apj, 756, L11,
  \dodoi{10.1088/2041-8205/756/1/L11}

\bibitem[{{Lithwick} {et~al.}(2012){Lithwick}, {Xie}, \&
  {Wu}}]{LithwickXieWu2012}
{Lithwick}, Y., {Xie}, J., \& {Wu}, Y. 2012, \apj, 761, 122,
  \dodoi{10.1088/0004-637X/761/2/122}

\bibitem[{{Loyd} {et~al.}(2020){Loyd}, {Shkolnik}, {Schneider},
  {Richey-Yowell}, {Barman}, {Peacock}, \& {Pagano}}]{Loyd2020}
{Loyd}, R.~O.~P., {Shkolnik}, E.~L., {Schneider}, A.~C., {et~al.} 2020, \apj,
  890, 23, \dodoi{10.3847/1538-4357/ab6605}

\bibitem[{{Mandel} \& {Agol}(2002)}]{MandelAgol2002}
{Mandel}, K., \& {Agol}, E. 2002, \apj, 580, L171, \dodoi{10.1086/345520}

\bibitem[{{Marcy} {et~al.}(2014){Marcy}, {Isaacson}, {Howard}, {Rowe},
  {Jenkins}, {Bryson}, {Latham}, {Howell}, {Gautier}, {Batalha}, {Rogers},
  {Ciardi}, {Fischer}, {Gilliland}, {Kjeldsen}, {Christensen-Dalsgaard},
  {Huber}, {Chaplin}, {Basu}, {Buchhave}, {Quinn}, {Borucki}, {Koch}, {Hunter},
  {Caldwell}, {Van Cleve}, {Kolbl}, {Weiss}, {Petigura}, {Seager}, {Morton},
  {Johnson}, {Ballard}, {Burke}, {Cochran}, {Endl}, {MacQueen}, {Everett},
  {Lissauer}, {Ford}, {Torres}, {Fressin}, {Brown}, {Steffen}, {Charbonneau},
  {Basri}, {Sasselov}, {Winn}, {Sanchis-Ojeda}, {Christiansen}, {Adams},
  {Henze}, {Dupree}, {Fabrycky}, {Fortney}, {Tarter}, {Holman}, {Tenenbaum},
  {Shporer}, {Lucas}, {Welsh}, {Orosz}, {Bedding}, {Campante}, {Davies},
  {Elsworth}, {Handberg}, {Hekker}, {Karoff}, {Kawaler}, {Lund}, {Lundkvist},
  {Metcalfe}, {Miglio}, {Silva Aguirre}, {Stello}, {White}, {Boss}, {Devore},
  {Gould}, {Prsa}, {Agol}, {Barclay}, {Coughlin}, {Brugamyer}, {Mullally},
  {Quintana}, {Still}, {Thompson}, {Morrison}, {Twicken}, {D{\'e}sert},
  {Carter}, {Crepp}, {H{\'e}brard}, {Santerne}, {Moutou}, {Sobeck}, {Hudgins},
  {Haas}, {Robertson}, {Lillo-Box}, \& {Barrado}}]{Marcy2014}
{Marcy}, G.~W., {Isaacson}, H., {Howard}, A.~W., {et~al.} 2014, \apjs, 210, 20,
  \dodoi{10.1088/0067-0049/210/2/20}

\bibitem[{{Masuda}(2014)}]{Masuda2014}
{Masuda}, K. 2014, \apj, 783, 53, \dodoi{10.1088/0004-637X/783/1/53}

\bibitem[{{Masuda} {et~al.}(2020){Masuda}, {Winn}, \& {Kawahara}}]{Masuda2020}
{Masuda}, K., {Winn}, J.~N., \& {Kawahara}, H. 2020, \aj, 159, 38,
  \dodoi{10.3847/1538-3881/ab5c1d}

\bibitem[{{Mayor} \& {Queloz}(1995)}]{Mayor1995}
{Mayor}, M., \& {Queloz}, D. 1995, \nat, 378, 355, \dodoi{10.1038/378355a0}

\bibitem[{{Mazeh} \& {Faigler}(2010)}]{Mazeh2010}
{Mazeh}, T., \& {Faigler}, S. 2010, \aap, 521, L59,
  \dodoi{10.1051/0004-6361/201015550}

\bibitem[{{Mazeh} {et~al.}(2016){Mazeh}, {Holczer}, \& {Faigler}}]{Mazeh2016}
{Mazeh}, T., {Holczer}, T., \& {Faigler}, S. 2016, \aap, 589, A75,
  \dodoi{10.1051/0004-6361/201528065}

\bibitem[{{Millholland} \& {Laughlin}(2019)}]{Millholland2019}
{Millholland}, S., \& {Laughlin}, G. 2019, Nature Astronomy, 3, 424,
  \dodoi{10.1038/s41550-019-0701-7}

\bibitem[{{Millholland} {et~al.}(2021){Millholland}, {He}, {Ford}, {Ragozzine},
  {Fabrycky}, \& {Winn}}]{Millholland2021}
{Millholland}, S.~C., {He}, M.~Y., {Ford}, E.~B., {et~al.} 2021, arXiv
  e-prints, arXiv:2106.15589.
\newblock \doarXiv{2106.15589}

\bibitem[{{Mills} \& {Fabrycky}(2017)}]{MillsFabrycky2017}
{Mills}, S.~M., \& {Fabrycky}, D.~C. 2017, \aj, 153, 45,
  \dodoi{10.3847/1538-3881/153/1/45}

\bibitem[{{Mills} {et~al.}(2019){Mills}, {Howard}, {Weiss}, {Steffen},
  {Isaacson}, {Fulton}, {Petigura}, {Kosiarek}, {Hirsch}, \&
  {Boisvert}}]{MillsEtAl2019}
{Mills}, S.~M., {Howard}, A.~W., {Weiss}, L.~M., {et~al.} 2019, \aj, 157, 145,
  \dodoi{10.3847/1538-3881/ab0899}

\bibitem[{{Murray} \& {Dermott}(1999)}]{SSD1999}
{Murray}, C.~D., \& {Dermott}, S.~F. 1999, {Solar system dynamics}

\bibitem[{{Ofir} \& {Dreizler}(2013)}]{OfirDreizler2013}
{Ofir}, A., \& {Dreizler}, S. 2013, \aap, 555, A58,
  \dodoi{10.1051/0004-6361/201219877}

\bibitem[{{Ofir} {et~al.}(2014){Ofir}, {Dreizler}, {Zechmeister}, \&
  {Husser}}]{Ofir2014}
{Ofir}, A., {Dreizler}, S., {Zechmeister}, M., \& {Husser}, T.-O. 2014, \aap,
  561, A103, \dodoi{10.1051/0004-6361/201220935}

\bibitem[{{Ofir} {et~al.}(2018){Ofir}, {Xie}, {Jiang}, {Sari}, \&
  {Aharonson}}]{OfirEtAl2018}
{Ofir}, A., {Xie}, J.-W., {Jiang}, C.-F., {Sari}, R., \& {Aharonson}, O. 2018,
  The Astrophysical Journal Supplement Series, 234, 9,
  \dodoi{10.3847/1538-4365/aa9f2b}

\bibitem[{{Owen} \& {Wu}(2017)}]{Owen2017}
{Owen}, J.~E., \& {Wu}, Y. 2017, \apj, 847, 29,
  \dodoi{10.3847/1538-4357/aa890a}

\bibitem[{Pastore \& Calcagnì(2019)}]{Pastore2019}
Pastore, M., \& Calcagnì, A. 2019, Frontiers in Psychology, 10, 1089,
  \dodoi{10.3389/fpsyg.2019.01089}

\bibitem[{{Ragozzine} \& {Holman}(2010)}]{RagozzineHolman2010}
{Ragozzine}, D., \& {Holman}, M.~J. 2010, ArXiv e-prints, arXiv:1006.3727.
\newblock \doarXiv{1006.3727}

\bibitem[{{Rauer} {et~al.}(2014){Rauer}, {Catala}, {Aerts}, {Appourchaux},
  {Benz}, {Brandeker}, {Christensen-Dalsgaard}, {Deleuil}, {Gizon}, {Goupil},
  {G{\"u}del}, {Janot-Pacheco}, {Mas-Hesse}, {Pagano}, {Piotto}, {Pollacco},
  {Santos}, {Smith}, {Su{\'a}rez}, {Szab{\'o}}, {Udry}, {Adibekyan}, {Alibert},
  {Almenara}, {Amaro-Seoane}, {Eiff}, {Asplund}, {Antonello}, {Barnes},
  {Baudin}, {Belkacem}, {Bergemann}, {Bihain}, {Birch}, {Bonfils}, {Boisse},
  {Bonomo}, {Borsa}, {Brand{\~a}o}, {Brocato}, {Brun}, {Burleigh}, {Burston},
  {Cabrera}, {Cassisi}, {Chaplin}, {Charpinet}, {Chiappini}, {Church},
  {Csizmadia}, {Cunha}, {Damasso}, {Davies}, {Deeg}, {D{\'\i}az}, {Dreizler},
  {Dreyer}, {Eggenberger}, {Ehrenreich}, {Eigm{\"u}ller}, {Erikson}, {Farmer},
  {Feltzing}, {de Oliveira Fialho}, {Figueira}, {Forveille}, {Fridlund},
  {Garc{\'\i}a}, {Giommi}, {Giuffrida}, {Godolt}, {Gomes da Silva}, {Granzer},
  {Grenfell}, {Grotsch-Noels}, {G{\"u}nther}, {Haswell}, {Hatzes},
  {H{\'e}brard}, {Hekker}, {Helled}, {Heng}, {Jenkins}, {Johansen},
  {Khodachenko}, {Kislyakova}, {Kley}, {Kolb}, {Krivova}, {Kupka}, {Lammer},
  {Lanza}, {Lebreton}, {Magrin}, {Marcos-Arenal}, {Marrese}, {Marques},
  {Martins}, {Mathis}, {Mathur}, {Messina}, {Miglio}, {Montalban}, {Montalto},
  {Monteiro}, {Moradi}, {Moravveji}, {Mordasini}, {Morel}, {Mortier},
  {Nascimbeni}, {Nelson}, {Nielsen}, {Noack}, {Norton}, {Ofir}, {Oshagh},
  {Ouazzani}, {P{\'a}pics}, {Parro}, {Petit}, {Plez}, {Poretti}, {Quirrenbach},
  {Ragazzoni}, {Raimondo}, {Rainer}, {Reese}, {Redmer}, {Reffert},
  {Rojas-Ayala}, {Roxburgh}, {Salmon}, {Santerne}, {Schneider}, {Schou},
  {Schuh}, {Schunker}, {Silva-Valio}, {Silvotti}, {Skillen}, {Snellen}, {Sohl},
  {Sousa}, {Sozzetti}, {Stello}, {Strassmeier}, {{\v{S}}vanda}, {Szab{\'o}},
  {Tkachenko}, {Valencia}, {Van Grootel}, {Vauclair}, {Ventura}, {Wagner},
  {Walton}, {Weingrill}, {Werner}, {Wheatley}, \& {Zwintz}}]{Rauer2014}
{Rauer}, H., {Catala}, C., {Aerts}, C., {et~al.} 2014, Experimental Astronomy,
  38, 249, \dodoi{10.1007/s10686-014-9383-4}

\bibitem[{{Ricker} {et~al.}(2010){Ricker}, {Latham}, {Vanderspek}, {Ennico},
  {Bakos}, {Brown}, {Burgasser}, {Charbonneau}, {Clampin}, {Deming}, {Doty},
  {Dunham}, {Elliot}, {Holman}, {Ida}, {Jenkins}, {Jernigan}, {Kawai},
  {Laughlin}, {Lissauer}, {Martel}, {Sasselov}, {Schingler}, {Seager},
  {Torres}, {Udry}, {Villasenor}, {Winn}, \& {Worden}}]{RickerEtAl2010}
{Ricker}, G.~R., {Latham}, D.~W., {Vanderspek}, R.~K., {et~al.} 2010, in
  American Astronomical Society Meeting Abstracts \#215, Vol. 215, 450.06

\bibitem[{{Rivera} {et~al.}(2005){Rivera}, {Lissauer}, {Butler}, {Marcy},
  {Vogt}, {Fischer}, {Brown}, {Laughlin}, \& {Henry}}]{RiveraEtAl2005}
{Rivera}, E.~J., {Lissauer}, J.~J., {Butler}, R.~P., {et~al.} 2005, \apj, 634,
  625, \dodoi{10.1086/491669}

\bibitem[{{Rogers}(2015)}]{Rogers2015}
{Rogers}, L.~A. 2015, \apj, 801, 41, \dodoi{10.1088/0004-637X/801/1/41}

\bibitem[{Schwarz(1978)}]{Schwartz1978}
Schwarz, G. 1978, The Annals of Statistics, 6, 461 ,
  \dodoi{10.1214/aos/1176344136}

\bibitem[{{Shahaf} {et~al.}(2021){Shahaf}, {Mazeh}, {Zucker}, \&
  {Fabrycky}}]{Shahaf2021}
{Shahaf}, S., {Mazeh}, T., {Zucker}, S., \& {Fabrycky}, D. 2021, \mnras, 505,
  1293, \dodoi{10.1093/mnras/stab1359}

\bibitem[{{Smith} {et~al.}(2012){Smith}, {Stumpe}, {Van Cleve}, {Jenkins},
  {Barclay}, {Fanelli}, {Girouard}, {Kolodziejczak}, {McCauliff}, {Morris}, \&
  {Twicken}}]{Smith2012}
{Smith}, J.~C., {Stumpe}, M.~C., {Van Cleve}, J.~E., {et~al.} 2012, \pasp, 124,
  1000, \dodoi{10.1086/667697}

\bibitem[{Sorokhtin {et~al.}(2011)Sorokhtin, Chilingar, \&
  Sorokhtin}]{SOROKHTIN201113}
Sorokhtin, O., Chilingar, G., \& Sorokhtin, N. 2011, in Developments in Earth
  and Environmental Sciences, Vol.~10, Evolution of Earth and its Climate:
  Birth, Life and Death of Earth, ed. O.~Sorokhtin, G.~Chilingarian, \&
  N.~Sorokhtin (Elsevier), 13--60,
  \dodoi{https://doi.org/10.1016/B978-0-444-53757-7.00002-7}

\bibitem[{{Steffen} {et~al.}(2012){Steffen}, {Fabrycky}, {Ford}, {Carter},
  {D{\'e}sert}, {Fressin}, {Holman}, {Lissauer}, {Moorhead}, {Rowe},
  {Ragozzine}, {Welsh}, {Batalha}, {Borucki}, {Buchhave}, {Bryson}, {Caldwell},
  {Charbonneau}, {Ciardi}, {Cochran}, {Endl}, {Everett}, {Gautier},
  {Gilliland}, {Girouard}, {Jenkins}, {Horch}, {Howell}, {Isaacson}, {Klaus},
  {Koch}, {Latham}, {Li}, {Lucas}, {MacQueen}, {Marcy}, {McCauliff}, {Middour},
  {Morris}, {Mullally}, {Quinn}, {Quintana}, {Shporer}, {Still}, {Tenenbaum},
  {Thompson}, {Twicken}, \& {Van Cleve}}]{Steffen2012}
{Steffen}, J.~H., {Fabrycky}, D.~C., {Ford}, E.~B., {et~al.} 2012, \mnras, 421,
  2342, \dodoi{10.1111/j.1365-2966.2012.20467.x}

\bibitem[{{Stumpe} {et~al.}(2014){Stumpe}, {Smith}, {Catanzarite}, {Van Cleve},
  {Jenkins}, {Twicken}, \& {Girouard}}]{Stumpe2014}
{Stumpe}, M.~C., {Smith}, J.~C., {Catanzarite}, J.~H., {et~al.} 2014, \pasp,
  126, 100, \dodoi{10.1086/674989}

\bibitem[{{Stumpe} {et~al.}(2012){Stumpe}, {Smith}, {Van Cleve}, {Twicken},
  {Barclay}, {Fanelli}, {Girouard}, {Jenkins}, {Kolodziejczak}, {McCauliff}, \&
  {Morris}}]{Stumpe2012}
{Stumpe}, M.~C., {Smith}, J.~C., {Van Cleve}, J.~E., {et~al.} 2012, \pasp, 124,
  985, \dodoi{10.1086/667698}

\bibitem[{ter Braak \& Vrugt(2008)}]{BraakVrugt2008}
ter Braak, C. J.~F., \& Vrugt, J.~A. 2008, Statistics and Computing, 18, 435

\bibitem[{{Vissapragada} {et~al.}(2020){Vissapragada}, {Jontof-Hutter},
  {Shporer}, {Knutson}, {Liu}, {Thorngren}, {Lee}, {Chachan}, {Mawet},
  {Millar-Blanchaer}, {Nilsson}, {Tinyanont}, {Vasisht}, \&
  {Wright}}]{Vissapragada2020}
{Vissapragada}, S., {Jontof-Hutter}, D., {Shporer}, A., {et~al.} 2020, \aj,
  159, 108, \dodoi{10.3847/1538-3881/ab65c8}

\bibitem[{Vrieze(2012)}]{Vrieze2012}
Vrieze, S.~I. 2012, Psychological Methods, 17, 228, \dodoi{10.1037/a0027127}

\bibitem[{{Weiss} {et~al.}(2017){Weiss}, {Deck}, {Sinukoff}, {Petigura},
  {Agol}, {Lee}, {Becker}, {Howard}, {Isaacson}, {Crossfield}, {Fulton},
  {Hirsch}, \& {Benneke}}]{Weiss2018}
{Weiss}, L.~M., {Deck}, K.~M., {Sinukoff}, E., {et~al.} 2017, \aj, 153, 265,
  \dodoi{10.3847/1538-3881/aa6c29}

\bibitem[{{Weiss} {et~al.}(2018){Weiss}, {Marcy}, {Petigura}, {Fulton},
  {Howard}, {Winn}, {Isaacson}, {Morton}, {Hirsch}, {Sinukoff}, {Cumming},
  {Hebb}, \& {Cargile}}]{Weiss2018a}
{Weiss}, L.~M., {Marcy}, G.~W., {Petigura}, E.~A., {et~al.} 2018, \aj, 155, 48,
  \dodoi{10.3847/1538-3881/aa9ff6}

\bibitem[{{Wolszczan} \& {Frail}(1992)}]{Wolszczan1992}
{Wolszczan}, A., \& {Frail}, D.~A. 1992, \nat, 355, 145,
  \dodoi{10.1038/355145a0}

\bibitem[{{Xie}(2014)}]{Xie2014}
{Xie}, J.-W. 2014, \apjs, 210, 25, \dodoi{10.1088/0067-0049/210/2/25}

\bibitem[{{Xie} {et~al.}(2016){Xie}, {Dong}, {Zhu}, {Huber}, {Zheng}, {De Cat},
  {Fu}, {Liu}, {Luo}, \& {Wu}}]{XieEtAl2016}
{Xie}, J.-W., {Dong}, S., {Zhu}, Z., {et~al.} 2016, Proceedings of the National
  Academy of Science, 113, 11431, \dodoi{10.1073/pnas.1604692113}

\bibitem[{{Xu} \& {Fabrycky}(2019)}]{XuFabrycky2019}
{Xu}, W., \& {Fabrycky}, D. 2019, arXiv e-prints, arXiv:1904.02290.
\newblock \doarXiv{1904.02290}

\bibitem[{{Yoffe} {et~al.}(2021){Yoffe}, {Ofir}, \& {Aharonson}}]{Yoffe2021}
{Yoffe}, G., {Ofir}, A., \& {Aharonson}, O. 2021, \apj, 908, 114,
  \dodoi{10.3847/1538-4357/abc87a}

\end{thebibliography}
\bibliographystyle{aasjournal}


\end{document}